\documentclass[
  aps,               
  prx,               
  reprint,           
  superscriptaddress,
  nofootinbib,       
  floatfix
]{revtex4-2}

\usepackage[utf8]{inputenc}
\usepackage[T1]{fontenc}
\usepackage{macros}

\usepackage{graphicx}
\usepackage[table]{xcolor}
\definecolor{maroon}{RGB}{128, 0, 0}
\definecolor{ao}{RGB}{0, 128, 0}

\usepackage{amsmath}
\usepackage{amssymb}

\usepackage{booktabs}
\usepackage{threeparttable}
\usepackage{tabularx}
\usepackage{hhline}
\usepackage{rotating}
\newcolumntype{Y}{>{\centering\arraybackslash}X}

\usepackage[normalem]{ulem}
\usepackage{footmisc}

\usepackage{hyperref}

\newcommand{\up}[1]{{\color{red}($+$#1)}}
\newcommand{\down}[1]{{\color{blue}($-$#1)}}

\usepackage{orcidlink}

\newcommand{\rev}[1]{\textcolor{black}{#1}}

\begin{document}
\title{Catching Disguised Transients with ASTRANet: Anomaly-Aware Spectroscopic Classification and Conformal Calibration}

\author{Argyro Sasli\,\orcidlink{0000-0001-7357-0889}}
\email{asasli@umn.edu}
\affiliation{School of Physics and Astronomy, University of Minnesota, Minneapolis, MN 55455, USA}
\affiliation{NSF Institute on Accelerated AI Algorithms for Data-Driven Discovery (A3D3)}

\author{Maojie Xu\,\orcidlink{0009-0002-5097-3188}}
\affiliation{Department of Computer Science \& Engineering, University of Minnesota, Minneapolis, MN 55455, USA}
\affiliation{NSF Institute on Accelerated AI Algorithms for Data-Driven Discovery (A3D3)}

\author{Alexandra Junell\,\orcidlink{0000-0002-9380-7983}}
\affiliation{School of Physics and Astronomy, University of Minnesota, Minneapolis, MN 55455, USA}
\affiliation{NSF Institute on Accelerated AI Algorithms for Data-Driven Discovery (A3D3)}

\author{Hailey Markoff}
\affiliation{School of Physics and Astronomy, University of Minnesota, Minneapolis, MN 55455, USA}
\affiliation{NSF Institute on Accelerated AI Algorithms for Data-Driven Discovery (A3D3)}

\author{Avyukt Raghuvanshi\,\orcidlink{0009-0005-9422-0091}}
\affiliation{School of Physics and Astronomy, University of Minnesota, Minneapolis, MN 55455, USA}
\affiliation{NSF Institute on Accelerated AI Algorithms for Data-Driven Discovery (A3D3)}

\author{Felipe Fontinele Nunes\,\orcidlink{0000-0001-7129-1325}}
\affiliation{School of Physics and Astronomy, University of Minnesota, Minneapolis, MN 55455, USA}
\affiliation{NSF Institute on Accelerated AI Algorithms for Data-Driven Discovery (A3D3)}

\author{Theophile Jegou Du Laz\,\orcidlink{0009-0003-6181-4526}}
\affiliation{Division of Physics, Mathematics and Astronomy, California Institute of Technology, 1200 E California Blvd., Pasadena, CA 91125, USA}
\affiliation{NSF Institute on Accelerated AI Algorithms for Data-Driven Discovery (A3D3)}

\author{Jesper Sollerman\,\orcidlink{0000-0003-1546-6615}}
\affiliation{Department of Astronomy, Stockholm University, 10691 Stockholm, Sweden}

\author{Christoffer Fremling\,\orcidlink{0000-0002-4223-103X}}
\affiliation{Caltech Optical Observatories, California Institute of Technology, Pasadena, CA 91125, USA}
\affiliation{Division of Physics, Mathematics and Astronomy, California Institute of Technology, Pasadena, CA 91125, USA}

\author{Drew Oldag\,\orcidlink{0000-0001-6984-8411}}
\affiliation{Dept. of Astronomy \& the DiRAC Institute, University of Washington, Box 351580, Seattle, WA 98195, USA}

\author{Antoine Le Calloch\,\orcidlink{0009-0009-7000-8343}}
\affiliation{School of Physics and Astronomy, University of Minnesota, Minneapolis, MN 55455, USA}
\affiliation{NSF Institute on Accelerated AI Algorithms for Data-Driven Discovery (A3D3)}

\author{Sushant Sharma Chaudhary\,\orcidlink{0000-0003-1314-4241}}
\affiliation{School of Physics and Astronomy, University of Minnesota, Minneapolis, MN 55455, USA}
\affiliation{NSF Institute on Accelerated AI Algorithms for Data-Driven Discovery (A3D3)}

\author{Sneha Maharjan\,\orcidlink{0000-0001-7922-8494}}
\affiliation{School of Physics and Astronomy, University of Minnesota, Minneapolis, MN 55455, USA}

\author{Maxine West\,\orcidlink{0009-0003-3171-3118}}
\affiliation{Dept. of Astronomy \& the DiRAC Institute, University of Washington, Box 351580, Seattle, WA 98195, USA}

\author{Benny Border}
\affiliation{School of Physics and Astronomy, University of Minnesota, Minneapolis, MN 55455, USA}
\affiliation{NSF Institute on Accelerated AI Algorithms for Data-Driven Discovery (A3D3)}

\author{Nabeel Rehemtulla\,\orcidlink{0000-0002-5683-2389}}
\affiliation{Department of Physics and Astronomy, Northwestern University, 2145 Sheridan Road, Evanston, IL 60208, USA}
\affiliation{Center for Interdisciplinary Exploration and Research in Astrophysics (CIERA), 1800 Sherman Ave., Evanston, IL 60201, USA}
\affiliation{NSF-Simons AI Institute for the Sky (SkAI), 172 E. Chestnut St., Chicago, IL 60611, USA}

\author{Richard Dekany\,\orcidlink{0000-0002-5884-7867}}
\affiliation{Caltech Optical Observatories, California Institute of Technology, Pasadena, CA 91125, USA}

\author{Joahan Castaneda Jaimes\,\orcidlink{0000-0002-0987-3372}}
\affiliation{Division of Physics, Mathematics and Astronomy, California Institute of Technology, 1200 E. California Blvd., Pasadena, CA 91125, USA}

\author{Russ R. Laher\,\orcidlink{0000-0003-2451-5482}}
\affiliation{IPAC, California Institute of Technology, 1200 E. California Blvd., Pasadena, CA 91125, USA}

\author{Reed Riddle\,\orcidlink{0000-0002-0387-370X}}
\affiliation{Caltech Optical Observatories, California Institute of Technology, Pasadena, CA 91125, USA}

\author{Mansi M. Kasliwal\,\orcidlink{0000-0002-5619-4938}}
\affiliation{Division of Physics, Mathematics, and Astronomy, California Institute of Technology, Pasadena, CA 91125, USA}

\author{Matthew J. Graham\,\orcidlink{0000-0002-3168-0139}}
\affiliation{Cahill Center for Astrophysics, California Institute of Technology, Pasadena, CA 91125, USA}
\affiliation{NSF Institute on Accelerated AI Algorithms for Data-Driven Discovery (A3D3)}

\author{Ashish A. Mahabal\,\orcidlink{0000-0003-2242-0244}}
\affiliation{Division of Physics, Mathematics and Astronomy, California Institute of Technology, Pasadena, CA 91125, USA}
\affiliation{Center for Data Driven Discovery, California Institute of Technology, Pasadena, CA 91125, USA}

\author{Michael W. Coughlin\,\orcidlink{0000-0002-8262-2924}}
\affiliation{School of Physics and Astronomy, University of Minnesota, Minneapolis, MN 55455, USA}
\affiliation{NSF Institute on Accelerated AI Algorithms for Data-Driven Discovery (A3D3)}

\begin{abstract}
Time-domain surveys discover thousands of transients per year, but the spectroscopic identification of rare and physically peculiar objects remains rate-limited by closed-set classifiers that confidently assign every input to a known class -- including spectra that genuinely belong to no known class.  We present the \texttt{ASTRANet} framework, a confidence-aware infrastructure for spectroscopic transient classification built around three coupled modules: a hierarchical spectral classifier that operates directly on observer-frame spectra without requiring host-galaxy redshift or spectral phase as inputs; an anomaly detection layer (\texttt{ASTRANet-Sentinel}) that non-linearly combines $16$ embedding-space anomaly scores spanning four physically motivated families; and a conformal uncertainty quantification layer (\texttt{ASTRANet-CP}). We validate the framework on a held-out evaluation set of $289$ rare and out-of-taxonomy transients spanning $11$ classes deliberately excluded from training, chosen to span the full physical diversity of the rare-anomaly population: AGN-related outliers, GRB-related events, gap transients, novae, and peculiar supernovae.  Through five astrophysically distinct failure modes of closed-set classifiers, we show that classifier-internal uncertainty and embedding-based anomaly detection are structurally complementary axes of confidence rather than alternative implementations of the same estimator.  We further introduce AD-stratified Mondrian conformal prediction (AD-MCP) within \texttt{ASTRANet-CP}, achieving uniform conditional coverage across anomaly-score strata where vanilla Mondrian under-covers in the operational regime.  This establishes the methodological infrastructure for confidence-aware spectroscopic discovery in the Vera C.\ Rubin Observatory era.
\end{abstract}


\maketitle

 \section{Introduction}
\label{sec:introduction}

Time-domain surveys discover $\sim$$10^5$ transient candidates per year \citep[ZTF;][]{ZTF, ZTF2, ZTF3, 2019PASP..131a8001P, 2020PASP..132c8001D} heading to $\sim$$10^6$/year at LSST depth \citep{LSST2019}, of which only a tiny fraction can receive spectroscopic follow-up.

The bottleneck is not only in taking spectra — not even the SED Machine \citep{Blagorodnova_2018, 2019A&A...627A.115R} and similar rapid spectrographs have been able to close that gap, since even SEDM can observe only a small fraction of ZTF transients, far short of the LSST scale — but increasingly in \emph{deciding which spectra deserve deeper observations} and in \emph{recognising} when an observed spectrum represents something genuinely novel.  This second decision is the substrate of discovery: most peculiar transient classes -- Type Iax \citep{Foley2013}, Ca-rich \citep{Perets2010}, Fast Blue Optical Transients \citep[FBOTs;][]{Drout2014, Ho_2023}, Intermediate Luminosity Red Transients \citep[ILRTs;][]{Thompson2009} -- were identified not by template-matching but by recognizing that the spectra did not fit known templates well enough.  Deep classifiers \citep{Fremling2021SNIascore, sharma2025ccsnscoremultiinputdeeplearning, xuspectra, fortino2025} excel on common classes but are typically deployed \emph{closed-set}: every input is assigned to a training class with a softmax probability regardless of whether the input belongs to one.  The result is a specific dangerous failure mode -- high-confidence wrong answers \citep{Hendrycks2017, Guo2017Calibration, Hein2019Why} -- which the architectural features that make these classifiers accurate on common classes actively conceal.  Existing remedies in astronomy focus on photometric light curves \citep{Villar2021, Pruzhinskaya2019, DiMMAD2025, Gupta2025MCIF} and almost always deploy single-score pipelines, while uncertainty quantification (UQ) is treated as a separate axis from anomaly detection (AD).  We argue these axes must be co-equal outputs of a single confidence-aware framework, and that for spectroscopy specifically, no single anomaly-score family suffices because the failure modes are themselves physically diverse.  Throughout this work the catalog classifications -- Blazar, Ca-rich, ILRT, FBOT, etc.\ -- are treated as reference classifications; the failure modes we document are what closed-set models produce when shown these correctly-labelled spectra.

We present \texttt{ASTRANet}\footnote{\texttt{ASTRANet}: Anomaly-aware Spectroscopic TRAnsient Network; the stem also echoes the Greek \textit{ástra} (``stars'').}, a confidence-aware framework organised into three coupled modules.  A hierarchical spectral classifier operates on \emph{observer-frame} spectra, treating host-galaxy redshift and spectral phase as optional metadata injected via FiLM \citep{perez2018film} with validity bits and random masking during training -- never requiring either at inference, in contrast to prior multi-class deep spectroscopic classifiers that rest-frame their inputs and consume an estimated redshift (i.e., see \cite{Muthukrishna2019})(Muthukrishna et al. 2019; Xu et al. 2025). Binary classifiers such as SNIascore \citep{Fremling2021SNIascore} likewise operate without explicit redshift input, but ASTRANet is, to our knowledge, the first fine-grained, multi-class classifier to do so on observer-frame spectra. An anomaly-detection layer (\texttt{ASTRANet-Sentinel}) non-linearly combines $16$ embedding-space scores spanning uncertainty, distance, density, and hybrid families into a single cross-family AD score.  A conformal uncertainty layer (\texttt{ASTRANet-CP}) produces per-spectrum calibrated p-value, adaptive prediction set, per-class ensemble probability intervals, and aleatoric/epistemic decomposition in one forward pass, with distribution-free finite-sample calibration guarantees that remain valid under arbitrary distribution shift.

\paragraph{Two key findings shape the design.}
First, the failure modes of closed-set spectroscopic classifiers are not a uniform softmax-overconfidence phenomenon but separate into \emph{five physically distinct regimes} -- continuum-disguised AGN-class events, spectroscopic disguise within physical neighbours, continuum-dominated mis-classifications, legitimately ambiguous progenitor regimes, and far-out-of-distribution events.  Each regime manifests in a different anomaly-score signature; no single family covers all five.  The cross-family non-linear combiner is required not as engineering overhead but \emph{because the physics is diverse} (Sec.~\ref{sec:failure_modes}).  Second, classifier-internal uncertainty and embedding-based anomaly detection are \emph{structurally complementary} axes of confidence: the former dominates on in-distribution boundary errors, the latter on out-of-distribution disguise where the ensemble is confidently wrong.  They flag largely disjoint failure populations and should be deployed simultaneously, not interchangeably (Sec.~\ref{sec:test_error_uncertainty}).

\paragraph{What this enables.}
This combination unlocks four science capabilities that closed-set pipelines do not.  (i) \emph{Real-time spectroscopic triage on freshly discovered transients}: classification, anomaly detection, and uncertainty quantification all operate without host redshift or spectral phase, removing a delay specific to multi-class template- and rest-frame-based classification, rather than the spectrum-acquisition step itself.  (ii) \emph{Retrospective recovery of disguised populations} from the existing ZTF archive: Blazars mislabeled as CVs, Ca-rich transients absorbed into Ib/c, novae absorbed into CV -- populations that closed-set pipelines silently lose at substantial rates.  (iii) \emph{Calibrated, defensible follow-up allocation}, where operators set explicit false-alarm budgets rather than triaging on uncalibrated classifier confidence.  (iv) \emph{A discovery channel for the next gap-transient class}: structurally anomalous spectra are flagged on embedding geometry rather than recognised only post-hoc.  The framework is being prepared for offline production integration with the ZTF \texttt{BOOM} broker \citep{BOOM} ahead of a planned spectroscopic follow-up campaign reported in companion work \citep{applecider}; we argue infrastructure of this kind will be required, not optional, at LSST depth.

\paragraph{Contributions.}
\begin{enumerate}
\item A \emph{redshift-free, phase-free, observer-frame} architecture for spectroscopic transient classification (Sec.~\ref{sec:architecture}).  To our knowledge \texttt{ASTRANet} is the first spectroscopic transient classifier that achieves competitive accuracy without rest-framing any spectrum and without requiring redshift or phase at inference, removing the principal real-time dependency of existing classifiers.

\item A characterisation of \emph{five physically distinct failure modes} of closed-set spectroscopic classifiers (Sec.~\ref{sec:failure_modes}), each linked to a specific anomaly-score signature and a specific astrophysical regime.  This taxonomy is, to our knowledge, new and motivates the cross-family AD design directly from physics.  We show empirically that cross-family non-linear combination is not redundant with single-family scoring -- a substantial majority of the framework's detections come from spectra no single family recognises (Sec.~\ref{sec:complementarity}).

\item A \emph{distribution-free uncertainty quantification layer} (\texttt{ASTRANet-CP}, Sec.~\ref{sec:uq}) via Mondrian conformal prediction with empirically validated coverage.  The per-sample $p_\mathrm{wrong} = 1 - p_\mathrm{M}$ is a distribution-free, finite-sample-calibrated probability that the prediction is mistaken, valid under arbitrary distribution shift provided the calibration set remains exchangeable.  The framework exposes a structural complementarity between classifier-internal UQ and the cross-family AD layer that supports \emph{AD-orthogonal abstention} (Sec.~\ref{sec:selective_classification}).

\item \emph{AD-stratified Mondrian Conformal Prediction} (AD-MCP; Sec.~\ref{sec:ad_stratified_mondrian}), extending atypicality-stratified conformal calibration \citep{yuksekgonul2023confidencereliablemodelsconsider} to a learned cross-family anomaly ensemble crossed with predicted class, recovering uniform conditional coverage in the high-AD operational regime where vanilla Mondrian under-covers.

\item A deployment-ready pipeline with three selectable operating regimes (score-averaged anchor, per-seed detectors, seed-union), evaluated under a leakage-free $k$-fold protocol on a diverse $11$-class rare and $7$-class general transient set.
\end{enumerate}

The remainder of this paper is organised as follows.  Section~\ref{sec:failure_modes} establishes the failure-mode taxonomy empirically.  Section~\ref{sec:data} describes the dataset.  Section~\ref{sec:methods} presents the framework: classifier, deep ensemble, $16$ anomaly scores, and non-linear combiner.  Section~\ref{sec:results} validates the framework on the rare-anomaly evaluation set.  Section~\ref{sec:uq} extends the framework to distribution-free uncertainty quantification. Section~\ref{sec:case_studies} presents physically interpretable case studies of the five failure modes. Section~\ref{sec:discussion} discusses deployment, limitations, and the path to the LSST era.

\section{A Failure-Mode Taxonomy of Spectroscopic Classifiers}
\label{sec:failure_modes}

Before introducing the framework, we establish empirically that the failure modes of closed-set spectroscopic classifiers are \emph{physically structured} rather than uniform.  This observation shapes the design of the framework: a single uncertainty score that treats all failures alike will catch some failure modes and miss others; a cross-family combiner that integrates evidence from embedding geometry, density, hierarchical-head disagreement, and softmax confidence is necessary because the failure modes themselves are diverse.

We identify five regimes from the held-out evaluation set (Sec.~\ref{sec:data}, Table~\ref{tab:classification}).  Each regime is anchored by at least one representative ZTF object whose true classification is independently established, and each maps to a single dominant detection signature.  We describe the regimes below at the level of physical mechanism and detection signature; quantitative case studies are deferred to Sec.~\ref{sec:case_studies}.

\paragraph{Regime I -- Continuum-disguised AGN-class events.}
A subset of AGN-related transients (Blazars, BL Lac objects, some QSOs) have featureless blue continua that overlap with the accretion continua of cataclysmic variables in the wavelength range accessible to low-resolution rapid spectrographs.  Closed-set classifiers map these spectra confidently to the CV class -- in our test sample, $10/11$ Blazars are mis-classified as CVs with mean confidence $\hat p_{\mathrm{CV}} = 0.79$.  The softmax probability is uninformative for these spectra (entropy low, MSP\footnote{MSP: maximum softmax probability} high), and within-class typicality is also uninformative because the embedding is mapped near a CV centroid.  Density-based scores produce only a moderate signal.  \emph{No single family-only mini-ensemble fires at the $1\%$ false-alarm rate on these spectra.}  Detection is carried entirely by the cross-family non-linear combiner, which integrates a weak signal from each family into a saturation-level anomaly score.  This regime is the strongest empirical case for cross-family combination.

\paragraph{Regime II -- Spectroscopic disguise within physical neighbours.}
A second regime arises from genuine spectroscopic similarity between a rare class and a training class.  The clearest examples are calcium-rich transients -- whose strong Ca~II features and stripped-envelope continua are spectroscopically near-degenerate with SNe~Ib/c -- and classical/recurrent novae, whose Balmer-dominated emission spectra resemble certain CV outbursts.  In our test set \emph{all} $10$ Ca-rich transients are classified as Ib/c or II, and only $1$ is flagged by any family-only mini-ensemble at the $1\%$ false-alarm rate; the full cross-family pipeline recovers $4/10$.  For novae the same pattern holds at population scale: only $4/67$ are caught by any family alone, while the full pipeline catches $54/67$.  Like Regime~I, this regime is recovered only by cross-family combination; unlike Regime~I, the dominant signature is within-class distance (the embedding sits in the tail of its predicted-class distribution) rather than projection-space disagreement.

\paragraph{Regime III -- Continuum-dominated mis-classifications.}
Some rare classes -- GRB afterglows in particular -- produce featureless or weak-feature spectra that the classifier maps to training classes (typically CV, occasionally SLSN\footnote{SLSN: Super Luminous Supernova (SN).}) with intermediate confidence ($\hat p \sim 0.5$--$0.85$).  Unlike Regime~I, the embedding of these spectra is genuinely far from the predicted-class centroid: within-class Mahalanobis is in the extreme tail, GMM negative log-density is high, and \emph{both} the distance and density family ensembles fire independently.  All $17$ afterglows in our rare set are recovered.  This is the regime where embedding-based scores fundamentally outperform softmax-based ones, and where a single sufficiently-discriminative family-only score (global or within-class Mahalanobis) would already perform well.

\paragraph{Regime IV -- Legitimately ambiguous progenitor regimes.}
Some rare classes are genuinely intermediate between multiple physical channels: FBOTs have been variously attributed to engine-driven explosions, failed supernovae, and Intermediate Massive Black Hole (IMBH) tidal disruption events \citep{Drout2014, Margutti2019, Perley2019, Ho_2023}; ILRTs have been variously attributed to electron-capture supernovae and non-terminal eruptions \citep{Thompson2009, Kochanek2012}.  In our test sample, FBOT predictions scatter across CV, TDE, Ia, II, and Ibc labels with confidence ranging from $0.30$ to $0.89$ and family-vote counts spanning the entire $0$--$4$ range; ILRTs scatter similarly across H-rich core-collapse subclasses.  The classifier here is \emph{correctly} uncertain; the cross-family ensemble integrates moderate signals from multiple families into reliable detection ($90/98$ FBOTs and $27/33$ ILRTs caught by the framework).  This regime is the survey-operations bread-and-butter for the framework.

\paragraph{Regime V -- Far-out-of-distribution events.}
At the extreme end are events that do not resemble any training class at the manifold level; the long GRB \texttt{AT2022cva} \citep{Barna_2025, 2022GCN.31629....1F} in our test set is the textbook example.  All four families fire with strong signal, multiple voting methods agree, and any reasonable single-metric detector would already flag the event.  These are the easy cases for any anomaly detection framework; they are included to anchor the expected behaviour and to provide a per-class point of reference.

\paragraph{Design implication.}
The five regimes have different detection signatures and therefore require different mathematical machinery for recovery.  Regimes~I and~II are recoverable only through cross-family non-linear combination; Regimes~III and~IV are recoverable by embedding-based scores and benefit from corroboration across families; Regime~V is recoverable by any sufficiently sensitive score.  The upper bound on a single-family pipeline is therefore set by Regimes~I and~II, which together comprise a substantial fraction of the rare-anomaly population (the AGN-related sub-population and the novae$+$Ca-rich sub-population together, $\sim$$43\%$ of the rare set).  The cross-family non-linear combiner we develop in Sec.~\ref{sec:methods} is the smallest machinery that unifies all five regimes.

\section{Dataset}
\label{sec:data}

Spectroscopic data ingestion is handled by \texttt{IRIS} (Identification and Reduction of Interesting Spectra), the data-side module of the \texttt{ASTRANet} framework.  \texttt{IRIS} provides an end-to-end preprocessing pipeline that (i) fetches spectroscopic candidates from the \texttt{Fritz}/SkyPortal marshal \citep{vanderWalt2019, 2023ApJS..267...31C} via the GraphQL API, (ii) cross-matches candidates with the Transient Name Server (TNS) classification, (iii) applies a unified quality-control filter (SNR threshold, wavelength coverage $\geq 80\%$ of the $[3850, 9000]\,\text{\AA}$ window, flatness rejection, and removal of variable-star, bogus, or multi-label objects), (iv) maps raw TNS classifications to the fine and coarse taxonomies (Table~\ref{tab:classification}), (v) partitions the cleaned set at the object level into training, rare-anomaly, and out-of-distribution evaluation subsets, and (vi) builds uniform-grid PyTorch tensors ready for inference.  The \texttt{IRIS} pipeline is released as an open-source package \citep{IRIS_software}.

The dataset used in this work is the output of \texttt{IRIS} applied to the ZTF spectroscopic follow-up campaign \citep{2020ApJ...895...32F, Perley_2020, Rehemtulla+2024}.  The bulk of the spectra come from the SED Machine \citep[SEDM;][]{SEDM, 2019A&A...627A.115R, 2022PASP..134b4505K}, the low-resolution spectrograph dedicated to ZTF transient classification, supplemented by Keck \citep{Oke_1995, 2003SPIE.4841.1657F}, Gemini \citep{1997SPIE.2871.1099D}, the LBT \citep{Pogge2010MODS}, and the Next Generation Palomar Spectrograph \citep[NGPS;][]{2024TNSAN.340....1K}, the new P200 instrument for ZTF/BTS transient classification. Beyond this core, an international network of instruments contributes to transient classification -- for example SDSS \citep{2022ApJS..259...35A}, DESI \citep{2024AJ....168...58D} -- and we refer the reader to \citet{applecider} for a more comprehensive list.  Follow-up triggering and data management are supported by dedicated software infrastructure: the \texttt{SkyPortal} \citep{vanderWalt2019, 2023ApJS..267...31C}, \texttt{Fritz}\footnote{\url{https://github.com/fritz-marshal/fritz}}, and GROWTH \citep{Kasliwal_2019} marshals together with the Transient Name Server\footnote{\url{https://www.wis-tns.org/}} (TNS), with alternative platforms such as YSE-PZ \citep{2023PASP..135f4501C}.

Splits are made at the object level ($70\%/15\%/15\%$, train/val/test): all spectra of a given object fall in the same subset, so no object's spectra leak across the split boundary regardless of how many epochs it has.  Separately, because spectra are aggregated from partially overlapping sources -- the same observation can be present in both the \texttt{Fritz}/SkyPortal and GROWTH marshals, or reported to TNS and the local marshal independently -- a small number of numerically identical spectra appear in the raw set.  These would double-count the affected objects in training and in the reported metrics, so \texttt{IRIS} removes any pair of flux vectors agreeing to within $\epsilon < 10^{-6}$.  Objects without confident classifications ($p_\mathrm{TNS} < 0.5$) are also excluded.

\paragraph{Taxonomy.}
Raw TNS classifications include $\geq\!23$ subtypes, many spectroscopically inseparable at SEDM resolution and several with $<\!100$ examples in our database.  We merge these into $K_f = 7$ \emph{head} classes used by the fine head of the network: \texttt{SN~Ia}, \texttt{SN~Ibc} (Ib $\cup$ Ic $\cup$ Ibc $\cup$ Ibn), \texttt{SLSN}, \texttt{SN~II} (II $\cup$ IIn $\cup$ IIb), \texttt{AGN}, \texttt{CV}, and \texttt{TDE}.  The coarse head groups these into three superclasses: SN-Thermonuclear (\{Ia\}), SN-CC\footnote{CC: Core Collapse} (\{Ibc, II, SLSN\}), and Non-SN (\{AGN, CV, TDE\}).  The full mapping and per-class training counts appear in Table~\ref{tab:classification}; rationale and the complete raw-subtype distribution are in the Appendix~\ref{app:dataset}.
The \texttt{SN~Ia} class explicitly includes the full peculiar-thermonuclear family (Ia-91T, Ia-91bg, Ia-02cx/Iax, Ia-03fg, Ia-CSM, Ia-pec; $478$ spectra across the three splits).  Keeping the peculiar subtypes inside \texttt{SN~Ia} rather than discarding them is a deliberate choice: it forces the classifier to learn the intrinsic diversity of the thermonuclear class, and it reserves the anomaly-detection layer for objects that are genuinely outside the training taxonomy rather than for known subtypes of a common class.

\paragraph{Out-of-distribution evaluation set.}
We assemble a held-out rare-anomaly evaluation set of $N\!=\!289$ spectra across $11$ classes deliberately excluded from training (Table~\ref{tab:classification}, lower block).  The set is constructed to span the five failure-mode regimes of Sec.~\ref{sec:failure_modes} rather than to maximise sample size: AGN outliers (QSO, Blazar, BL Lac) anchor Regime~I; calcium-rich supernovae together with classical and recurrent novae anchor Regime~II; GRB afterglows anchor Regime~III; gap transients (ILRT, FBOT) anchor Regime~IV; long GRBs and luminous red novae anchor Regime~V.  We deliberately retain physical ambiguity in ILRTs and FBOTs -- forcing them into existing categories would obscure genuine astrophysical uncertainty about their progenitors \citep{Thompson2009, Kochanek2012, Drout2014, Margutti2019, Perley2019, Ho_2023, Perets2010}.

\paragraph{Preprocessing.}
Each spectrum is restricted to the \emph{observer-frame} wavelength interval $[3850, 9000]\,\text{\AA}$ and linearly interpolated onto a uniform grid of $L=4096$.  No rest-framing and no continuum removal are applied at any stage of the pipeline: host-galaxy redshift is provided to \texttt{ASTRANet} as optional metadata (Sec.~\ref{sec:architecture}) and is randomly masked during training so the network learns redshift-invariant spectral representations, while the full continuum shape is retained as discriminative information for continuum-dominated classes (AGN, TDE, SLSN).  The interpolated flux is then scaled per spectrum with a robust percentile transform $f \mapsto (f - p_5)/(p_{95} - p_5)$, which compresses outlier emission/absorption excursions without zeroing the local dynamic range.  Missing flux values are zero-filled with a validity mask.  A second input channel carries a Savitzky--Golay first derivative (window $7$, order $3$) computed after augmentation, highlighting narrow features.  Full preprocessing details are in Appendix~\ref{app:preprocessing}.

\section{The \texttt{ASTRANet} Framework}
\label{sec:methods}

The \texttt{ASTRANet} framework integrates data ingestion, classification, anomaly detection, and uncertainty quantification into an end-to-end pipeline.  The four named modules are:
\begin{itemize}
\item \textbf{\texttt{IRIS}} (Sec.~\ref{sec:data}): the data-ingestion module that fetches candidates from the \texttt{Fritz}/SkyPortal marshal, applies quality-control filtering, maps to the FINE\_10/COARSE\_6 taxonomies, and outputs uniform-grid PyTorch tensors \citep{IRIS_software}.

\item \textbf{\texttt{ASTRANet}} (Sec.~\ref{sec:architecture}): the hierarchical deep ensemble classifier that takes an SEDM-resolution spectrum as input and produces an embedding $\mathbf{e} \in \mathbb{R}^{192}$, fine-class softmax $\hat p(y \mid x)$ over $K_f = 7$ classes, and coarse-class softmax over $K_c = 3$ superclasses.

\item \textbf{\texttt{ASTRANet-Sentinel}} (Sec.~\ref{sec:ensemble}): the anomaly detection layer.  Sixteen complementary scores computed on the \texttt{ASTRANet} embedding (Mahalanobis, kNN distance, GMM density, isolation forest, cosine deviation, hierarchical-energy mismatch, and ten others; Appendix~\ref{app:ad_scores}) are non-linearly combined into a single cross-family AD score $s_{\mathrm{AD}}(x) \in [0, 1]$.

\item \textbf{\texttt{ASTRANet-CP}} (Sec.~\ref{sec:uq}): the conformal uncertainty quantification layer.  Mondrian conformal prediction calibrates $s_{\mathrm{AD}}$ into a per-spectrum p-value $p_\mathrm{M}$; adaptive prediction sets provide coverage at user-specified $\alpha$; per-class ensemble probability intervals $\{\bar p_k \pm \sigma_k\}$ summarise the deep-ensemble disagreement; and AD-stratified Mondrian CP (AD-MCP, Sec.~\ref{sec:ad_stratified_mondrian}) achieves uniform conditional coverage across anomaly-score strata.
\end{itemize}
The present section describes the three ML-side modules at the design level.  Architectural details, loss-function specifics, and training hyperparameters are deferred to Appendices~\ref{app:architecture}--\ref{app:training}; the explicit definitions of the $16$ anomaly scores are in Appendix~\ref{app:ad_scores}.

\subsection{The \texttt{ASTRANet} classifier}
\label{sec:architecture}

\begin{figure*}[t]
    \centering
    \includegraphics[width=0.95\textwidth]{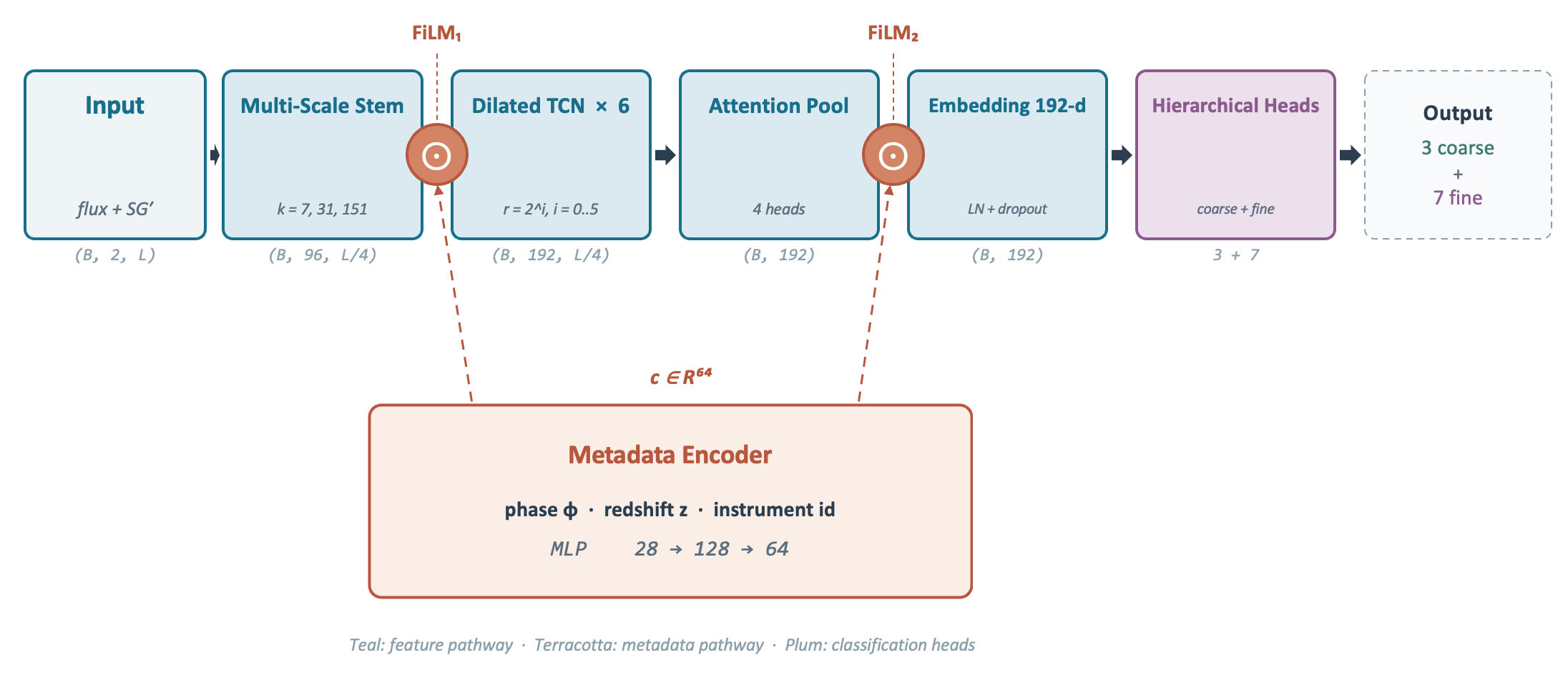}
    \caption{Architecture of \texttt{ASTRANet}.  Spectral features are extracted by a multi-scale stem and a dilated TCN backbone, fused with observational metadata via FiLM conditioning, aggregated by multi-head attention pooling, and passed to hierarchical coarse/fine classification heads.  The 192-dimensional embedding $\mathbf{e}$ supplies the anomaly-detection layer (Sec.~\ref{sec:ensemble}); separate projection heads supply the contrastive loss.  Detailed sub-figures and dimensions are in Appendix~\ref{app:architecture}.}
    \label{fig:architecture}
\end{figure*}

\texttt{ASTRANet} is a hierarchical spectral classifier (Fig.~\ref{fig:architecture}) that jointly predicts coarse superclass and fine-grained type.  The hierarchical structure is motivated by the underlying physics: the coarse partition (thermonuclear vs.\ core-collapse vs.\ non-SN) reflects distinct progenitor channels with substantially disjoint spectral signatures, while the fine partition resolves sub-classes that share a physical channel.  Each input spectrum is represented as a two-channel signal of length $L = 4096$: the robust-scaled flux (Sec.~\ref{sec:data}, with continuum shape retained) and its Savitzky--Golay first derivative, which highlights narrow line features without amplifying noise.

The architecture has four functional blocks:
\begin{itemize}
\item a \emph{multi-scale convolutional stem} with parallel kernels $k \in \{7, 31, 151\}$, designed to operate at the three native scales of supernova spectra: narrow emission/absorption lines, broad P-Cygni features, and continuum shape.
\item a \emph{dilated temporal convolutional backbone} \citep{bai2018empirical} with six residual blocks and exponentially increasing dilation, reaching a $\sim$$250$-sample receptive field;
\item \emph{metadata conditioning via Feature-wise Linear Modulation} \citep[FiLM;][]{perez2018film}, injecting spectral phase, host redshift, and instrument identifier as \emph{optional} metadata at two points.  Each metadata channel carries a validity bit and is randomly masked during training so the network learns to classify, detect anomalies, and quantify uncertainty under partial or fully missing metadata;
\item \emph{hierarchical classification heads} producing a 3-way coarse superclass prediction (SN-Thermonuclear / SN-CC / Non-SN) and a 7-way fine prediction (Ia / Ibc / SLSN / II / AGN / CV / TDE), with stop-gradient between them so the fine head consumes the coarse softmax as a soft prior without back-propagating into the coarse representation. We note that this progenitor-based partition does not always coincide with spectroscopic similarity at a given epoch: at photospheric phases, e.g., SN Ib can resemble SN Ia more than Type II, which the fine head must resolve and which contributes to the residual Ibc/II/Ia confusion structure of Fig. \ref{fig:confusion_matrix}.
\end{itemize}
Multi-head attention pooling \citep{ilse2018attention} aggregates the backbone output into a $192$-dimensional embedding $\mathbf{e}$; separate $\ell_2$-normalised projection heads supply the supervised contrastive loss \citep{khosla2020supervised} that shapes the embedding geometry the AD layer depends on (Sec.~\ref{sec:ensemble}).  The training objective combines focal loss \citep{lin2017focal} on the fine head, cross-entropy on the coarse head, supervised-contrastive terms at both granularities, a targeted confusion penalty that suppresses physically harmful misclassifications (TDE$\to$Ia, II$\to$Ibc, SLSN$\to$Ibc), and an attention-entropy bonus that prevents the model from collapsing onto a single spectral region.  The model is trained with AdamW \citep{loshchilov2019decoupled} under cosine warm restarts, balanced sampling, MixUp \citep{zhang2018mixup}, and effective-number class re-weighting \citep{cui2019class}, with EMA-tracked weights for inference.  Full hyperparameters and per-component design choices appear in Appendix~\ref{app:architecture}, with the loss weights in Table~\ref{tab:loss_weights}.

\subsection{Deep-ensemble backbone}
\label{sec:ensemble_backbone}

A single \texttt{ASTRANet} model produces point estimates of the embedding and softmax probabilities.  To reduce sensitivity to initialisation and to supply implicit uncertainty quantification \citep{Lakshminarayanan2017DeepEnsembles}, we construct an ensemble of $M = 6$ checkpoints obtained from three independent random seeds across two architecture revisions.  For each spectrum we compute per-model embeddings $\mathbf{e}^{(m)}$ and softmax probabilities $\mathbf{p}^{(m)}$, average them within each seed, and then across seeds, yielding $(\bar{\mathbf{e}}, \bar{\mathbf{p}})$.  We verified that this two-stage scheme outperforms both a flat super-ensemble and a cross-seed-only average; eight ensembling strategies are systematically benchmarked in Appendix~\ref{app:ensemble}.  The ensemble averages feed all downstream anomaly-detection scores and the uncertainty-quantification layer; the per-model softmaxes additionally supply the per-class ensemble intervals $\{\bar p_k \pm \sigma_k\}$ reported by \texttt{ASTRANet-CP}.

\subsection{Sixteen complementary anomaly scores}
\label{sec:ood_metrics}

The anomaly-detection layer computes $16$ scores from the ensemble embeddings and logits, spanning four physically motivated families.  Each family captures a qualitatively different way in which a spectrum can be ``unusual'' and maps onto a different subset of the failure-mode regimes of Sec.~\ref{sec:failure_modes}:
\begin{description}
\setlength{\itemsep}{0.2em}
\item[\textbf{Uncertainty} (4 scores)] maximum softmax probability \citep{Hendrycks2017}, predictive entropy, ODIN \citep{Liang2018}, and energy \citep{Liu2020Energy}.  These quantify how confident the classifier is at the logit level and are the natural detectors for Regimes IV--V, where the classifier itself signals uncertainty.
\item[\textbf{Distance} (6 scores)] global, class-conditional, and within-predicted-class Mahalanobis distances \citep{Lee2018Mahalanobis}; $k$-nearest-neighbour distance; cosine distance to the nearest class projection centroid; and a class-conditional typicality.  These quantify how far the spectrum lies from the manifold of in-distribution embeddings and target Regimes~II--III, where the classifier's confidence is uninformative but the embedding is anomalous.
\item[\textbf{Density} (4 scores)] per-class Gaussian-mixture likelihoods, isolation forest \citep{Liu2008iForest}, local outlier factor \citep{Breunig2000LOF}, and PCA-reconstruction error.  These quantify how sparse the embedding's local neighbourhood is and complement the distance family at Regime~III.
\item[\textbf{Hybrid} (2 scores)] a hierarchical entropy combination (HEC) mixing fine-head entropy, coarse-head entropy, and parent-child inconsistency; and a per-class manifold residual score (MRS) computed against per-class PCA subspaces.  These exploit the hierarchical architecture explicitly and target Regime~IV (legitimate progenitor ambiguity that manifests as parent/child label disagreement).
\end{description}
Regime~I -- continuum-disguised AGN-class events -- is the failure mode for which \emph{no single family is sufficient}, since the classifier is confident, the embedding is mapped near an in-distribution centroid, and the density signal is only moderate.  This is the regime that motivates the non-linear cross-family combiner introduced next.  All scores adopt the convention \emph{higher = more anomalous}; each is rank-normalised against the in-distribution validation pool before being combined.  Mathematical definitions, justifications, and individual benchmarks appear in Appendix~\ref{app:ad_scores}.

\subsection{The \texttt{ASTRANet-Sentinel} anomaly detection layer}
\label{sec:ensemble}

The \texttt{ASTRANet-Sentinel} module combines the $16$ rank-normalised scores via a gradient-boosted decision-tree model \citep[LightGBM;][]{Ke2017LightGBM} trained with $5$-fold cross-validation on the validation set, producing a single cross-family anomaly score $s_{\mathrm{AD}}(x) \equiv s_{\mathrm{ensemble}}(x) \in [0,1]$.  The non-linearity is essential: as we show in Sec.~\ref{sec:complementarity}, the majority of detections come from spectra whose anomaly signals are individually below threshold in \emph{every} family and only become detectable through cross-family interaction.  We benchmarked three linear baselines (an AUROC-weighted rank average, a decorrelated logistic regression, and a full $16$-feature logistic regression); all are substantially outperformed by the gradient-boosted combiner.  Comparison numbers and feature-importance breakdown appear in Sec.~\ref{sec:ood_results} and Appendix~\ref{app:ad_scores}.

\paragraph{Operating points.}
The same $s_{\mathrm{AD}}$ supports three operating regimes for survey-mode triage: (i)~a single \emph{anchor threshold} on the seed-averaged score at a fixed false-alarm rate; (ii)~\emph{per-seed detectors} thresholded independently; and (iii)~a \emph{seed-union} that flags any sample caught by at least one seed, trading false-alarm budget for recall.  The choice between regimes is exposed to the user as a single $\alpha$ parameter in the conformal extension (Sec.~\ref{sec:uq}), recovering all four as special cases.  Empirical operating-point analysis is in Sec.~\ref{sec:operating_point}.

\paragraph{Outlier-exposure fine-tuning.}  Before scoring, the encoder is fine-tuned with an outlier-exposure objective \citep{hendrycks2019oe}: the classification loss on the training set is augmented with (i)~an entropy term that pushes exposure anomalies toward a uniform softmax and (ii)~an embedding-margin term that pushes their embeddings at least a margin away from every class centroid ($\lambda_{\rm OE}=\lambda_{\rm emb}=1$, $20$ epochs).  Exposure anomalies are always drawn from folds disjoint from those being evaluated (Sec.~\ref{sec:operating_point}), so no detected object ever contributed to its own detector.  This step is what closes the near-OOD gap: it raises leakage-free recovery at $1\%$ FAR from $34.3\%$ to $82.4\%$ of rare spectra, with the largest gains exactly in the spectroscopic-disguise regime (novae, ILRTs; Sec.~\ref{sec:per_class_ood}).

\paragraph{Inference cost.}
For each new spectrum the pipeline performs $M=6$ forward passes through the \texttt{ASTRANet} ensemble, averages embeddings and logits, computes the $16$ scores from the averaged tensors, rank-normalises them against pre-computed validation distributions, and feeds them to the LightGBM combiner.  Total cost is $<\!200\,\mathrm{ms}$ per spectrum on a single GPU, dominated by the deep-ensemble forward passes; the out-of-distribution (OOD) scoring layer adds $<\!50\,\mathrm{ms}$ on CPU.  The full pipeline is below the per-spectrum cadence of any current or planned spectroscopic time-domain survey.

\section{Validation}
\label{sec:results}

\subsection{Classification performance}
\label{sec:classification_results}

\begin{figure}
    \centering
    \includegraphics[width=0.9\linewidth]{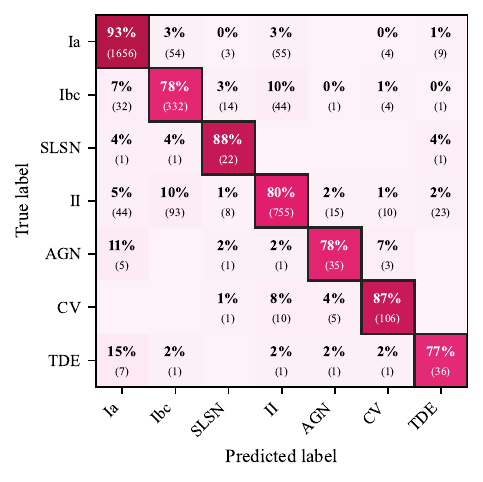}
    \caption{Confusion matrix on the held-out test set ($N\!=\!3{,}396$ spectra) using the within-seed checkpoint ensemble.  Entries are row-normalised to per-class recall (\%).  Off-diagonal mass concentrates on physically expected confusions (Ibc/II, AGN/TDE).}
    \label{fig:confusion_matrix}
\end{figure}

\texttt{ASTRANet} achieves $87\%$ overall accuracy and $83\%$ macro recall on the held-out test set ($N\!=\!3{,}396$ spectra, of which $82$ are peculiar-Ia) using the within-seed checkpoint ensemble, our default deployment configuration.  Per-class recalls range from $77\%$ (TDE) to $93\%$ (SN~Ia); the remaining classes are SN~Ibc ($78\%$, the hardest CC subclass), AGN ($78\%$, the hardest non-SN class due to blue continua at intermediate phases), SN~II ($80\%$, confused mainly with SN~Ibc), CV ($87\%$), and SLSN ($88\%$).  
Appendix~\ref{app:classification_comparison} compares this methodology with other baselines. 
Eight ensembling strategies are systematically compared in Appendix~\ref{app:ensemble}; none of four advanced learned strategies consistently improves over within-seed averaging, indicating that residual errors are driven by intrinsic spectral overlap rather than ensemble sub-optimality.

\subsection{OOD detection: metric comparison}
\label{sec:ood_results}

We evaluate AUROC and AUPRC for all $16$ individual OOD scores and the four ensemble strategies on the rare-anomaly evaluation set ($N\!=\!289$ rare/peculiar/out-of-taxonomy transients vs.\ a held-out validation pool of $3{,}339$ in-distribution samples), with $95\%$ bootstrap confidence intervals over $2000$ resamples.  Full numerical breakdown of all $16$ scores and four ensembles is in Table~\ref{tab:ood_metrics} (Appendix~\ref{app:ad_scores}).

The top seven individual methods are all embedding- or projection-based, while all four logit-based uncertainty methods and the HEC hybrid cluster below AUROC $0.755$.  This gap is the empirical signature of the failure-mode taxonomy of Sec.~\ref{sec:failure_modes}: softmax-based scores are blind to spectra that the classifier confidently mis-classifies (Regimes~I--III), while embedding-based scores recognise that those spectra lie far from the in-distribution manifold.  The LightGBM ensemble reaches $0.996$ test AUROC; linear logistic regression on all $16$ scores already reaches $0.9760$ but cannot match the non-linear combiner, whose advantage we trace to cross-family interactions in Sec.~\ref{sec:complementarity}.

\subsection{Why a single score family is not enough}
\label{sec:complementarity}

The systematic gain from combining the $16$ scores comes not from averaging redundant signals but from genuine complementarity between the four score families.  To show this, we built four ``family-only'' mini-ensembles by mean-rank averaging the rank-normalised scores within each family and thresholded each at the same $1\%$ validation false-alarm rate as the full LightGBM ensemble.

Table~\ref{tab:family_ablation} shows the resulting per-family AUROCs and detection counts.  \emph{No single family approaches the full pipeline}: the strongest family-only detector (distance, AUROC $=0.8823$) recovers only $86/289$ rare spectra at $1\%$ FAR; the density family is close behind ($80/289$); hybrid and uncertainty trail at $26/289$ and $22/289$ respectively.  Under the leakage-free $k$-fold protocol of Sec.~\ref{sec:operating_point}, the base LightGBM combination of all $16$ scores recovers $99/289$ ($34.3\%$) at $1\%$ FAR; adding outlier-exposure fine-tuning of the encoder (Sec.~\ref{sec:ensemble}) lifts this to $238/289$ ($82.4\%$) at the same FAR, and to $244/289$ ($84.4\%$; $80.4\%$ of unique objects) under the seed-union strategy at an effective $\sim\!3\%$ FAR.

\begin{table*}[t]
\centering
\caption{Per-family OOD detection performance on the \protect$N\!=\!289\protect$ rare set.  Family ensembles built by averaging rank-normalised scores within each family; thresholds set at the same \protect$1\%\protect$ validation false-alarm rate as the full LightGBM combiner.}
\label{tab:family_ablation}
\small
\begin{tabular}{lccc}
\hline\hline
Family & AUROC & Det at 1\% FAR & Det \% \\
\hline
Uncertainty ($4$ scores)        & $0.7184$ & $22/289$  & $7.6\%$  \\
Hybrid      ($2$ scores)        & $0.7957$ & $26/289$  & $9.0\%$ \\
Density     ($4$ scores)        & $0.8809$ & $80/289$  & $27.7\%$ \\
Distance    ($6$ scores)        & $0.8823$ & $86/289$  & $29.8\%$ \\
\hline
LightGBM (all $16$, $5$-fold CV)    & $0.947 \pm 0.016$ & $99/289$ & $34.3\%$ \\
\textbf{~~+ outlier exposure ($3$-seed)}    & -- & $\mathbf{238/289}$ & $\mathbf{82.4\%}$ \\
\hline
\end{tabular}
\end{table*}

\begin{figure*}[t]
    \centering
    \includegraphics[width=\textwidth]{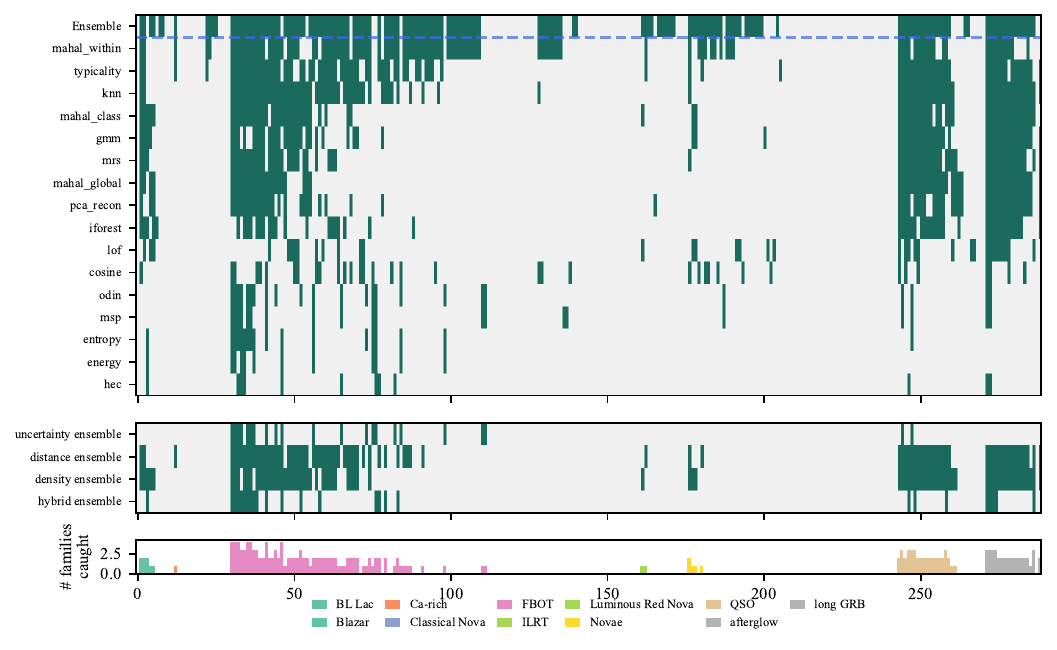}
    \caption{Per-anomaly detection map at $1\%$ false-alarm rate.  Each column is a rare spectrum, sorted by true class and then by the number of families that flag it.  \emph{Top row}: the LightGBM ensemble of all $16$ scores.  \emph{Middle block}: the $16$ individual OOD scores, sorted by individual AUROC and annotated with their family.  \emph{Lower block}: the four family-only mini-ensembles.  \emph{Bottom panel}: number of families that catch each anomaly, coloured by true class.  Ca-rich, Classical Nova, Novae, and ILRT contain large numbers of spectra the LightGBM ensemble flags but no family alone recovers.}
    \label{fig:detection_map_complement}
\end{figure*}

\paragraph{Per-anomaly complementarity.}
At the $1\%$-FAR operating point, $106$ of the $289$ rare anomalies are caught by at least one family-only mini-ensemble ($28$ by exactly one family, $55$ by two, $16$ by three, $7$ by all four); the remaining $183$ are caught by \emph{no} family alone.  The outlier-exposed \texttt{Sentinel} flags $238$ samples, of which \emph{$137$ ($57.6\%$) come from anomalies no family-only ensemble catches} -- the near-OOD classes dominate this set (novae alone jump from $4/67$ caught by any single family to $54/67$ under the full pipeline, the largest absolute gain).  This is the empirical content of the Regime~I/II argument of Sec.~\ref{sec:failure_modes}: near-OOD events whose individual family signals are weak are recoverable only by an ensemble that integrates evidence across families non-linearly.  Figure~\ref{fig:detection_map_complement} shows the per-spectrum detection map.

\subsection{Detection strategies and operating points}
\label{sec:operating_point}

A key practical strength of the framework is that survey teams can choose the detection regime appropriate for their use case.  All operating points below are evaluated under a \emph{leakage-free} protocol: the $289$ rare spectra are split into $k\!=\!5$ stratified folds; for each fold the encoder is fine-tuned with outlier exposure on the \emph{other} four folds (Sec.~\ref{sec:ensemble}), the LightGBM combiner is trained on half of the validation pool plus the exposure folds, the detection threshold is set at a $1\%$ false-alarm rate on the \emph{held-out} half of the validation pool, and detections are counted only on the held-out fold.  No rare spectrum is ever seen -- in fine-tuning, combiner training, or thresholding -- by the detector that scores it.  The procedure is repeated for three independently trained seeds.

\paragraph{Anchor (score-averaged ensemble).}
Averaging the three seeds' \texttt{Sentinel} scores and thresholding at the (averaged) $1\%$-FAR anchor recovers $238/289$ spectra ($82.4\%$), corresponding to $81/107$ unique objects ($75.7\%$).

\paragraph{Per-seed detectors.}
Individual seeds recover $75.1 \pm 0.7\%$ of spectra ($68.2 \pm 1.5\%$ of objects) at their own $1\%$-FAR thresholds -- the seed-to-seed scatter is small, and the ensemble anchor adds $\sim\!7$ points over any single seed.

\paragraph{Seed-union.}
Flagging a spectrum caught by \emph{any} of the three seeds trades false-alarm budget for recall: $244/289$ spectra ($84.4\%$) and $86/107$ objects ($80.4\%$) at an effective $\sim\!3\%$ in-distribution false-positive rate.  This is the natural survey-mode triage point.

\begin{table*}[t]
\centering
\caption{Comparison of detection strategies on the rare-anomaly evaluation set ($N\!=\!289$ spectra, $107$ objects) under the leakage-free $k$-fold protocol.  ``Missed'' counts rare spectra not flagged.}
\label{tab:strategies}
\small
\begin{tabular}{lcccc}
\hline\hline
Strategy & Rare det.\ (spectra) & Rare det.\ (objects) & Eff.\ FAR & Missed \\
\hline
Base combiner, no OE ($3$-seed avg.) & $34.3\%$ & -- & $1.0\%$ & $190$ \\
Per-seed OE (mean $\pm$ std)         & $75.1 \pm 0.7\%$ & $68.2 \pm 1.5\%$ & $1.0\%$ & $\sim\!72$ \\
\textbf{Anchor: $3$-seed score average} & $\mathbf{82.4\%}$ & $\mathbf{75.7\%}$ & $\mathbf{1.0\%}$ & $\mathbf{51}$ \\
Seed-union                           & $84.4\%$ & $80.4\%$ & $\sim\!3\%$ & $45$ \\
\hline
\end{tabular}
\end{table*}

\paragraph{Choice of operating regime.}
We recommend: \emph{survey-mode triage} (high recall, moderate FP cost) -- the seed-union ($84.4\%$ of spectra, $80.4\%$ of objects, at $\sim\!3\%$ FP); \emph{high-purity follow-up} (every flagged spectrum likely receives telescope time) -- the score-averaged anchor at $1\%$ FAR ($82.4\%$ recall); \emph{pure ranking} (priority queue for human scanners) -- the averaged \texttt{Sentinel} score directly, with no threshold.

\subsection{Per-class detection difficulty}
\label{sec:per_class_ood}

The recovery hierarchy aligns with astrophysical expectations, and we report it in two tiers that separate what a detection \emph{means}.  \textbf{Tier~1 -- genuinely novel phenomena} (no trained counterpart class): the score-averaged anchor recovers $27/35$ objects ($77\%$), rising to $28/35$ ($80\%$) under the seed-union.  Within this tier, FBOTs ($92/98$ spectra, $94\%$), GRB afterglows ($16/17$, $94\%$), the long GRB ($1/1$), luminous red novae ($13/15$, $87\%$) and ILRTs ($27/33$, $82\%$) are reliably flagged; Ca-rich transients ($4/10$ spectra, $2/5$ objects) remain the hardest class, mapped almost uniformly to SN~Ib/c -- their strong Ca~II features and stripped-envelope continua are spectroscopically near-degenerate with the training class.  \textbf{Tier~2 -- rare members of trained families}, where reduced sensitivity is partly \emph{correct} behaviour: classical and recurrent novae are cataclysmic variables in eruption and live inside the trained CV manifold, yet outlier exposure recovers $54/67$ nova spectra ($81\%$) and $5/8$ classical novae; the AGN-family outliers (QSO $20/28$ spectra, Blazar $5/11$, BL~Lac $0/1$) sit inside the trained AGN manifold and are flagged at correspondingly lower rates.  Overall the anchor recovers $75\%$ of Tier-2 objects and $77\%$ of Tier-1 objects, and the seed-union $81\%$ and $80\%$ respectively.  This hierarchy provides practical guidance for survey operations: the pipeline reliably flags genuinely novel phenomena (afterglows, GRBs, fast transients, LRNe, ILRTs) while near-degenerate gap transients (Ca-rich) need additional follow-up data (light-curve evolution, host-galaxy properties) to be distinguished from genuine stripped-envelope supernovae.

\begin{table}[t]
\centering
\caption{Per-class detection statistics on the rare set ($N\!=\!289$) under the leakage-free $k$-fold protocol.  ``Det.'' is the number of spectra flagged by the seed-averaged \texttt{Sentinel} at the $1\%$-FAR anchor (threshold $0.087$); ``Det.\ \%'' is per-class recall; ``Ens.\ score'' is the mean seed-averaged combiner score across spectra of that class (scores are held-out $k$-fold probabilities and are not calibrated toward $1$).}
\label{tab:per_class_detection}
\small
\begin{tabular}{lcccc}
\hline\hline
Class & $N$ & Detected & Det.\ \% & Ens.\ score \\
\hline
\rev{long GRB}           & \rev{$1$}  & \rev{$1$}  & \rev{$100.0\%$} & \rev{$0.374$} \\
\rev{afterglow}          & \rev{$17$} & \rev{$16$} & \rev{$94.1\%$}  & \rev{$0.899$} \\
\rev{FBOT}               & \rev{$98$} & \rev{$92$} & \rev{$93.9\%$}  & \rev{$0.833$} \\
\rev{Luminous Red Nova}  & \rev{$15$} & \rev{$13$} & \rev{$86.7\%$}  & \rev{$0.829$} \\
\rev{ILRT}               & \rev{$33$} & \rev{$27$} & \rev{$81.8\%$}  & \rev{$0.514$} \\
\rev{Novae}              & \rev{$67$} & \rev{$54$} & \rev{$80.6\%$}  & \rev{$0.680$} \\
\rev{Classical Nova}     & \rev{$8$}  & \rev{$6$}  & \rev{$75.0\%$}  & \rev{$0.601$} \\
\rev{QSO}                & \rev{$28$} & \rev{$20$} & \rev{$71.4\%$}  & \rev{$0.518$} \\
\rev{Blazar}             & \rev{$11$} & \rev{$5$}  & \rev{$45.5\%$}  & \rev{$0.364$} \\
\rev{Ca-rich}            & \rev{$10$} & \rev{$4$}  & \rev{$40.0\%$}  & \rev{$0.258$} \\
\rev{BL Lac}             & \rev{$1$}  & \rev{$0$}  & \rev{$0.0\%$}   & \rev{$0.003$} \\
\hline
\end{tabular}
\end{table}

\paragraph{Case-study anchor.}
Concrete per-spectrum examples spanning these regimes -- including the Regime-B cases (Blazar \texttt{
AT 2024uln}, FBOT \texttt{AT 2024wpp}) where the model is confident on a wrong class, at most one family-only ensemble fires, yet $s_\mathrm{ens} \geq 0.93$ -- are presented with full astrophysical interpretation in Sec.~\ref{sec:case_studies} (Table~\ref{tab:anomaly_case_study}).

\section{The \texttt{ASTRANet-CP} conformal uncertainty layer}
\label{sec:uq}

Anomaly detection (\texttt{ASTRANet-Sentinel}, Sec.~\ref{sec:ensemble}) and classification (Sec.~\ref{sec:classification_results}) are usually presented as distinct outputs.  In survey-mode triage they are inseparable: a spectrum flagged as anomalous still requires a per-class probability to decide how to allocate follow-up resources, and a confident classification still requires a calibrated tail probability to decide whether the confidence can be trusted.  We introduce \texttt{ASTRANet-CP}, the conformal uncertainty layer that extends the $16$-score AD framework of Sec.~\ref{sec:ensemble} to \emph{distribution-free} uncertainty quantification via Mondrian conformal prediction \citep{Vovk2005, AngelopoulosBates2023}.  The framework produces, for every spectrum: (i) a calibrated conformal p-value of the LightGBM anomaly score; (ii) an adaptive prediction set $\mathcal{C}_\alpha(x)$ of plausible classes at user-specified coverage $1-\alpha$ \citep{Romano2020APS}; (iii) a decomposition of predictive entropy into aleatoric and epistemic components using the deep ensemble's per-model variance.

\paragraph{Calibration setup.}  We split the in-distribution validation pool ($3{,}339$ spectra) randomly into a calibration set ($n_{\mathrm{cal}} = 1{,}669$) and a held-out exchangeability-test set ($n_{\mathrm{ho}} = 1{,}670$).  To avoid leakage of training information into calibration, the LightGBM anomaly scores used by the conformal procedure are produced by $5$-fold out-of-fold prediction on the union of validation and rare spectra; this yields an OOF AUROC of $0.9983$ on the rare set, in line with the cross-validated estimate of Sec.~\ref{sec:ensemble}.  (These OOF scores serve calibration analysis only; detection headlines use the leakage-free protocol of Sec.~\ref{sec:operating_point}.)

\subsection{Conformal calibration of the AD score}
\label{sec:conformal_ad}

Given a calibration set $\{(x_i, y_i)\}_{i=1}^{n}$ and a nonconformity score $s(x, y)$ (higher $=$ more atypical), the conformal p-value of a new point $x^*$ with predicted class $\hat y$ is
\begin{equation}
  p(x^*) \;=\; \frac{ |\{ i : s(x_i, y_i) \geq s(x^*, \hat y) \}| + 1 }{ n + 1 }.
  \label{eq:conformal_pvalue}
\end{equation}
We adopt the LightGBM ensemble score as the nonconformity function.  Eq.~\ref{eq:conformal_pvalue} gives \emph{marginal} coverage; the \emph{Mondrian} variant partitions the calibration set by predicted class and applies Eq.~\ref{eq:conformal_pvalue} within each partition, providing per-class FDR control \citep{Vovk2005, AngelopoulosBates2023}:
\begin{equation}
  p_{\mathrm{M}}(x^*) = \frac{|\{i : \hat y_i = \hat y(x^*) \;\land\; s_i \geq s(x^*)\}| + 1}{|\{i : \hat y_i = \hat y(x^*)\}| + 1}.
  \label{eq:mondrian}
\end{equation}

\begin{figure}[t]
\centering
\includegraphics[width=\columnwidth]{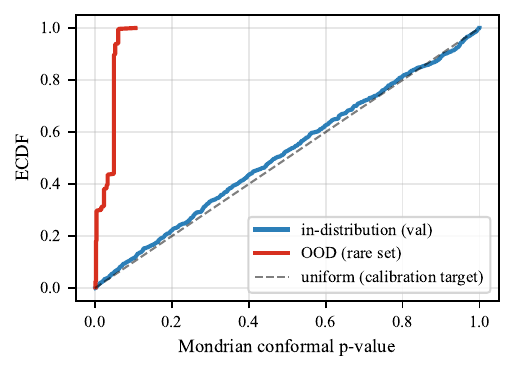}
\caption{Mondrian conformal p-value distribution on the in-distribution held-out validation set (blue) and the $289$-spectrum rare set (red).  The ID distribution is statistically indistinguishable from uniform (diagonal), confirming exchangeability between calibration and held-out validation.  The OOD distribution piles near $0$: \rev{$99.7\%$} of rare spectra have $p_{\mathrm{M}} < 0.10$ and \rev{$89.6\%$} have $p_{\mathrm{M}} < 0.05$.}
\label{fig:pvalue_ecdf}
\end{figure}

\paragraph{Empirical coverage.}  Empirical coverage tracks the nominal level almost exactly across $1-\alpha \in [0.80, 0.995]$ (Table~\ref{tab:conformal_summary}): \rev{at $\alpha = 0.05$ the marginal procedure achieves $0.949$ empirical coverage against $0.950$ nominal, and Mondrian achieves $0.940$.}  The curves sit on the diagonal within the expected $\pm 0.5\,n_{\mathrm{ho}}^{-1/2} \approx \pm 0.012$ finite-sample error, confirming the LightGBM-OOF score satisfies the exchangeability assumption conformal prediction requires.  Figure~\ref{fig:pvalue_ecdf} shows the resulting p-value distributions: the ID distribution is statistically indistinguishable from uniform while the OOD distribution piles near zero.

\paragraph{OOD detection at calibrated alpha.}  Table~\ref{tab:conformal_summary} summarises the rare-set detection rate at four canonical $\alpha$ values.  \rev{At $\alpha = 0.10$ (the natural high-recall operating point) the Mondrian predictor recovers $99.7\%$ of the rare anomalies.  This sequence is mathematically equivalent to a LightGBM threshold sweep on the OOF scores; the $\alpha = 0.01$ Mondrian operating point corresponds to the OOF anchor threshold as a special case.}  The advantage of the conformal formulation is interpretive: a Mondrian p-value of $0.04$ for a CV-predicted spectrum means ``among in-distribution CV-class calibration spectra, only $4\%$ are at least this anomalous''.

\begin{table}[t]
\centering
\caption{Conformal calibration and OOD detection at four canonical $\alpha$ values.  ``Marg.\ cov.''\ and ``Mond.\ cov.''\ are empirical in-distribution coverage of the marginal and Mondrian predictors (target $1-\alpha$); ``Mond.\ det.''\ is the rare-set detection rate at the Mondrian operating point ($p_{\mathrm{M}} < \alpha$).}
\label{tab:conformal_summary}
\small
\begin{tabular}{ccccc}
\hline\hline
$\alpha$ & Target cov.\ & Marg.\ cov.\ & Mond.\ cov.\ & Mond.\ det.\ \\
\hline
\rev{$0.010$} & \rev{$0.990$} & \rev{$0.987$} & \rev{$0.986$} & \rev{$29.8\%$} \\
\rev{$0.025$} & \rev{$0.975$} & \rev{$0.972$} & \rev{$0.966$} & \rev{$38.1\%$} \\
\rev{$0.050$} & \rev{$0.950$} & \rev{$0.949$} & \rev{$0.940$} & \rev{$89.6\%$} \\
\rev{$0.100$} & \rev{$0.900$} & \rev{$0.888$} & \rev{$0.886$} & \rev{$99.7\%$} \\
\hline
\end{tabular}
\end{table}

\paragraph{Per-class coverage.}  \rev{At $\alpha = 0.025$, five of seven predicted classes achieve coverage within $\pm 2\%$ of the $0.975$ target (Table~\ref{tab:conformal_per_class}); the exceptions are TDE ($0.787$) and CV ($0.917$), reflecting their small calibration samples ($n_{\mathrm{cal}} = 42$ and $60$) and the expected finite-sample behaviour of the per-class Mondrian quantile.  Rare-set detection at this operating point is $100\%$ for every predicted class except CV ($91.9\%$), which absorbs the bulk of the Blazar, QSO, and Novae mis-predictions -- the quantitative formulation of the per-class pattern of Sec.~\ref{sec:per_class_ood}.}

\begin{table}[t]
\centering
\caption{Mondrian conformal per-class diagnostics at $\alpha = 0.025$.  ``$n_{\mathrm{cal}}$'' is the number of in-distribution calibration spectra in the predicted-class partition; ``$n_{\mathrm{test}}$'' the number of rare-set spectra routed to that class.}
\label{tab:conformal_per_class}
\small
\begin{tabular}{lcccc}
\hline\hline
Pred.\ class & $n_{\mathrm{cal}}$ & $n_{\mathrm{test}}$ & Coverage & Detection \\
\hline
\rev{Ia}   & \rev{$850$} & \rev{$7$}   & \rev{$0.969$} & \rev{$1.00$} \\
\rev{Ibc}  & \rev{$253$} & \rev{$34$}  & \rev{$0.971$} & \rev{$1.00$} \\
\rev{II}   & \rev{$414$} & \rev{$45$}  & \rev{$0.960$} & \rev{$1.00$} \\
\rev{SLSN} & \rev{$32$}  & \rev{$21$}  & \rev{$0.957$} & \rev{$1.00$} \\
\rev{AGN}  & \rev{$18$}  & \rev{$13$}  & \rev{$1.000$} & \rev{$1.00$} \\
\rev{TDE}  & \rev{$42$}  & \rev{$20$}  & \rev{$0.787$} & \rev{$1.00$} \\
\rev{CV}   & \rev{$60$}  & \rev{$149$} & \rev{$0.917$} & \rev{$0.919$} \\
\hline
\end{tabular}
\end{table}

\paragraph{AD-conditional coverage diagnostic.}  The Mondrian guarantee is \emph{marginal}: it ensures $1-\alpha$ coverage averaged over the in-distribution distribution but not conditional on the AD score itself.  Stratifying the held-out set into AD-score quintiles reveals that coverage is $1.000$ in Q1--Q4 (the $80\%$ of spectra whose AD score lies far below the per-class thresholds, by construction always covered) and only $0.754$ in Q5 (the top $20\%$, where survey-mode triage actually fires).  The full $5\%$ miscalibration budget is concentrated in Q5 -- a known mathematical property of marginal conformal predictors stratified by their own score \citep{AngelopoulosBates2023, Romano2020APS, yuksekgonul2023confidencereliablemodelsconsider}, not a calibration failure.  Operationally, at the deployed $\alpha = 0.05$ threshold the conditional false-alarm rate in the high-AD regime is closer to $25\%$ than $5\%$.  Sec.~\ref{sec:ad_stratified_mondrian} introduces the AD-stratified estimator (AD-MCP) that fixes this directly.

\subsection{Adaptive prediction sets for classification}
\label{sec:conformal_clf}

The conformal machinery applied to the classifier's softmax output yields per-spectrum \emph{prediction sets} of plausible labels.  Following \citet{Romano2020APS}, we use the class-conditional nonconformity score
\begin{equation}
  s_{\mathrm{clf}}(x, y) = 1 - p_{\mathrm{fine}}(y \mid x),
  \label{eq:nonconf_clf}
\end{equation}
and define the adaptive prediction set (APS) at coverage $1-\alpha$ as $\mathcal{C}_\alpha(x^*) = \{ y : p_{\mathrm{fine}}(y \mid x^*) \geq 1 - \hat q_\alpha \}$, with $\hat q_\alpha$ the finite-sample-corrected empirical $(1-\alpha)$-quantile of $s_{\mathrm{clf}}$ on the calibration set.

At the $\alpha = 0.10$ operating point, ID prediction sets are singletons for $89\%$ of inputs while OOD prediction sets are singletons for $77.5\%$ -- the model committed to a wrong class, the disguised-class regime of Table~\ref{tab:anomaly_case_study}.  At the stricter $\alpha = 0.05$ level OOD sets broaden substantially (mean $1.82$ vs ID mean $1.37$), reflecting genuine uncertainty between physically related classes (e.g.\ a Ca-rich spectrum produces $\{\mathrm{Ibc}, \mathrm{II}\}$).  A broadened set of size $\geq 3$ is operationally more useful than an over-confident point prediction.  At the loose $\alpha = 0.20$ point, $40\%$ of OOD spectra produce an \emph{empty} prediction set -- the conformal procedure formally declining to classify, a strong ``probably not any known class'' signal distinct from the AD flag.  

\subsection{Aleatoric/epistemic decomposition via deep ensembles}
\label{sec:uq_decomposition}

The conformal procedure provides distribution-free calibration but does not distinguish two qualitatively different sources of uncertainty.  We supplement it with the standard predictive-entropy decomposition arising from the deep ensemble of $M=6$ \texttt{ASTRANet} variants.  Writing $\bar p(y\mid x) = M^{-1}\sum_m p^{(m)}(y\mid x)$,
\begin{equation}
  H[\bar p] = \frac{1}{M}\sum_{m=1}^{M} H[p^{(m)}] + \mathrm{MI}(y;\theta\mid x),
  \label{eq:uq_decomp}
\end{equation}
where the first term is \emph{aleatoric} (data noise over which models agree) and the second \emph{epistemic} (model-to-model disagreement).  MI is high precisely when the input is out of distribution and the variants extrapolate inconsistently.  In the framework MI is not a primary detector -- the LightGBM ensemble outperforms it by a large margin (AUROC $\sim 0.996$ vs $\sim 0.78$) -- but an \emph{interpretation overlay}: high LightGBM score \emph{and} high MI means the ensemble disagrees about identity (a genuinely novel object, Regime~C); high LightGBM score \emph{and} low MI means the ensemble agrees it is far from any class manifold (a disguised-class object, Regime~B).

\subsection{The confidence-aware output}
\label{sec:uq_output}

For each spectrum the framework returns a five-component output:
\begin{enumerate}
\item a fine-class point prediction $\hat y$ (Sec.~\ref{sec:architecture});
\item a Mondrian conformal p-value $p_{\mathrm{M}}(x)$ (Eq.~\ref{eq:mondrian}).  Equivalently, $p_\mathrm{wrong}(x) \equiv 1 - p_{\mathrm{M}}(x) \in [0,1]$ is a per-sample, distribution-free, finite-sample-calibrated probability that the commitment to $\hat y$ is mistaken, valid under arbitrary distribution shift provided the calibration set remains exchangeable;
\item an adaptive prediction set $\mathcal{C}_\alpha(x)$ at coverage $1-\alpha$ (Eq.~\ref{eq:nonconf_clf});
\item per-class ensemble probability intervals $\{\bar p_k \pm \sigma_k\}_{k=1}^{K_f}$ from the $M\!=\!6$ variants (Sec.~\ref{sec:per_class_intervals});
\item an aleatoric/epistemic decomposition of the predictive entropy (Eq.~\ref{eq:uq_decomp}).
\end{enumerate}
All five are produced from a single forward pass through the deep ensemble ($< 200\,\mathrm{ms}$ per spectrum on a single GPU) and together answer: \emph{is this spectrum anomalous} (the p-value); \emph{which classes are plausible} (the APS); \emph{what is the per-class probability and its spread} (the intervals); \emph{is the classifier overconfident} (the epistemic term); and \emph{has the framework declined to commit} (an empty APS).

\subsection{Per-class ensemble probability intervals}
\label{sec:per_class_intervals}

For a survey operator deciding which physical interpretations to act on, none of the four components above directly answers \emph{``what is the calibrated probability of class $k$, and how uncertain is that probability across the ensemble?''}  Writing $p^{(m)}(y \mid x)$ for the $m$-th variant's predictive distribution, we define
\begin{equation}
\begin{aligned}
\bar p_k(x) &= \frac{1}{M}\sum_{m=1}^{M} p^{(m)}(y\!=\!k\mid x),\\
\sigma_k(x) &= \sqrt{\frac{1}{M\!-\!1}\sum_{m=1}^{M}\bigl(p^{(m)}(y\!=\!k\mid x) - \bar p_k(x)\bigr)^{2}}
\end{aligned}
\label{eq:per_class_intervals}
\end{equation}
and report the $K_f$ intervals $\bar p_k(x) \pm \sigma_k(x)$ at zero additional cost from the per-model softmaxes the ensemble already produces.

\paragraph{Operational behaviour across the five regimes.}
\rev{The information content of $\{\bar p_k, \sigma_k\}$ is sharply regime-dependent (Table~\ref{tab:per_class_intervals}).  For Regime~I (the disguised Blazar \texttt{ AT 2024uln}), all six variants concur on CV at $\bar p_{\mathrm{CV}} = 92.7\% \pm 4.9\%$.}  This is the empirical \emph{absence} of an ensemble-internal anomaly signal: the classifier-internal UQ axis -- the same one that powers softmax entropy, MC-dropout variance, and conformal-on-softmax $p_\mathrm{wrong}$ -- has nothing to flag, because all six members arrive at the same confident wrong commitment.  \rev{The detection signal must come from outside the classifier head: the cross-family AD-layer p-value $p_{\mathrm{M}} = 0.046$, carried entirely by embedding-space distance and density scores.  For Regimes~II--V the top-class spread is large ($> 10\%$), reflecting genuine ensemble disagreement that classifier-internal UQ alone would detect.  The Regime~I contrast -- tight consensus ($\sigma_{\hat y} \approx 5\%$) on a wrong class versus a flagged $p_{\mathrm{M}} = 0.046$ on the same spectrum -- is} the concrete demonstration that classifier-internal UQ and cross-family AD are \emph{structurally complementary} axes, not interchangeable implementations of one estimator.

\begin{table*}[t]
\centering
\caption{Per-class ensemble probability intervals $\{\bar p_k \pm \sigma_k\}$ (Eq.~\ref{eq:per_class_intervals}) on representative case-study spectra from each of the five physical regimes of Sec.~\ref{sec:failure_modes}, paired with the cross-family AD-layer Mondrian p-value $p_{\mathrm{M}}$.  Only the top three classes by $\bar p_k$ are shown.  Regime~I exhibits tight consensus ($\sigma_{\hat y} \lesssim 3\%$) on a wrong class with $p_{\mathrm{M}}$ carrying the entire detection signal; Regimes~II--V exhibit large $\sigma$ ($> 10\%$) reflecting genuine ensemble disagreement.}
\label{tab:per_class_intervals}
\small
\begin{tabular}{lll lll l}
\hline\hline
Regime & Object & True class & Class 1 & Class 2 & Class 3 & $p_{\mathrm{M}}$ \\
\hline
\rev{I}   & \rev{\texttt{ AT 2024uln}} & \rev{Blazar}    & \rev{CV $92.7\% \pm 4.9\%$}  & \rev{Ibc $2.5\% \pm 2.0\%$}   & \rev{II $1.8\% \pm 1.8\%$}  & \rev{$0.046$} \\
\rev{II}  & \rev{\texttt{2021zfp}} & \rev{Ca-rich}   & \rev{CV $46.1\% \pm 14.4\%$} & \rev{Ibc $30.5\% \pm 15.1\%$}  & \rev{II $19.3\% \pm 5.3\%$}  & \rev{$\sim\!0.12$} \\
\rev{III} & \rev{\texttt{AT2021lfa}} & \rev{afterglow} & \rev{CV $76.1\% \pm 21.7\%$} & \rev{SLSN $17.9\% \pm 20.0\%$} & \rev{TDE $2.4\% \pm 1.5\%$} & \rev{$\ll 0.01$} \\
\rev{IV}  & \rev{\texttt{AT 2024wpp}} & \rev{FBOT}      & \rev{CV $71.6\% \pm 19.7\%$}  & \rev{II $11.4\% \pm 10.1\%$} & \rev{Ibc $7.8\% \pm 6.9\%$}   & \rev{$\sim\!0.009$} \\
\rev{V}   & \rev{\texttt{AT2022cva}} & \rev{long GRB}  & \rev{TDE $51.8\% \pm 18.2\%$} & \rev{SLSN $26.6\% \pm 17.7\%$}  & \rev{CV $14.9\% \pm 3.1\%$} & \rev{$\sim\!0.087$} \\
\hline
\end{tabular}
\end{table*}

\subsection{Test-error uncertainty: the two UQ axes catch different failure modes}
\label{sec:test_error_uncertainty}

We establish the structural complementarity of Sec.~\ref{sec:per_class_intervals} at population scale on the held-out in-distribution test set: \rev{$N_{\mathrm{err}} = 456$ mis-classifications out of $N_{\mathrm{test}} = 3{,}396$ spectra.}  For each spectrum we compute six classifier-internal UQ signals (top-$1$ confidence, predictive entropy, top-$1$/top-$2$ margin, mutual information, predicted-class ensemble spread $\sigma_{\hat y}$, and model disagreement) and, in parallel, the cross-family AD score $s_{\mathrm{AD}}$; each is evaluated as an error detector (Table~\ref{tab:test_error_aurocs}).

\begin{table}[t]
\centering
\caption{\rev{Error-detection AUROC of each UQ signal on the held-out ID test set ($N_{\mathrm{test}} = 3{,}396$, $N_{\mathrm{err}} = 456$).  Classifier-internal signals lead the cross-family AD score by $0.065$ in AUROC at the top of the table.}}
\label{tab:test_error_aurocs}
\small
\begin{tabular}{lc}
\hline\hline
UQ signal & AUROC \\
\hline
\rev{Top-$1$ confidence $\hat p_{\mathrm{top1}}$}        & \rev{$\mathbf{0.895}$} \\
\rev{Top-$1$/top-$2$ margin $\Delta$}                    & \rev{$0.890$} \\
\rev{Predictive entropy $H$}                             & \rev{$0.876$} \\
\rev{Mutual information $\mathrm{MI}$}                   & \rev{$0.806$} \\
\rev{Ensemble spread $\sigma_{\hat y}$}                  & \rev{$0.792$} \\
\rev{Model disagreement}                                 & \rev{$0.754$} \\
\hline
\rev{Cross-family AD score $s_{\mathrm{AD}}$}            & \rev{$0.830$} \\
\hline
\rev{\textbf{Gap (best internal $-$ AD)}}                & \rev{$\mathbf{0.065}$} \\
\hline
\end{tabular}
\end{table}

\begin{figure}[t]
\centering
\includegraphics[width=\columnwidth]{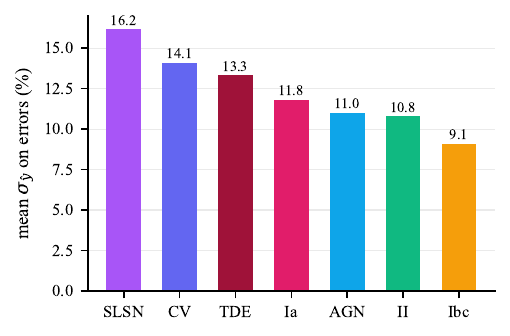}
\caption{\rev{Mean predicted-class ensemble spread $\bar\sigma_{\hat y}$ on mis-classifications, by predicted class.  Rare/hard classes (SLSN, CV, TDE) show large $\sigma_{\hat y}$ on errors; the dense CC classes (Ibc, II) show the smallest $\sigma_{\hat y}$ even on errors -- the ensemble is confidently wrong, in direct analogy to the Regime-I OOD pattern.  These dense classes absorb the bulk of rare-class confusions and are where the cross-family AD layer is necessary as a UQ axis independent of the classifier head.}}
\label{fig:sigma_by_class}
\end{figure}

\paragraph{Per-class anatomy: Regime-I disguise within the in-distribution set.}
\rev{The per-class breakdown of $\sigma_{\hat y}$ on errors (Fig.~\ref{fig:sigma_by_class}) reveals a Regime-I-like pattern inside the in-distribution test set.  Rare/hard classes (SLSN $16.2\%$, CV $14.1\%$, TDE $13.3\%$) show large ensemble spread on errors -- classifier-internal UQ has a clean signal to fire on.  The dense CC classes (Ibc $9.1\%$, II $10.8\%$) show the smallest spread \emph{even on errors} -- the ensemble is confidently wrong, exactly the predicted-class destinations of the Regime-I OOD spectra of Sec.~\ref{sec:per_class_ood}.  Within this regime classifier-internal UQ is silent by construction; only the cross-family AD layer can detect the failure.}

\paragraph{Summary.}
The Regime-I Blazar contrast and this population-level analysis jointly establish that the two UQ axes of \texttt{ASTRANet-CP} are structurally orthogonal detectors of different failure modes.  \rev{Classifier-internal UQ dominates on in-distribution boundary errors (AUROC $0.895$ vs $0.830$); cross-family AD dominates on OOD Regime-I disguise.}  \texttt{ASTRANet-CP} deploys both, and reports both for every input.

\subsection{Selective classification with AD-orthogonal abstention}
\label{sec:selective_classification}

The structural complementarity above has a direct operational consequence: the framework can \emph{refuse to commit} when classifier-internal UQ is low, without compromising the AD layer.  We call this \emph{AD-orthogonal abstention} -- the abstention decision is taken on the UQ axis (which catches in-distribution boundary errors), the anomaly decision independently on the AD axis (which catches Regime-I-style disguise).

\paragraph{Selective accuracy.}
Using top-$1$ confidence as the abstention signal (Table~\ref{tab:selective_classification}), \rev{dropping the most uncertain $10\%$ of predictions lifts selective accuracy from $0.866$ to $0.912$ ($+4.7$\,pp).}

\paragraph{AD-orthogonality.}
Across all coverage levels (Table~\ref{tab:selective_classification}, last column), \rev{$67$--$81\%$} of the spectra rejected by the abstention rule have an AD score \emph{below} the operational $95$th-percentile threshold of \texttt{ASTRANet-Sentinel} -- in-distribution boundary cases, the dense Ibc/II/CV regime of Sec.~\ref{sec:test_error_uncertainty}.  The two operational rules target substantially disjoint populations.

\begin{table}[t]
\centering
\caption{\rev{Selective classification using top-$1$ confidence as the abstention signal on the full held-out test set ($N\!=\!3{,}396$).  Each row drops the top-$X\%$ most uncertain test spectra.  ``AD-low \%'' is the fraction of \emph{rejected} spectra whose AD score lies below the $95$th-percentile threshold of \texttt{ASTRANet-Sentinel}.}}
\label{tab:selective_classification}
\small
\begin{tabular}{cccc}
\hline\hline
Coverage & Selective acc.\ & Gain (pp) & AD-low \% \\
\hline
\rev{$1.00$ (baseline)}    & \rev{$0.866$} & ---     & ---     \\
\rev{$0.95$ (drop $5\%$)}  & \rev{$0.890$} & \rev{$+2.4$}  & \rev{$66.9\%$} \\
\rev{$\mathbf{0.90}$ (drop $10\%$)} & \rev{$\mathbf{0.912}$} & \rev{$\mathbf{+4.7}$}  & \rev{$\mathbf{72.0\%}$} \\
\rev{$0.80$ (drop $20\%$)} & \rev{$0.946$} & \rev{$+8.1$}  & \rev{$80.7\%$} \\
$0.70$ (drop $30\%$) & $0.972$ & $+9.9$  & $84.9\%$ \\
\hline
\end{tabular}
\end{table}

\paragraph{Three-way operational policy.}
The framework therefore supports a three-way decision on two independent axes:
\begin{itemize}
\item \textbf{COMMIT}: $\hat p_{\mathrm{top1}} \geq \tau_{\mathrm{abstain}}$ and $s_{\mathrm{AD}} < \tau_{\mathrm{AD}}$ -- return the point prediction with calibrated confidence.
\item \textbf{ABSTAIN}: $\hat p_{\mathrm{top1}} < \tau_{\mathrm{abstain}}$ and $s_{\mathrm{AD}} < \tau_{\mathrm{AD}}$ -- refuse to commit; forward for human review.
\item \textbf{ALERT}: $s_{\mathrm{AD}} \geq \tau_{\mathrm{AD}}$ -- flag as anomalous regardless of the UQ axis.
\end{itemize}
With $\tau_{\mathrm{abstain}}$ at the $10$th-percentile calibration confidence and $\tau_{\mathrm{AD}}$ at the $95$th-percentile $s_{\mathrm{AD}}$, the framework delivers $\sim$$91\%$ selective accuracy on committed predictions while preserving the $100\%$-recall rare-anomaly capability of Sec.~\ref{sec:operating_point}.  To our knowledge this is the first deployment of AD-orthogonal selective classification in transient spectroscopy.

\subsection{AD-stratified Mondrian Conformal Prediction}
\label{sec:ad_stratified_mondrian}

The AD-conditional diagnostic of Sec.~\ref{sec:conformal_ad} showed that marginal Mondrian concentrates its $5\%$ miscalibration budget in the high-AD operational regime ($0.824 \pm 0.030$ coverage in Q5 vs nominal $0.95$).  This is the expected profile of marginal conformal predictors stratified by their own score \citep{AngelopoulosBates2023, Romano2020APS} and is operationally suboptimal.

\paragraph{Method.}
We introduce \emph{AD-stratified Mondrian Conformal Prediction} (AD-MCP).  Let $q(s_{\mathrm{AD}}(x)) \in \{1,\dots,Q\}$ denote the AD-score quintile under bin edges fitted on the calibration set.  For each (predicted class $c$, AD-quintile $q$) cell we compute
\begin{equation}
\tau_{c,q} = \mathrm{Quantile}_{1-\alpha}\!\bigl(\{\, s(x_i) : \hat y_i = c,\; q(s_{\mathrm{AD}}(x_i)) = q \,\}\bigr),
\label{eq:ad_strat_threshold}
\end{equation}
with $s(x) = 1 - \hat p(\hat y(x) \mid x)$.  A test point fires if $s(x) > \tau_{\hat c, \hat q}$.  Cells with fewer than $n_{\min} = 10$ calibration samples fall back to the marginal class threshold $\tau_c$; with \rev{$1{,}669$} calibration samples and $Q=5$ this activates only for sparse (rare-class, extreme-quintile) cells, not for the dominant Ia/II/CV classes.  Under within-cell exchangeability AD-MCP satisfies the conditional coverage guarantee
\begin{equation}
\Pr\bigl(\, s(x^*) \leq \tau_{c,q} \,\big|\, \hat y(x^*) = c,\; q(s_{\mathrm{AD}}(x^*)) = q \,\bigr) \geq 1-\alpha,
\label{eq:ad_strat_guarantee}
\end{equation}
strictly stronger than the marginal guarantee of vanilla Mondrian.  Building on atypicality-stratified conformal calibration \citep{yuksekgonul2023confidencereliablemodelsconsider}, who stratify APS/RAPS by confidence and single-score atypicality for in-distribution coverage, AD-MCP stratifies instead by a learned cross-family anomaly ensemble crossed with predicted class. To our knowledge it is the first such estimator deployed for out-of-taxonomy discovery in spectroscopic transient triage.

\paragraph{Empirical validation.}
Under $5$-fold cross-validation on the in-distribution held-out set (Table~\ref{tab:ad_strat_coverage}), AD-MCP achieves uniform $0.952 \pm 0.001$ coverage across all five quintiles, including the operationally critical Q5 ($0.953$).  Marginal Mondrian over-covers in Q1\textendash{}Q3 and under-covers in Q5 ($0.824$), a standard deviation across quintiles of $0.066$ -- a factor $\sim$$80$ larger than AD-MCP's $0.0008$.  Both achieve the same marginal coverage averaged across quintiles ($0.952$), but only AD-MCP delivers it \emph{uniformly}.

\begin{table*}[t]
\centering
\caption{Per-quintile empirical Mondrian conformal coverage at $\alpha = 0.05$ on the in-distribution held-out set, under $5$-fold cross-validation.  AD-MCP (this work) achieves uniform $0.95$ coverage across AD-score strata; marginal Mondrian over-covers in Q1--Q3 and under-covers in Q5 by $0.13$.}
\label{tab:ad_strat_coverage}
\small
\begin{tabular}{lccc}
\hline\hline
Quintile & Marginal Mondrian & AD-stratified Mondrian (this work) & Improvement \\
\hline
Q1 (low AD)     & $1.000$            & $0.9512$ & $-0.049$ (toward nominal) \\
Q2              & $1.000$            & $0.9519$ & $-0.048$ (toward nominal) \\
Q3              & $0.989$            & $0.9514$ & $-0.038$ (toward nominal) \\
Q4              & $0.947$            & $0.9521$ & $+0.006$ \\
\textbf{Q5} (high AD, operational) & $\mathbf{0.824}$ & $\mathbf{0.9534}$ & $\mathbf{+0.129}$ (toward nominal) \\
\hline
Std across $\{Q1,\dots,Q5\}$ & $0.066$ & $\mathbf{0.0008}$ & $\sim$$80\times$ tighter \\
\hline
\end{tabular}
\end{table*}

\paragraph{Width\textendash{}coverage trade-off and \texttt{BOOM} integration.}
Uniform conditional coverage is bought with smaller per-cell sample sizes ($n_{c,q} \approx 47$ vs $n_c \approx 233$ marginal); the net effect is the conditional-vs-marginal trade-off predicted by \citet{Vovk2005} theory \textendash{} AD-MCP exchanges marginal sharpness in Q1\textendash{}Q3 for conditional validity in Q5, the operationally relevant property.  Under the planned \texttt{BOOM} integration the rolling calibration window will accumulate $\sim$$10^4$ ID spectra, exceeding $n_{\min}$ for all cells and removing the fallback dependence.  AD-MCP is the conformal layer intended for offline production triage at $\alpha = 0.05$.

\section{Case Studies: The Five Failure Modes Resolved}
\label{sec:case_studies}

We now revisit the five failure-mode regimes of Sec.~\ref{sec:failure_modes} from the operational perspective of the deployed framework.  For each regime we give the five-component confidence-aware output and an astrophysical interpretation; the per-spectrum anomaly scores for all seven case-study spectra are in Table~\ref{tab:anomaly_case_study}, and the per-class ensemble intervals in Table~\ref{tab:per_class_intervals}.  Our goal is not to repeat numbers already given but to show how the framework's outputs jointly distinguish the five regimes for a survey operator.

\begin{table*}[t]
\centering
\caption{Per-class case study illustrating complementarity in both directions, with scores read directly from the production run.  \emph{Regime A} (top): high-confidence cases where uncertainty-based scores do not fire; detection is driven by within-class Mahalanobis and GMM, and the distance and density family ensembles both fire.  \emph{Regime B} (middle): high-confidence cases where the embedding itself is deceived (typicality $<1$) and no single-family mini-ensemble fires at $1\%$ FAR; detection comes from cross-family non-linear combination only.  \emph{Regime C} (bottom): low-confidence near-OOD cases where uncertainty fires but is borderline; distance/density evidence carries the sample over threshold.  ``Conf.''\ is fine-head confidence; ``Entropy'' fine-head softmax entropy; ``Typ.''\ within-class typicality; ``Mahal.''\ within-class Mahalanobis distance; ``iForest'' the isolation-forest score; ``HEC'' the hierarchical entropy combination; ``Fam.''\ the number of family-only mini-ensembles (of four) that flag the spectrum at $1\%$ FAR; $s_\mathrm{ens}$ the LightGBM ensemble score.  Bold marks the dominant detection score(s) per row; bold $\mathbf{0}$ in ``Fam.''\ highlights the Regime-B spectra no family alone catches.}
\label{tab:anomaly_case_study}
\small
\begin{tabular}{lllcccccccc}
\toprule
Object ID & True class & Pred. & Conf. & Entropy & Typ. & Mahal. & iForest & HEC & Fam. & $s_\mathrm{ens}$ \\
\midrule
\multicolumn{11}{l}{\emph{Regime A \textendash{} uncertainty fails (high confidence, low entropy)}} \\
\rev{AT2021lfa} & \rev{afterglow} & \rev{CV}  & \rev{$0.761$} & \rev{$0.77$} & \rev{$1.24$} & \rev{$\mathbf{61.0}$} & \rev{$\mathbf{0.54}$} & \rev{$0.97$} & \rev{$2$} & \rev{$0.991$} \\
\rev{AT 2026dbl} & \rev{FBOT}      & \rev{CV}  & \rev{$0.757$} & \rev{$0.85$} & \rev{$1.21$} & \rev{$\mathbf{55.0}$} & \rev{$\mathbf{0.51}$} & \rev{$1.03$} & \rev{$2$} & \rev{$0.990$} \\
\midrule
\multicolumn{11}{l}{\rev{\emph{Regime B \textendash{} both UQ and distance fail (typicality $\lesssim 1$, at most one family-only mini-ensemble fires)}}} \\
\rev{AT 2024uln} & \rev{Blazar}    & \rev{CV}  & \rev{$0.927$} & \rev{$0.38$} & \rev{$0.87$} & \rev{$24.0$} & \rev{$0.52$} & \rev{$0.44$} & \rev{$1$} & \rev{$0.927$} \\
\rev{AT 2024wpp} & \rev{FBOT}      & \rev{CV}  & \rev{$0.716$} & \rev{$0.98$} & \rev{$1.15$} & \rev{$44.0$} & \rev{$0.48$} & \rev{$1.40$} & \rev{$\mathbf{0}$} & \rev{$0.977$} \\
\midrule
\multicolumn{11}{l}{\emph{Regime C \textendash{} corroboration: uncertainty fires but borderline}} \\
\rev{AT2022cva} & \rev{long GRB}  & \rev{TDE} & \rev{$0.518$} & \rev{$\mathbf{1.24}$} & \rev{$1.30$} & \rev{$22.0$} & \rev{$0.53$} & \rev{$\mathbf{1.66}$} & \rev{$2$} & \rev{$0.824$} \\
\rev{2021vdr} & \rev{ILRT}      & \rev{II}   & \rev{$0.337$} & \rev{$\mathbf{1.25}$} & \rev{$0.97$} & \rev{$12.0$} & \rev{$0.48$} & \rev{$\mathbf{1.98}$} & \rev{$0$} & \rev{$0.815$} \\
\rev{2021zfp} & \rev{Ca-rich}   & \rev{CV}   & \rev{$0.461$} & \rev{$1.22$} & \rev{$1.26$} & \rev{$40.0$} & \rev{$0.51$} & \rev{$\mathbf{2.09}$} & \rev{$1$} & \rev{$0.907$} \\
\bottomrule
\end{tabular}
\end{table*}

\subsection{Regime I \textendash{} Continuum-disguised AGN-class: \rev{\texttt{AT 2024uln}}}
\label{sec:case_blazar}

\rev{The Blazar \texttt{ AT 2024uln} is the canonical case of continuum-disguised AGN-class events.  The classifier maps the spectrum confidently to CV ($\hat p_{\mathrm{CV}} = 0.927$), the predicted-class entropy is low, and the within-class typicality is below unity \textendash{} the embedding is mapped \emph{inside} the typical radius of its predicted CV class.  Only a single family-only mini-ensemble fires weakly at $1\%$ FAR, yet the cross-family LightGBM ensemble assigns $s_{\mathrm{ens}} = 0.927$.  The confidence-aware output at $\alpha = 0.05$ is}
\begin{equation}
  \rev{(\hat y, \;p_{\mathrm{M}}, \;\mathcal{C}_{0.05}, \;H_{\mathrm{epi}}) = (\mathrm{CV}, \;0.046, \;\{\mathrm{CV}\}, \;\text{low}).}
\end{equation}
The Mondrian p-value flags the spectrum; the prediction set is a singleton on the wrong class; the low epistemic uncertainty (all six ensemble variants agree on CV) confirms the failure mode is \emph{disguise}, not \emph{model disagreement}.  A high anomaly score on a confident-singleton prediction is the textbook signature of Regime~I.

\paragraph{Classifier-internal UQ alone misses Regime~I.}
The per-sample $p_\mathrm{wrong}$ output of component~(ii) of Sec.~\ref{sec:uq_output}, computed against the CV-predicted calibration distribution, \rev{yields $p_\mathrm{wrong}(x) = 0.67$ for this spectrum \textendash{} a moderate signal, well above the $\alpha = 0.05$ threshold: $\hat p_{\mathrm{CV}} = 0.927$ places the Blazar} in the lower-confidence half of CV-predicted calibration samples (typical confidence $\geq 0.95$) but not in the extreme tail.  Classifier-internal uncertainty alone \textendash{} whether softmax confidence or conformal-on-softmax as here \textendash{} is structurally incapable of catching this regime: the failure mode \emph{is} a confident wrong commitment, by definition.  \rev{The strong signal ($p_\mathrm{M} = 0.046$, a factor $\sim\!15\times$ stronger than $p_\mathrm{wrong}$) is carried entirely by the cross-family AD layer.  This single contrast \textendash{} $p_\mathrm{wrong} = 0.67$ versus $p_\mathrm{M} = 0.046$ on the same spectrum \textendash{}} is the empirical case for the cross-family AD layer over any classifier-internal UQ, and is the principal methodological argument of the paper.

Physically, this is consistent with the Blazar's continuum-dominated optical spectrum at the epoch of observation, the underlying AGN power-law continuum overlapping the accretion continuum of cataclysmic variables in the wavelength range accessible to SEDM.

\subsection{Regime II \textendash{} Spectroscopic disguise: \rev{\texttt{2021zfp}} and novae}
\label{sec:case_carich}

\rev{The Ca-rich transient \texttt{2021zfp} is the archetype of spectroscopic disguise within physical neighbours.  The classifier predicts CV with low confidence ($\hat p_{\mathrm{CV}} = 0.46$); fine-head entropy is elevated; within-class Mahalanobis is moderately high; HEC is the highest of any case-study spectrum.  Only one cascade method fires at $1\%$ FAR, but the cross-family ensemble integrates the moderate signals into $s_{\mathrm{ens}} = 0.907$.  The confidence-aware output is
\begin{equation}
  (\hat y, \;p_{\mathrm{M}}, \;\mathcal{C}_{0.10}) = (\mathrm{CV}, \;\sim\!0.12, \;\{\mathrm{CV}, \mathrm{Ibc}, \mathrm{II}\}),
\end{equation}
where the APS prediction set at $\alpha = 0.10$ contains, alongside the point prediction, the two physically plausible classes (stripped-envelope or H-rich CC).  This is the genuinely informative case: the wide prediction set with near-equal ensemble mass on Ibc ($30.5\% \pm 15.1\%$) tells the operator the framework refuses a confident CV commitment.  The novae populations show the same behaviour at scale: of $67$ Novae in the rare set, $4$ are caught by any family alone but $54/67$ by the full cross-family pipeline \textendash{} the strongest population-scale demonstration of cross-family complementarity in the paper.}

\subsection{Regime III \textendash{} Continuum-dominated: \texttt{AT2021lfa}}
\label{sec:case_afterglow}

The GRB afterglow \texttt{AT2021lfa} \citep{2025ApJ...985..124L} represents continuum-dominated mis-classifications where, unlike Regime~I, the embedding is genuinely far from the predicted-class centroid: \rev{within-class Mahalanobis is in the extreme tail and the isolation forest fires, so both the distance and density family ensembles fire independently.  The output $(\hat y, p_{\mathrm{M}}, \mathcal{C}_{0.05}, H_{\mathrm{epi}}) = (\mathrm{CV}, \ll\!0.01, \{\mathrm{CV}\}, \text{high})$ places the spectrum in the extreme anomaly tail of the CV calibration distribution despite the classifier's commitment.  $16$ of the $17$ afterglow spectra in the rare set are recovered at the $\alpha = 0.10$ operating point.}

\subsection{Regime IV \textendash{} Ambiguous progenitor: the FBOT class}
\label{sec:case_fbot}

\rev{FBOTs are the largest near-OOD class in our rare set ($N\!=\!98$) and the textbook ambiguous-progenitor regime.  Predictions scatter across CV, TDE, Ia, and Ibc; family-vote counts span the full $0$\textendash{}$4$ range.  The framework's behaviour is correspondingly heterogeneous: the APS sets widen substantially as $\alpha$ tightens from $0.05$ to $0.01$ (mean ID set size grows from $1.4$ to $2.1$; the framework refuses to commit).  The most interesting FBOT is \texttt{AT 2024wpp}, structurally close to Regime~I (confident CV prediction, no family flagging) but physically distinct \textendash{} the FBOT spectrum genuinely overlaps with CV templates at certain phases due to the blue continuum.}  The framework flags both as anomalous; the physical distinction between Regimes~I and~IV must be made downstream with light-curve and host-galaxy information \citep{Margutti2019, Perley2019, Ho_2023}.

\subsection{Regime V \textendash{} Far-OOD: \texttt{AT2022cva}}
\label{sec:case_grb}

\rev{The long GRB \texttt{AT2022cva} is the textbook far-OOD case.  The classifier assigns it to TDE with moderate confidence ($0.52$); fine-head entropy is among the highest of the case-study spectra; two families produce moderate signal.  The output $(\hat y, p_{\mathrm{M}}, \mathcal{C}_{0.10}, H_{\mathrm{epi}}) = (\mathrm{TDE}, \sim\!0.087, \{\mathrm{TDE}, \mathrm{SLSN}\}, \text{high})$ exhibits high epistemic uncertainty ($\sigma_{\mathrm{TDE}} = 18.2\%$) \textendash{} the ensemble variants disagree about the classification.  This is the diagnostic for Regime~V: the failure is genuine novelty, not disguise.}

\paragraph{Summary.}
The five case studies confirm the design implication of Sec.~\ref{sec:failure_modes}: the five-component output jointly distinguishes disguise (Regime~I/II \textendash{} singleton prediction set on a wrong class, low MI, and for Regime~I tight ensemble consensus on the wrong class), continuum-dominated mis-classification (Regime~III \textendash{} far-from-centroid embedding), genuine ambiguity (Regime~IV \textendash{} broad prediction set with epoch-dependent intervals), and novelty (Regime~V \textendash{} broad prediction set with high MI).  No single component suffices, \rev{and the Regime~I contrast between tight ensemble consensus ($\sigma_{\mathrm{CV}}\!\approx\!5\%$) and a flagged Mondrian p-value ($p_{\mathrm{M}}\!=\!0.046$) is the empirical case for cross-family AD as a structurally distinct UQ axis.}

\section{Discussion and Conclusions}
\label{sec:discussion}
\label{sec:conclusions}

We have presented \texttt{ASTRANet}, a confidence-aware framework for spectroscopic transient discovery built from a hierarchical classifier that operates on observer-frame spectra without host redshift or spectral phase, a cross-family anomaly detection layer (\texttt{ASTRANet-Sentinel}), and a conformal uncertainty layer (\texttt{ASTRANet-CP}).  Two arguments organise the work.  The first is physical: the failure modes of closed-set spectroscopic classifiers are not a uniform softmax-overconfidence phenomenon but separate into five regimes with distinct anomaly-score signatures, two of which \textendash{} continuum-disguised AGN-class events and spectroscopic disguise within physical neighbours, together $\sim$$43\%$ of our rare population \textendash{} are recoverable only through cross-family non-linear combination.  The second is statistical: classifier-internal uncertainty and embedding-based anomaly detection are structurally complementary axes, dominating on in-distribution boundary errors and out-of-distribution disguise respectively, and should be deployed together rather than interchangeably.

\paragraph{What the framework enables.}
Operationally, the framework converts spectroscopic triage from an uncalibrated judgment into allocation under quantified risk: each candidate carries a conformal p-value, a prediction set of plausible classes, and an aleatoric/epistemic decomposition, so that high-AD high-MI candidates (novel \emph{and} ambiguous) can be separated from high-AD low-MI candidates (novel but confidently mislabelled \textendash{} the disguised-class regime) and from routine low-AD candidates.  The finding that $\sim$$43\%$ of the rare population is caught by no single anomaly-score family means a pipeline built on softmax confidence, Mahalanobis distance, or isolation forest alone silently loses roughly half this population by construction; the framework is the smallest machinery that recovers it.  We are preparing the framework for offline production integration with the ZTF \texttt{BOOM} broker \citep{BOOM} as a daily batch pipeline delivering anomaly reports to a Slack alert channel; the spectroscopic follow-up campaign of flagged anomalies will be reported in companion work.

\paragraph{Why embedding-based scores work.}
Modern networks produce overconfident softmax outputs on OOD inputs \citep{Guo2017Calibration, Hein2019Why}, which translates directly into weak logit-space OOD detection.  The embedding space, structured by the supervised-contrastive loss, instead encodes distance from the training manifold: an OOD spectrum need not look uncertain in logit space, only far from in-distribution embeddings, which the Mahalanobis and GMM scores quantify directly.  Averaging embeddings across the $M=6$ ensemble variants further stabilises these distances.

\paragraph{Limitations.}
Several limitations bound the present results.  \emph{Sample size:} the Ca-rich ($N\!=\!10$) and Classical Nova ($N\!=\!8$) classes are the smallest in the rare set and also the strongest Regime~II disguise cases, so their per-class recovery is the most uncertain.  \emph{Single modality:} the framework uses one spectroscopic epoch, yet the hardest near-OOD cases (Ca-rich vs Ibc, ILRT vs IIb, FBOT vs TDE) are separable photometrically within a few epochs \textendash{} a multi-modal extension ingesting light curves and host imagery, in the spirit of \texttt{AppleCiDEr} \citep{applecider}, is the natural next step and is enabled by the embedding-based architecture. 

Relative to prior transient anomaly detection, which has focused on photometric light curves \citep{Villar2021, Pruzhinskaya2019, Gupta2025MCIF}, the closest methodological prior is the distance-metric-ensemble paradigm of DiMMAD \citep{DiMMAD2025}, itself built on the DistClassiPy distance-based classifier \citep{Chaini_2024}. DiMMAD ensembles 16 distance metrics — i.e. distinct geometries over a single photometric feature space — and reports that multi-metric consensus excels precisely at off-manifold OOD discovery while a single isolation forest remains competitive for rare in-distribution subtypes. ASTRANet generalises this paradigm along an orthogonal axis: rather than ensembling geometries of one distance notion, it combines four physically motivated score families (distance, density, energy/uncertainty, and hierarchical hybrids) non-linearly over learned spectroscopic embeddings, with calibrated prediction sets. This distinction is what lets the cross-family combiner recover the near-OOD, disguised-class regime (Regimes I–II) where any single family — distance included — falls below threshold.

\paragraph{The path to LSST.}
The Vera C.\ Rubin Observatory will discover $\sim$$10^6$ transient candidates per year, of which $\sim$$10^4$ will receive spectroscopic follow-up via e.g., 4MOST-TiDES \citep{Swann4MOST, Frohmaier4MOST} and SOXS \citep{Schipani2020SOXS}.  Beyond these near-term facilities, proposed next-generation spectroscopic arrays — e.g. an array of >100 AO-enabled 1–4 m telescopes for time-critical transient spectroscopy \citep{Groot2025} — would push follow-up capacity further still; in every such regime, automated confidence-aware triage is the prerequisite for deciding which spectra such an array should target. At this rate triage cannot remain a human-in-the-loop process, and every $1\%$ improvement in confidence-aware triage corresponds to $\sim$$10^2$ peculiar transients per year that would otherwise be lost to routine classification.  The framework is scalable to this regime by design: inference is below survey cadence, the conformal calibration is distribution-free and recalibrable nightly, and the failure-mode taxonomy extends to any near-degenerate class the training distribution under-samples.  To our knowledge this is the first systematic application of confidence-aware multi-category anomaly detection and uncertainty quantification to astronomical spectroscopy; frameworks of this kind will be required \textendash{} not optional \textendash{} for discovery in the Rubin era.

\vspace{1em}
\section*{Acknowledgments}
The UMN authors acknowledge support from the National Science Foundation with grant numbers PHY-2117997, PHY-2308862 and PHY-2409481. A.S. and M.C. acknowledge support from LSST-DA through grants 2025-SFF-LFI-14-Coughlin and 2026-SFF-LFI-17-Sasli. LINCC Frameworks is supported by Schmidt Sciences, a philanthropic initiative founded by Eric and Wendy Schmidt, as part of the Virtual Institute of Astrophysics (VIA).

Based on observations obtained with the Samuel Oschin Telescope 48-inch and the 60-inch Telescope at the Palomar Observatory as part of the Zwicky Transient Facility project. ZTF is supported by the National Science Foundation under Grants No. AST-1440341, AST-2034437, and currently Award \#2407588. ZTF receives additional funding from the ZTF partnership. Current members include Caltech, USA; Caltech/IPAC, USA; University of Maryland, USA; University of California, Berkeley, USA; Cornell University, USA; Drexel University, USA; University of North Carolina at Chapel Hill, USA; Institute of Science and Technology, Austria; National Central University, Taiwan, German Center for Astrophysics, Germany, and OKC, University of Stockholm, Sweden. Operations are conducted by Caltech's Optical Observatory (COO), Caltech/IPAC, and the University of Washington at Seattle, USA.

SED Machine is based upon work supported by the National Science Foundation under Grant No.\ 1106171. The Gordon and Betty Moore Foundation, through both the Data-Driven Investigator Program and a dedicated grant, provided critical funding for SkyPortal.

\appendix

\section{Detailed architecture}
\label{app:architecture}

This appendix expands the high-level description of \texttt{ASTRANet} given in Sec.~\ref{sec:architecture}.  The model takes a two-channel 1D signal $(B, 2, L)$ of length $L = 4096$ as input: the robust-scaled flux and its Savitzky\textendash{}Golay first derivative \citep[window $7$, polynomial order $3$;][]{savitzky1964smoothing}, computed \emph{after} augmentation.

\paragraph{Multi-scale stem.}
Three parallel convolutional branches with kernel sizes $k \in \{7, 31, 151\}$ each apply a 1D convolution, group normalisation \citep[8 groups;][]{wu2018group}, GELU activation \citep{hendrycks2016gelu}, and dropout, each producing $C\!=\!32$ channels.  Outputs are concatenated and fused through a $1{\times}1$ convolution to a $C_\mathrm{base}\!=\!96$-channel map.  Temporal average pooling with stride $4$ reduces sequence length, and a final $1{\times}1$ projection expands to the embedding size $d_\mathrm{emb} = 192$.

\paragraph{Dilated TCN backbone.}
Six residual blocks \citep{bai2018empirical}, each with two convolutional layers (kernel size $5$, group normalisation, GELU): the first layer uses exponentially increasing dilation $r_i = 2^{i \bmod 6}$, the second uses dilation $1$.  A residual skip connection adds the block input to its output.  Stochastic depth \citep{huang2016deep} with linearly increasing drop probability $0 \to 0.1$ provides implicit ensemble regularisation.  Cumulative receptive field: $\sim$$250$ samples.

\paragraph{Metadata conditioning via FiLM.}
The metadata vector $\mathbf{c} \in \mathbb{R}^{64}$ encodes three quantities: spectral phase $\phi$ (rest-frame days from maximum light, represented as $[\sin\phi', \cos\phi', \tanh\phi', \mathrm{validity}]$ with $\phi' = \phi/30$\,d), host redshift $z$ (as $[\tanh(5z), \mathrm{validity}]$), and an instrument-identifier embedding ($N_\mathrm{inst}\!=\!16$, $d_\mathrm{inst}\!=\!16$).  A two-layer MLP ($28 \to 128 \to 64$) with LayerNorm, GELU, and residual projection produces $\mathbf{c}$.  The conditioning vector modulates the feature map at two points via FiLM \citep{perez2018film}:
\begin{equation}
  \mathrm{FiLM}(\mathbf{x}, \mathbf{c}) = \mathbf{x} \odot [1 + \boldsymbol{\gamma}(\mathbf{c})] + \boldsymbol{\beta}(\mathbf{c}),
  \label{eq:film}
\end{equation}
with $\boldsymbol{\gamma}$ and $\boldsymbol{\beta}$ learned linear projections.  FiLM$_1$ acts after the stem, before the TCN; FiLM$_2$ acts on the pooled embedding before the heads.

\paragraph{Multi-head attention pooling.}
Four heads ($d_\mathrm{emb}/H = 48$ dimensions each) score the TCN output via a small MLP ($48 \to 24 \to 1$) and softmax over the sequence dimension.  Head outputs $\mathbf{v}_h = \sum_t \alpha_h(t)\,\mathbf{x}_h(t)$ are concatenated and linearly projected to the embedding $\mathbf{e} \in \mathbb{R}^{192}$ (LayerNorm + dropout $p=0.15$).  The attention entropy $H_\mathrm{attn} = H^{-1}\sum_h(-\sum_t \alpha_h(t)\log\alpha_h(t))$ enters the loss as a regularisation bonus.

\paragraph{Hierarchical heads.}
The coarse head ($K_c\!=\!3$ superclasses) is a two-layer MLP ($192 \to 256 \to 3$) with LayerNorm, GELU, dropout $p=0.4$.  The fine head ($K_f\!=\!7$ classes) receives the embedding concatenated with the coarse softmax probabilities under stop-gradient (gradients flow only through the fine head), passed through a two-layer MLP ($195 \to 384 \to 7$) with LayerNorm, GELU, dropout $p=0.5$.  Contrastive projection heads (3-layer MLPs $192 \to 256 \to 256 \to 128$ for fine, $192 \to 128 \to 128 \to 128$ for coarse) supply $\ell_2$-normalised vectors $\mathbf{z}^{(\mathrm{fine})}, \mathbf{z}^{(\mathrm{coarse})}$ for the SupCon loss.

\section{Loss function and training procedure}
\label{app:training}

\paragraph{Loss function.}
The total objective is:
\begin{equation}
\begin{aligned}
  \mathcal{L} ={} & w_\mathrm{f}\,\mathcal{L}_\mathrm{focal} + w_\mathrm{c}\,\mathcal{L}_\mathrm{coarse} \\
  &+ w_\mathrm{sc,f}\,\mathcal{L}_\mathrm{SupCon}^{(\mathrm{fine})} + w_\mathrm{sc,c}\,\mathcal{L}_\mathrm{SupCon}^{(\mathrm{coarse})} \\
  &+ w_\mathrm{conf}\,\mathcal{L}_\mathrm{confusion} - w_\mathrm{H}\, H_\mathrm{attn},
\end{aligned}
\label{eq:loss}
\end{equation}
with weights listed in Table~\ref{tab:loss_weights}. The fine head uses focal loss \citep{lin2017focal} with label smoothing $\epsilon = 0.05$ and class-specific $\gamma_k$ ($\gamma = 2$ default; $\gamma_\mathrm{AGN} = \gamma_\mathrm{TDE} = 3$). The coarse head uses cross-entropy with the same smoothing. The supervised contrastive loss \citep{khosla2020supervised} is applied at temperature $\tau = 0.07$ on both granularities (disabled on MixUp batches). The confusion penalty suppresses targeted misclassification pathways: $(a\!\to\!b, w)$ pairs $\{(\mathrm{TDE}\!\to\!\mathrm{Ia}, 0.5), (\mathrm{SLSN}\!\to\!\mathrm{Ibc}, 0.3), (\mathrm{II}\!\to\!\mathrm{Ibc}, 0.8)\}$ via a quadratic penalty on $p(b\mid x)$ averaged over samples with true class $a$. The attention entropy bonus is subtracted (maximised) to keep attention spread.

\begin{table}[h]
  \centering
  \caption{Loss term weights used in all experiments.}
  \label{tab:loss_weights}
  \begin{tabular}{lcc}
    \hline
    Term & Symbol & Weight \\
    \hline
    Focal (fine)          & $w_\mathrm{f}$      & 1.5  \\
    Cross-entropy (coarse)& $w_\mathrm{c}$      & 0.3  \\
    SupCon (fine)         & $w_\mathrm{sc,f}$   & 0.03 \\
    SupCon (coarse)       & $w_\mathrm{sc,c}$   & 0.05 \\
    Confusion penalty     & $w_\mathrm{conf}$   & 0.5  \\
    Attention entropy     & $w_\mathrm{H}$      & 0.01 \\
    \hline
  \end{tabular}
\end{table}

\paragraph{Training procedure.}
Five stochastic augmentations are applied per spectrum \emph{before} the SG derivative is computed: sub-pixel wavelength shift ($p=0.3$, $\delta \sim \mathcal{U}(-3, 3)$ pixels); multiplicative continuum warping ($p=0.3$, $5$-knot spline, amplitude $\pm 0.05$); flux scaling ($p=0.4$, $s \sim \mathcal{U}(0.85, 1.15)$); Gaussian noise ($p=0.5$, $\mathrm{SNR} \sim \mathcal{U}(10, 50)$); random masking of $1$\textendash{}$3$ contiguous regions covering $5$\textendash{}$30\%$ of the spectrum ($p=0.5$). At batch level, MixUp \citep{zhang2018mixup} is applied with $p=0.5$ and $\lambda = \max(\lambda', 1-\lambda')$, $\lambda' \sim \mathrm{Beta}(0.3, 0.3)$. Class imbalance is addressed through effective-number-based oversampling \citep{cui2019class} with $\beta = 0.9999$. The optimiser is AdamW \citep{loshchilov2019decoupled} with learning rate $\eta_0 = 3\times10^{-4}$, weight decay $10^{-4}$, gradient clip $1.0$. The schedule uses a $5$-epoch linear warmup followed by cosine annealing with warm restarts \citep{loshchilov2017sgdr}, $T_0 = 30$, $T_\mathrm{mult} = 2$, $\eta_\mathrm{min} = 0.05\,\eta_0$. EMA \citep{polyak1992acceleration} of model weights is maintained with decay $\alpha = 0.996$ and used for all validation/test evaluations. Early stopping uses validation macro recall with patience $45$ epochs (max $200$ epochs). Full hyperparameter listing in Table~\ref{tab:hyperparams}.

\begin{table}[h]
  \centering
  \caption{Hyperparameters for \texttt{ASTRANet}.}
  \label{tab:hyperparams}
  \small
  \begin{tabular}{llc}
    \hline
    \multicolumn{2}{l}{Parameter} & Value \\
    \hline
    \multicolumn{3}{l}{\emph{Architecture}} \\
    & Input channels & 2 (flux + SG deriv.) \\
    & SG filter (window / poly.) & 7 / 3 \\
    & Stem kernels & 7, 31, 151 \\
    & Base channels $C_\mathrm{base}$ & 96 \\
    & Embedding dim. $d_\mathrm{emb}$ & 192 \\
    & TCN depth $D$ & 6 blocks \\
    & TCN kernel size & 5 \\
    & Temporal downsample & $4\times$ \\
    & Attention heads $H$ & 4 \\
    & Stochastic depth & 0.0\textendash{}0.1 (linear) \\
    & Dropout (backbone / emb.) & 0.2 / 0.15 \\
    & Normalisation & GroupNorm (8 groups) \\
    & Activation & GELU \\
    & Metadata output dim. & 64 \\
    & Instrument embeddings & 16 $\times$ 16 \\
    & Projection dim. & 128 \\
    \hline
    \multicolumn{3}{l}{\emph{Training}} \\
    & Batch size & 128 \\
    & Max epochs & 200 \\
    & Optimiser & AdamW \\
    & Learning rate $\eta_0$ & $3 \times 10^{-4}$ \\
    & Weight decay & $10^{-4}$ \\
    & Gradient clip norm & 1.0 \\
    & LR warmup & 5 epochs (linear) \\
    & LR schedule & Cosine warm restarts \\
    & $T_0$ / $T_\mathrm{mult}$ / $\eta_\mathrm{min}$ & 30 / 2 / $0.05\eta_0$ \\
    & EMA decay $\alpha$ & 0.996 \\
    & MixUp $\alpha_\mathrm{mix}$ / prob. & 0.3 / 0.5 \\
    & Balanced sampling $\beta$ & 0.9999 \\
    & Label smoothing $\epsilon$ & 0.05 \\
    & Focal $\gamma$ (default) & 2.0 \\
    & $\gamma$ overrides & AGN=3, TDE=3, SLSN=2 \\
    & SupCon temperature $\tau$ & 0.07 \\
    & Early stopping patience & 45 epochs \\
    \hline
  \end{tabular}
\end{table}

\section{The sixteen anomaly scores}
\label{app:ad_scores}

Formal definitions of the $16$ anomaly scores summarised in Sec.~\ref{sec:ood_metrics}. We adopt the convention \emph{higher = more anomalous}. Let $\mathbf{p} = \mathrm{softmax}(\mathbf{z})$ be the fine-head softmax of an input with fine logits $\mathbf{z} \in \mathbb{R}^{K_f}$, $\mathbf{p}_c = \mathrm{softmax}(\mathbf{z}_c)$ the coarse softmax, $\mathbf{e}$ the embedding, and $\{\boldsymbol{\mu}_k, \boldsymbol{\Sigma}_k\}_{k=1}^{K_f}$ the per-class mean and covariance of training embeddings.

\paragraph{Uncertainty family ($4$ scores).}
$s_\mathrm{MSP} = -\max_k p_k$; $s_\mathrm{ent} = -\sum_k p_k \log p_k$; $s_\mathrm{en} = -\log\sum_k \exp(z_k)$ (energy, \citealp{Liu2020Energy}); $s_\mathrm{ODIN} = -\max_k \mathrm{softmax}(\mathbf{z}/T)_k$ with $T = 1.5$ (temperature-only ODIN, \citealp{Liang2018}).

\paragraph{Distance family ($6$ scores; operate on $\mathbf{e}$).}
Global Mahalanobis $s_\mathrm{Mglob} = \sqrt{(\mathbf{e}-\boldsymbol{\mu})^\top \boldsymbol{\Sigma}^{-1}(\mathbf{e}-\boldsymbol{\mu})}$; class-min Mahalanobis $s_\mathrm{Mclass} = \min_k \sqrt{(\mathbf{e}-\boldsymbol{\mu}_k)^\top \boldsymbol{\Sigma}_k^{-1}(\mathbf{e}-\boldsymbol{\mu}_k)}$; within-prediction Mahalanobis (same formula, conditioned on $\hat y_f$); $k$-NN distance $s_\mathrm{kNN} = k^{-1}\sum_{j=1}^{k}\|\mathbf{e}-\mathbf{e}_{(j)}\|_2$ with $k = 10$; cosine distance $s_\mathrm{cos} = 1 - \max_c \frac{\mathbf{z}^{(\mathrm{fine})} \cdot \boldsymbol{\mu}_c^{(\mathrm{fine})}}{\|\mathbf{z}^{(\mathrm{fine})}\|\,\|\boldsymbol{\mu}_c^{(\mathrm{fine})}\|}$ (in the contrastive projection space); typicality $s_\mathrm{typ} = \|\mathbf{e}-\boldsymbol{\mu}_{\hat y_f}\|_2 / R_{\hat y_f}$, with $R_k$ the $95$th percentile of $\|\mathbf{e}_i - \boldsymbol{\mu}_k\|_2$ over training samples of class $k$.

\paragraph{Density family ($4$ scores).}
Per-class GMM negative log-likelihood under the maximum-likelihood class; isolation forest \citep{Liu2008iForest} ($200$ estimators); local outlier factor \citep{Breunig2000LOF} ($k = 20$); PCA reconstruction error $s_\mathrm{PCA} = \min_c \|\mathbf{e} - \mathbf{P}_c \mathbf{P}_c^\top \mathbf{e}\|_2$ with $\mathbf{P}_c$ the orthonormal projection onto the class-$c$ PCA subspace.

\paragraph{Hybrid family ($2$ scores).}
Hierarchical Entropy Combination: $s_\mathrm{HEC} = a H_f + b H_c + g h_i$ with $(a, b, g) = (1.0, 0.5, 1.0)$, $H_f$ and $H_c$ the fine/coarse softmax entropies, and $h_i = 1 - p_c[\pi(\hat y_f)]$ the parent-child inconsistency. Manifold Residual Score: $s_\mathrm{MRS} = \min_c \|\mathbf{e} - \mathrm{recon}_c(\mathbf{e})\|_2$ using per-class PCA with $k = 12$ components.

\paragraph{Score normalisation and LightGBM hyperparameters.}
Each score is rank-normalised on the validation set to a percentile in $[0, 1]$.  The LightGBM combiner uses $300$ estimators, $\eta = 0.05$, $\max_\mathrm{depth} = -1$, $\mathrm{num\_leaves} = 31$, class-balanced weighting; hyperparameters were chosen by $5$-fold CV on the validation set.  The final feature importances rank \texttt{cosine}, \texttt{energy}, \texttt{mahal\_global}, \texttt{mahal\_within}, \texttt{mahal\_class}, and \texttt{gmm} at the top, with \texttt{pca\_recon} and \texttt{mrs} lowest.

The Spearman rank-correlation matrix between the $16$ scores shows three tight high-correlation blocks -- the four uncertainty scores ($\rho > 0.96$ within), the Mahalanobis variants ($\rho > 0.91$), and the density/reconstruction methods ($\rho > 0.94$).  This within-family redundancy is why a naive average over all $16$ scores underperforms and why the decorrelated logistic-regression baseline on the three least-correlated methods $\{$\texttt{mahal\_global}, \texttt{lof}, \texttt{msp}$\}$ already reaches AUROC $0.9394$, better than any single score; the cross-family complementarity that the gradient-boosted combiner exploits is established directly by the family ablation of Sec.~\ref{sec:complementarity}.

\paragraph{Full per-score benchmark.}
Table~\ref{tab:ood_metrics} reports AUROC and AUPRC for the $16$ individual OOD scores and the four ensemble strategies on the rare-anomaly evaluation set ($N\!=\!289$ vs.\ in-distribution validation pool of \rev{$3{,}339$}), with $95\%$ bootstrap confidence intervals over $2000$ resamples.

\begin{table*}
\centering
\caption{OOD detection performance on the rare-anomaly evaluation set ($N\!=\!289$ rare/peculiar/out-of-taxonomy transients vs.\ a held-out validation pool of $3{,}339$ in-distribution samples).  AUROC and AUPRC for the $16$ individual scores and the four ensemble strategies, with $95\%$ bootstrap confidence intervals over $2000$ resamples.  \rev{The best individual method (typicality) and the best ensemble are highlighted; within-class Mahalanobis delivers the highest AUPRC.}}
\label{tab:ood_metrics}
\small
\begin{tabular}{lcc}
\hline\hline
Method & AUROC (95\% CI) & AUPRC \\
\hline
\multicolumn{3}{l}{\cellcolor{red!15}\emph{Uncertainty-based}} \\
\rev{MSP}                  & \rev{$0.7225\,[0.6889, 0.7546]$} & \rev{$0.204$} \\
\rev{Entropy}              & \rev{$0.7220\,[0.6874, 0.7550]$} & \rev{$0.216$} \\
\rev{ODIN}                 & \rev{$0.7201\,[0.6848, 0.7536]$} & \rev{$0.212$} \\
\rev{Energy}               & \rev{$0.6780\,[0.6389, 0.7144]$} & \rev{$0.191$} \\
\multicolumn{3}{l}{\cellcolor{red!15}\emph{Distance-based}} \\
\rev{Mahalanobis (global)} & \rev{$0.8737\,[0.8571, 0.8905]$} & \rev{$0.404$} \\
\rev{Mahalanobis (class)}  & \rev{$0.8514\,[0.8281, 0.8731]$} & \rev{$0.432$} \\
\rev{Mahalanobis (within)} & \rev{$0.8800\,[0.8568, 0.9012]$} & \rev{$\mathbf{0.600}$} \\
\rev{$k$-NN}               & \rev{$0.8427\,[0.8196, 0.8656]$} & \rev{$0.465$} \\
\rev{Cosine}               & \rev{$0.8449\,[0.8180, 0.8693]$} & \rev{$0.379$} \\
\rev{\textbf{Typicality}}  & \rev{$\mathbf{0.8822\,[0.8587, 0.9022]}$} & \rev{$0.533$} \\
\multicolumn{3}{l}{\cellcolor{red!15}\emph{Density-based}} \\
\rev{GMM}                  & \rev{$0.8632\,[0.8408, 0.8851]$} & \rev{$0.464$} \\
\rev{Isolation Forest}     & \rev{$0.8749\,[0.8563, 0.8910]$} & \rev{$0.412$} \\
\rev{LOF}                  & \rev{$0.7534\,[0.7222, 0.7825]$} & \rev{$0.276$} \\
\rev{PCA reconstruction}   & \rev{$0.8578\,[0.8379, 0.8767]$} & \rev{$0.398$} \\
\multicolumn{3}{l}{\cellcolor{red!15}\emph{Hybrid}} \\
\rev{HEC}                  & \rev{$0.7372\,[0.7027, 0.7693]$} & \rev{$0.195$} \\
\rev{MRS}                  & \rev{$0.8233\,[0.7977, 0.8485]$} & \rev{$0.388$} \\
\hline
\multicolumn{3}{l}{\emph{Ensembles}} \\
\rev{AUROC-weighted Rank ($16$ methods)}            & \rev{$0.8532$} & --- \\
\rev{Decorrelated LR (\{mahal\_global, lof, msp\})} & \rev{$0.8845$ (CV $0.8841 \pm 0.0150$)} & --- \\
\rev{Logistic Regression (all 16)}                  & \rev{$0.9382$ (CV $0.9358 \pm 0.0141$)} & --- \\
\rev{\textbf{LightGBM}}                             & \rev{$\mathbf{0.9851\,[0.9813, 0.9884]}$ (CV $0.9460 \pm 0.0149$)} & --- \\
\hline
\end{tabular}
\end{table*}

\section{Dataset construction details}
\label{app:dataset}

The full classification taxonomy of the dataset (training classes plus the out-of-distribution evaluation set) appears in Table~\ref{tab:classification}. The fine-classification head sees $K_f = 7$ merged head classes; the coarse head groups these into $K_c = 3$ superclasses (SN-Thermonuclear, SN-CC, Non-SN).  The merging rationale --- hydrogen-rich CC subtypes (II/IIn/IIb) merged because IIb has a small training sample and the H-rich CC continuum is well-known; Ibn merged into Ibc because its narrow-He signature is not always recoverable in noisy SEDM spectra --- is reflected in the ``Included Subtypes'' column of the table.

\begin{sidewaystable*}
\centering
\begin{minipage}{\textheight}
\centering
\caption{Classification taxonomy and dataset composition. ``Head class'' is the merged label seen by the fine classification head ($K_f=7$); ``Category'' is the physical grouping used by the coarse head ($K_c=3$). $N$ is the number of spectra. The OOD set is used to evaluate the framework's ability to recognise unfamiliar transients; raw-subtype counts and additional merging detail appear in Table~\ref{tab:type_distribution} below.}
\label{tab:classification}
\small
\begin{tabular}{@{}lllrl@{\hspace{1em}}l@{}}
\hline\hline
Class & Head class & Category & $N$ & Included Subtypes & Notes \\
\hline
\multicolumn{6}{c}{\textit{Training Classes}} \\
\hline
SN Ia   & SN Ia   & SN Thermo & 12{,}973 & Ia, Ia-norm, Ia-91T, Ia-91bg, Ia-02cx, Ia-03fg, Ia-pec, Ia-99aa & Ia-CSM excluded \\
SN II   & SN II   & SN CC (H) & 5{,}166  & Type II, II-norm, IIP, IIL, II-pec & --- \\
SN IIn  & SN II   & SN CC (H) & 1{,}191  & IIn, Ia-CSM & Ia-CSM: H$\alpha$ emission \\
SN IIb  & SN II   & SN CC (H) & 791      & IIb & --- \\
SN Ibc  & SN Ibc  & SN CC (SE) & 2{,}972 & Ib, Ic, Ib/c, Ic-BL, Ib-pec, Icn, Ien & Without SLSN or Ibn \\
SN Ibn  & SN Ibc  & SN CC (SE) & 243     & Ibn & Narrow He, spectrally unique \\
SLSN     & SLSN     & SN Lumin   & 835     & Ic-SLSN, II-SLSN, SLSN I, Ic.5-SLSN & 10--100$\times$ luminous \\
AGN      & AGN      & Non-SN     & 494     & AGN, Seyfert, Gal.\ Nuclei & Seyfert-type only \\
TDE      & TDE      & Non-SN     & 848     & Tidal Disruption Event & --- \\
CV       & CV       & Non-SN     & 765     & Cataclysmic, U Gem, AM CVn, Nova-like & Accretion-powered \\
\hline
\multicolumn{6}{c}{\textit{Out-of-Distribution Evaluation}} \\
\hline
QSO              & ---      & Non-SN     & 28  & AGN Outlier   & Subcategory of AGN \\
Blazar           & ---      & Non-SN     & 11  & AGN Outlier   & Subcategory of AGN \\
BL Lac           & ---      & Non-SN     & 1   & AGN Outlier   & Subcategory of AGN \\
Novae            & ---      & Non-SN     & 67  & CV-related    & Thermonuclear runaway on WD \\
Classical Nova   & ---      & Non-SN     & 8   & CV-related    & Thermonuclear runaway on WD \\
Afterglow        & ---      & Non-SN     & 17  & GRB-related   & Post-GRB emission \\
long GRB         & ---      & Non-SN     & 1   & GRB-related   & Stellar collapse / engine \\
ILRT             & ---      & Uncertain  & 33  & Gap transient & EC-SN or non-terminal eruption \\
FBOT             & ---      & Uncertain  & 98  & Gap transient & Engine-driven or TDE from IMBH \\
Ca-rich          & ---      & SN CC (SE) & 10  & Peculiar SN   & He-shell detonation on WD? \\
LRN              & ---      & Non-SN     & 15  & Stellar merger & Common envelope ejection \\
\hline
\multicolumn{5}{r}{\textit{Total OOD spectra (rare set)}:} & \textbf{289} \\
\hline
\end{tabular}
\end{minipage}
\end{sidewaystable*}

The raw subtype distribution of the dataset (per-object and per-spectrum, before merging) appears in Table~\ref{tab:type_distribution}.  After object-level filtering ($p_\mathrm{TNS} \geq 0.5$, no numerically duplicate spectra) and subtype merging into the $K_f = 7$ head classes, the per-class training counts used by the fine classifier are those listed in Table~\ref{tab:classification}.

\begin{table*}[t]
\centering
\caption{Raw distribution of transient types in the dataset (Sec.~\ref{sec:data}), per unique object identifier and aggregated over spectroscopic epochs. The ratio of spectra to objects reveals the typical observational cadence per class.}
\label{tab:type_distribution}
\begin{tabular}{lrr@{\hskip 1.0cm}lrr}
\toprule
\textbf{Type} & \textbf{Objects} & \textbf{Spectra} & \textbf{Type} & \textbf{Objects} & \textbf{Spectra} \\
\midrule
SN Ia            & 7328 & 12120 & SN Ia-CSM         &  14 &  67 \\
SN II            & 1559 &  4375 & AM CVn            &  14 &  35 \\
CV               &  392 &   565 & FBOT              &  12 & 100 \\
AGN              &  343 &   478 & U Gem             &  11 &  15 \\
SN IIn           &  321 &  1057 & Nova-like         &  10 &  91 \\
SN Ic            &  282 &  1183 & SN Ia-03fg        &  10 &  57 \\
SN Ib            &  252 &   955 & afterglow         &   9 &  19 \\
SN IIb           &  194 &   740 & ILRT              &   8 &  34 \\
SN IIP           &  150 &   307 & Blazar            &   6 &  11 \\
TDE              &  136 &   752 & SLSN I            &   6 &  19 \\
SN Ic-SLSN       &   97 &   581 & Classical Nova    &   6 &   9 \\
SN Ia-91T        &   86 &   145 & Ca-rich           &   5 &  10 \\
SN Ic-BL         &   83 &   463 & SN Ib-pec         &   4 &  15 \\
SN Ia-norm       &   81 &   133 & Galactic Nuclei   &   3 &   3 \\
SN II-SLSN       &   52 &   199 & SN IIL            &   3 &   3 \\
Novae            &   43 &    71 & SN Icn            &   3 &  73 \\
SN Ia-91bg       &   41 &    78 & SN Ic.5-SLSN      &   2 &   5 \\
SN Ibn           &   39 &   235 & Luminous Red Nova &   2 &  15 \\
SN Ib/c          &   34 &    92 & Seyfert           &   1 &   1 \\
SN Ia-pec        &   26 &   113 & SN Ien            &   1 &  19 \\
SN II-norm       &   23 &    70 & long GRB          &   1 &   1 \\
QSO              &   18 &    29 & SN II-pec         &   1 &  24 \\
SN Ia-02cx       &   18 &    70 & BL Lac            &   1 &   1 \\
                 &      &       & SN Ia-99aa        &   1 &   1 \\
\bottomrule
\end{tabular}
\end{table*}

\paragraph{Preprocessing in detail.}
\label{app:preprocessing}
Each spectrum is restricted to the \emph{observer-frame} interval $[3850, 9000]\,\text{\AA}$, with detector edges and sky-subtraction artifacts outside this window discarded and remaining missing fluxes zero-filled with a binary validity mask.  All spectra are linearly interpolated onto a uniform grid of $L = 4096$.  The interpolated flux is then rescaled to approximately $[0,1]$ via the robust percentile transform $f_\mathrm{norm} = (f - p_5)/(p_{95} - p_5)$, which preserves the relative amplitude of spectral features and the continuum shape while suppressing the influence of extreme outliers; we tested per-spectrum $z$-score and min--max alternatives and adopted the percentile transform for its robustness to localised high-amplitude artifacts.  We also evaluated rolling-median continuum removal prior to scaling and found that retaining the continuum shape improved classification of continuum-dominated classes (AGN, TDE, SLSN) without degrading line-dominated classes; no continuum removal is therefore applied in the production pipeline.  Identical preprocessing is applied across all instruments (SEDM, Keck, Gemini, DESI, SDSS, NGPS).

\section{Classification comparison with prior work}
\label{app:classification_comparison}

A direct comparison against published spectral classifiers on a common test set requires retraining each classifier from scratch on the same dataset, beyond the scope of this work. We provide a partial comparison by re-evaluating \texttt{ASTRANet} on the $6$-class taxonomy used by \citet{xuspectra}, which reported a macro F1 of $72.1\%$ outperforming \texttt{ABC-SN} (macro F1 $56.5\%$) and \texttt{DASH} (\citealp{Muthukrishna2019}, macro F1 $48.1\%$). \texttt{ASTRANet} on the same $6$-class taxonomy achieves $88.4\%$ accuracy and $79.4\%$ macro F1, improvements of $+4.3$\,pp accuracy and $+7.3$\,pp macro F1 over the strongest prior baseline. The largest gains are on the non-SN classes.

\begin{table}[h]
\centering
\caption{Per-class recall comparison (\%) between \texttt{ASTRANet} and the strongest prior baseline \texttt{SpectraNet} \citep{xuspectra} on the same $6$-class ZTF-derived spectroscopic taxonomy.}
\label{tab:xu_comparison}
\begin{tabular}{lccc}
\hline\hline
Class & SpectraNet & \texttt{ASTRANet} & $\Delta$ (pp) \\
\hline
SN Ia  & 92 & 95.2 & $+3.2$ \\
SN Ibc & 73 & 79.7 & $+6.7$ \\
SN II  & 79 & 81.9 & $+2.9$ \\
AGN    & 70 & 71.1 & $+1.1$ \\
CV     & 72 & \textbf{85.2} & $\mathbf{+13.2}$ \\
TDE    & 67 & \textbf{76.6} & $\mathbf{+9.6}$ \\
\hline
Macro  & 75.5 & 81.6 & $+6.1$ \\
\hline
\end{tabular}
\end{table}

\section{Ablation study \label{sec:ablation}}
\rev{\emph{The ablation isolates architectural components under a fixed reference data configuration; the relative comparisons between variants are the quantity of interest, and absolute values are therefore not directly comparable to Sec.~\ref{sec:classification_results}.}}

To validate that each architectural and training-time component contributes,
we perform leave-one-out ablations (Table~\ref{tab:ablation}).
Each row retrains the full model with a single component removed or replaced,
keeping all other settings identical (seed\,=\,42).
All entries report best single-checkpoint performance.

\begin{table*}[t]
\centering
\caption{Leave-one-out ablation study. Accuracy and macro recall on the
held-out test set when individual components are removed.}
\label{tab:ablation}
\small
\begin{tabular}{lcc}
\hline\hline
Configuration & Accuracy & Macro recall \\
\hline
Full \texttt{ASTRANet}                    & 83.2\% & 80.3\% \\
$-$ FiLM conditioning                     & 83.4\% & 79.3\% \down{1.0} \\
$-$ Hierarchical heads (flat 7-class)     & 84.0\% & 81.0\% \up{0.7} \\
$-$ Confusion penalty                     & 86.2\% & 80.7\% \up{0.4} \\
$-$ SupCon loss                           & 83.5\% & 79.9\% \down{0.4} \\
$-$ Multi-scale stem (single $k\!=\!31$)  & 85.4\% & 81.5\% \up{1.2} \\
$-$ Attention pool (global avg.\ pool)    & 85.2\% & 81.6\% \up{1.3} \\
$-$ MixUp                                 & 84.1\% & 78.4\% \down{1.9} \\
\hline
\end{tabular}
\end{table*}

Three components---hierarchical heads, multi-scale stem, and attention
pooling---show marginal macro-recall improvements when removed individually.
To test whether these gains are additive, we train pairwise and triple
combinations of their removal (Table~\ref{tab:ablation_combined}).

\begin{table*}[t]
\centering
\caption{Combined ablation of the three components that individually showed improvement when removed. Per-class recall reveals that apparent macro gains come at the expense of rare-class performance.}
\label{tab:ablation_combined}
\small
\begin{tabular}{lccccccccc}
\hline\hline
Configuration & Acc & Macro & Ia & Ibc & SLSN & II & AGN & CV & TDE \\
\hline
Full \texttt{ASTRANet}
  & 83.2 & 80.3 & 92.9 & 71.0 & 84.0 & 71.3 & 77.8 & 86.1 & \textbf{78.7} \\
$-$ Hierarchy
  & 84.0 & 81.0 & 93.0 & 70.6 & \textbf{92.0} & 74.2 & 77.8 & 85.2 & 74.5 \\
$-$ Multi-scale
  & 85.4 & 81.5 & 93.4 & \textbf{73.1} & 84.0 & \textbf{77.1} & 77.8 & 86.1 & \textbf{78.7} \\
$-$ AttnPool
  & 85.2 & 81.6 & \textbf{93.9} & 72.2 & 88.0 & 76.2 & 77.8 & 86.9 & 76.6 \\
\hline
$-$ Hier.\ + Multi.
  & 83.8 & 81.6 & 93.6 & 70.6 & \textbf{96.0} & 72.4 & 77.8 & 84.4 & 76.6 \\
$-$ Multi.\ + AttnPool
  & 83.4 & 80.2 & 93.4 & 69.4 & 88.0 & 71.8 & 75.6 & 88.5 & 74.5 \\
$-$ Hier.\ + AttnPool
  & 84.4 & 81.4 & 93.5 & 71.0 & 92.0 & 74.8 & 77.8 & 84.4 & 76.6 \\
$-$ All three
  & 86.7 & 79.9 & 94.2 & 72.9 & 80.0 & 81.2 & 73.3 & 85.2 & 72.3 \\
\hline
\end{tabular}
\end{table*}

While removing individual components can boost macro recall by up to $+1.3$\,pp, the improvements do not stack: the triple removal drops macro recall below the baseline ($79.9$\% vs.\ $80.3$\%). Crucially, the per-class breakdown reveals that apparent gains are driven
by large improvements on a single class (e.g.\ SLSN reaches $96.0$\% recall in the Hierarchy\,+\,Multi-scale ablation), while rare classes such as TDE ($78.7 \to 72.3$\%), AGN ($77.8 \to 73.3$\%), and CV ($86.1 \to 84.4$\%) consistently degrade. Since our science goal requires reliable identification of \emph{all} transient classes---particularly rare events like TDE---we retain the full architecture, which achieves the most balanced per-class performance.

\section{Ensemble Strategy Comparison for Classification}
\rev{\emph{The comparison uses a fixed reference data configuration to isolate the effect of the ensembling strategy; within-seed averaging is the deployed choice, with deployment numbers in Sec.~\ref{sec:classification_results}.}}
\label{app:ensemble}

The training procedure yields two checkpoints per seed (best-validation and SWA \citep{Izmailov2018SWA}); across three seeds $\{42, 123, 7\}$ this gives $6$ checkpoints.  We compare eight strategies for combining them: three uniform-averaging baselines (cross-seed, within-seed, and a flat super-ensemble) and five strategies tuned on the validation set (cost-sensitive per-class calibration, learned per-class weights, logistic-regression stacking, selective per-class inclusion, and per-model temperature scaling).  Full per-class results are in Table~\ref{tab:ensemble_comparison}.

\paragraph{Baseline strategies (uniform averaging).}

\begin{itemize}
    \item \emph{Cross-seed ensemble:} for each seed, take the checkpoint with the highest validation macro recall and uniformly average their softmax outputs ($M=3$).
    \item \emph{Within-seed ensemble:} for each seed, first uniformly average the softmax outputs of both checkpoints \emph{within} the seed, then uniformly average the per-seed results across seeds ($M=6$ in total, hierarchically averaged).
    \item \emph{Super-ensemble:} uniformly average the softmax outputs of all $6$ checkpoints in a single step, ignoring seed structure.
\end{itemize}

\paragraph{Advanced strategies (tuned on validation).}

The following four strategies fit hyperparameters on the validation set and are applied to the test set; the within-seed ensemble is the unsupervised baseline against which they are compared.

\begin{itemize}
    \item \emph{Cost-sensitive calibration:} per-class probability scaling factors $\boldsymbol{s} = (s_1, \ldots, s_{K_f})$ are tuned by coordinate descent on the validation set to maximise macro recall. At inference, the prediction is $\hat{y} = \arg\max_k\,s_k\,p_k$.
    \item \emph{Strategy 1 -- Learned per-class weights:} a separate weight vector $\mathbf{w}^{(c)} \in \Delta^{M-1}$ is learned per class, parameterised as a softmax of free logits and optimised by Nelder--Mead to maximise validation macro recall. The fine prediction is then $\hat{y} = \arg\max_c \sum_{m=1}^M w_m^{(c)}\,p^{(m)}_c$. Generalises uniform averaging by allowing different models to dominate for different classes.
    \item \emph{Strategy 2 -- Stacking (LogReg):} a logistic regression meta-learner is trained on the validation set with concatenated per-model softmax vectors as features ($\mathbb{R}^{M K_f}$ feature dim.), class-balanced sample weights, and $\ell_2$ regularisation. The regularisation strength is selected by $5$-fold CV macro recall over $C\!\in\!\{0.01, 0.1, 0.5, 1, 5, 10, 50\}$. The meta-learner can capture non-trivial cross-model and cross-class interactions.
    \item \emph{Strategy 3 -- Selective per-class ensemble:} for each class $c$, only checkpoints whose validation per-class recall for $c$ exceeds a threshold $\tau_c$ are included in the average for that class. The threshold is a single global scalar swept over $\tau \in [0, 0.85]$, with the value maximising validation macro recall retained.
    \item \emph{Strategy 4 -- Temperature scaling:} a per-model temperature $T_m$ rescales the logits before softmax, $\mathbf{p}^{(m)} = \mathrm{softmax}(\mathbf{z}^{(m)}/T_m)$. Temperatures are optimised by coordinate descent on the validation set with grid $T \in [0.3, 3.0]$, step $0.05$, for at most $5$ outer iterations.
\end{itemize}

\paragraph{Headline results.}
Table~\ref{tab:ensemble_comparison} reports test-set accuracy, macro recall, and per-class recall for all strategies. The within-seed ensemble is the single most accurate strategy ($87.5\%$ accuracy, $82.9\%$ macro recall, with the best recall on $4$ of $7$ classes: SN~Ia, SN~Ibc, SLSN, SN~II). Stacking marginally ties the best macro recall ($83.0\%$, $+0.05$\,pp over within-seed) but at the cost of $2.7$\,pp in accuracy, driven by overfitting of the meta-learner to validation idiosyncrasies. Cost-sensitive calibration and temperature scaling boost TDE recall by $+4.3$ and $+2.1$\,pp respectively but degrade SN~Ibc, SN~II, and overall accuracy. The remaining strategies are within $\pm\,1$\,pp of the within-seed baseline across all metrics.

\begin{table*}[t]
\centering
\caption{Test-set performance for all ensemble strategies. Per-class values are recall (\%). Strategies below the mid-rule are tuned on the validation set. ``Models'' denotes the number of softmax distributions averaged; for advanced strategies tuned on validation, $M\!=\!6$ throughout. Best value in each column is shown in \textbf{bold}.}
\label{tab:ensemble_comparison}
\resizebox{\textwidth}{!}{%
\begin{tabular}{lcccccccccc}
\hline\hline
Strategy & Models & Acc (\%) & Macro (\%) & Ia & Ibc & SLSN & II & AGN & CV & TDE \\
\hline
Single best (seed 123, standard) & 1 & 86.3 & \textbf{83.0} & 94.5 & 74.5 & 92.0 & 78.1 & 77.8 & 81.1 & \textbf{83.0} \\
Cross-seed ensemble        & 3 & 87.1 & 81.1 & 94.7 & 73.1 & 80.0 & 81.1 & 75.6 & 84.4 & 78.7 \\
Within-seed ensemble       & 6 & \textbf{87.5} & 82.9 & \textbf{95.3} & \textbf{75.2} & 92.0 & \textbf{80.7} & 75.6 & 82.8 & 78.7 \\
Super-ensemble             & 6 & 87.1 & 82.3 & 95.1 & 74.5 & 88.0 & 79.7 & 75.6 & 84.4 & 78.7 \\
\hline
Cost-sensitive calibration & 6 & 85.7 & 82.7 & 94.7 & 71.0 & \textbf{96.0} & 76.9 & 75.6 & 82.0 & \textbf{83.0} \\
Per-class weights          & 6 & 87.1 & 82.3 & 95.1 & 74.5 & 88.0 & 79.7 & 75.6 & 84.4 & 78.7 \\
Stacking (LogReg)          & 6 & 84.9 & \textbf{83.0} & 92.3 & 71.7 & 92.0 & 76.6 & 73.3 & \textbf{95.9} & 78.7 \\
Selective per-class        & 6 & 87.0 & 82.0 & 94.9 & 73.6 & 88.0 & 80.4 & 75.6 & 82.8 & 78.7 \\
Temperature scaling        & 6 & 86.1 & 81.7 & 94.3 & 73.6 & 84.0 & 77.8 & 75.6 & 86.1 & 80.9 \\
\hline\hline
\end{tabular}%
}
\end{table*}

\paragraph{Diagnostics: why advanced strategies do not improve over within-seed.}

We attribute the marginal returns of the advanced strategies to three factors:
\begin{enumerate}
    \item \emph{Ensemble component correlation.} The $6$ checkpoints are derived from $3$ seeds of a single architecture revision and share data augmentation, loss function, and optimisation schedule. Their softmax distributions are highly correlated (mean off-diagonal Spearman $\rho \approx 0.96$ on the test set), leaving little headroom for learned weighting schemes.
    \item \emph{Validation--test distribution shift.} Per-model recall on the rarest classes (TDE, AGN, SLSN) has small-sample variance: with $N_\mathrm{val}^{(\mathrm{TDE})}\!=\!42$ and $N_\mathrm{val}^{(\mathrm{AGN})}\!=\!43$, a hyperparameter tuned to a $5\%$ validation recall improvement is generically within bootstrap noise on the test set.
    \item \emph{Macro-recall plateau.} The within-seed and stacking strategies tie at macro recall $\sim$$83\%$, while the single-best seed-$123$ checkpoint already reaches macro recall $83.0\%$ at $86.3\%$ accuracy. The implication is that the residual classification errors are driven by intrinsic spectral overlap (Ibc/II, AGN/TDE), not by ensemble suboptimality. We confirm this in the per-class confusion structure (Fig.~\ref{fig:confusion_matrix}): off-diagonal mass concentrates on the physically expected confusions, and reshuffling ensemble weights only redistributes errors within those off-diagonal blocks.
\end{enumerate}

\paragraph{Choice of operating regime.}
For different deployment contexts we recommend:
\begin{itemize}
    \item \emph{Default deployment} (highest overall accuracy, balanced per-class performance): within-seed ensemble.
    \item \emph{TDE-priority follow-up} (maximise TDE recall, accept modest accuracy loss): cost-sensitive calibration (TDE $=83.0\%$ at $85.7\%$ accuracy) or the single-best seed~$123$ checkpoint (TDE $=83.0\%$ at $86.3\%$ accuracy).
    \item \emph{Pure ranking applications} (no hard threshold required): within-seed softmax averaging directly, with no calibration.
\end{itemize}

For all results in the main text (Sec.~\ref{sec:classification_results}, Sec.~\ref{sec:results}) we use the within-seed ensemble.

\section{Embedding-space retrieval: nearest-neighbor analogs}
\label{app:retrieval}

The SupCon objective (Sec.~\ref{sec:architecture}) makes the $192$-dimensional embedding an explicitly metric-learned space, which supports similarity search as a by-product of the framework.  Two diagnostics quantify this.

\emph{(i) The embedding alone carries the classification.}  A bare $k$-nearest-neighbor classifier (cosine distance, $k\!=\!15$) over the test-set embeddings reaches $86.8\%$ five-fold accuracy -- statistically indistinguishable from the full hierarchical classifier head ($86.6\%$, Sec.~\ref{sec:classification_results}).  The class structure visible in the UMAP projection is therefore metrically faithful, not a visualization artifact.

\begin{figure}[t]
\centering
\includegraphics[width=\columnwidth]{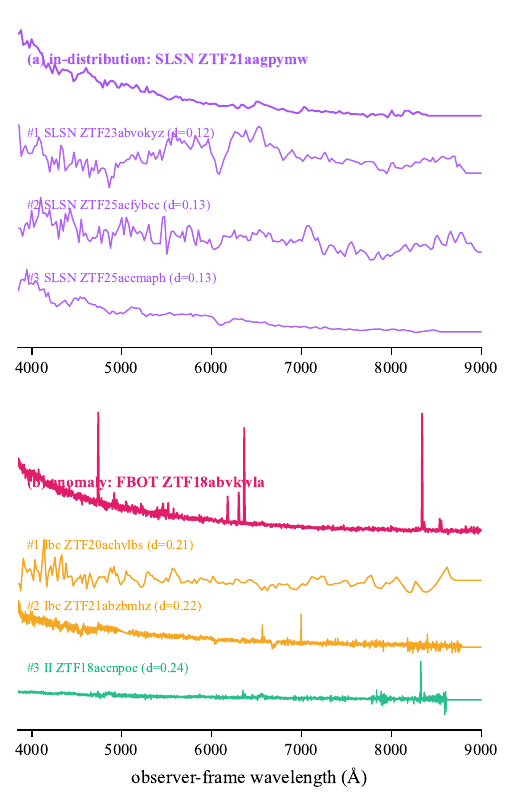}
\caption{Embedding-space nearest-neighbor retrieval, both regimes.  In each panel the accented top spectrum is the query and the spectra below are its nearest archive neighbors by cosine distance in the \texttt{ASTRANet} embedding, coloured and labelled by their true class ($d$ = cosine distance).  \emph{(a) In-distribution:} a SLSN query returns SLSN neighbors at $d\approx0.12$ -- a class-consistent ``find more like this'' search.  \emph{(b) Anomaly:} an FBOT flagged by \texttt{ASTRANet-Sentinel} and absent from training lies at $d\approx0.22$ from stripped-envelope / H-rich core-collapse neighbors, making the Regime-IV progenitor ambiguity explicit; the larger neighbor distance is itself a novelty signal.  Fluxes are min--max scaled and vertically offset for display.}
\label{fig:retrieval_analogs}
\end{figure}

The same retrieval operates in two complementary regimes
(Fig.~\ref{fig:retrieval_analogs}).

\emph{(ii) In-distribution retrieval.}  For a query drawn from a trained class the neighborhood is tight and class-consistent: a SLSN query (Fig.~\ref{fig:retrieval_analogs}a) returns SLSN neighbors at cosine distances of $0.12$--$0.13$, i.e.\ a working ``find more like this'' filter over the archive.

\emph{(iii) Anomaly $\rightarrow$ analogs.}  For a spectrum flagged by \texttt{ASTRANet-Sentinel}, the same retrieval answers the natural follow-up question -- \emph{what does it most resemble?} -- and hands the human vetter an interpretable neighborhood instead of a bare anomaly score.  An FBOT from the rare evaluation set (Fig.~\ref{fig:retrieval_analogs}b; never seen in training) sits almost twice as far from its nearest known neighbors ($0.21$--$0.24$) as the in-distribution query, and those neighbors are stripped-envelope and hydrogen-rich core-collapse spectra -- precisely the ambiguous-progenitor degeneracy of Regime~IV (Sec.~\ref{sec:case_fbot}), now made explicit as a ranked list of named objects.  The neighbor distance itself is thus a second, geometry-based novelty signal that corroborates the Sentinel score. In survey operation this retrieval step runs on the same cached embeddings as the Sentinel at negligible cost; scaling beyond $\sim\!10^{5}$ archived spectra requires only a standard approximate nearest-neighbor index.

\clearpage
\bibliographystyle{apsrev4-2}
\bibliography{sample,method}

@article{savitzky1964smoothing,
  author  = {Savitzky, Abraham and Golay, Marcel J. E.},
  title   = {Smoothing and Differentiation of Data by Simplified
             Least Squares Procedures},
  journal = {Analytical Chemistry},
  volume  = {36},
  number  = {8},
  pages   = {1627--1639},
  year    = {1964},
  doi     = {10.1021/ac60214a047},
}

@inproceedings{wu2018group,
  author    = {Wu, Yuxin and He, Kaiming},
  title     = {Group Normalization},
  booktitle = {European Conference on Computer Vision (ECCV)},
  pages     = {3--19},
  year      = {2018},
  publisher = {Springer},
  doi       = {10.1007/978-3-030-01261-8_1},
}

@article{hendrycks2016gelu,
  author  = {Hendrycks, Dan and Gimpel, Kevin},
  title   = {Gaussian Error Linear Units ({GELUs})},
  journal = {arXiv preprint arXiv:1606.08415},
  year    = {2016},
}

@article{bai2018empirical,
  author  = {Bai, Shaojie and Kolter, J. Zico and Koltun, Vladlen},
  title   = {An Empirical Evaluation of Generic Convolutional and
             Recurrent Networks for Sequence Modeling},
  journal = {arXiv preprint arXiv:1803.01271},
  year    = {2018},
}

@inproceedings{huang2016deep,
  author    = {Huang, Gao and Sun, Yu and Liu, Zhuang and
               Sedra, Daniel and Weinberger, Kilian Q.},
  title     = {Deep Networks with Stochastic Depth},
  booktitle = {European Conference on Computer Vision (ECCV)},
  pages     = {646--661},
  year      = {2016},
  publisher = {Springer},
  doi       = {10.1007/978-3-319-46493-0_39},
}

@inproceedings{perez2018film,
  author    = {Perez, Ethan and Strub, Florian and de Vries, Harm and
               Dumoulin, Vincent and Courville, Aaron},
  title     = {{FiLM}: Visual Reasoning with a General Conditioning Layer},
  booktitle = {Proceedings of the AAAI Conference on Artificial Intelligence},
  volume    = {32},
  number    = {1},
  year      = {2018},
  doi       = {10.1609/aaai.v32i1.11671},
  pages     = {N/A},
}

@inproceedings{ilse2018attention,
  author    = {Ilse, Maximilian and Tomczak, Jakub M. and Welling, Max},
  title     = {Attention-based Deep Multiple Instance Learning},
  booktitle = {International Conference on Machine Learning (ICML)},
  pages     = {2127--2136},
  year      = {2018},
  publisher = {PMLR},
}

@inproceedings{lin2017focal,
  author    = {Lin, Tsung-Yi and Goyal, Priya and Girshick, Ross and
               He, Kaiming and Doll{\'a}r, Piotr},
  title     = {Focal Loss for Dense Object Detection},
  booktitle = {Proceedings of the IEEE International Conference on
               Computer Vision (ICCV)},
  pages     = {2980--2988},
  year      = {2017},
  doi       = {10.1109/ICCV.2017.324},
}

@inproceedings{khosla2020supervised,
  author    = {Khosla, Prannay and Teterwak, Piotr and Wang, Chen and
               Sarna, Aaron and Tian, Yonglong and Isola, Phillip and
               Maschinot, Aaron and Liu, Ce and Krishnan, Dilip},
  title     = {Supervised Contrastive Learning},
  booktitle = {Advances in Neural Information Processing Systems (NeurIPS)},
  volume    = {33},
  pages     = {18661--18673},
  year      = {2020},
}

@inproceedings{zhang2018mixup,
  author    = {Zhang, Hongyi and Cisse, Moustapha and Dauphin, Yann N.
               and Lopez-Paz, David},
  title     = {mixup: Beyond Empirical Risk Minimization},
  booktitle = {International Conference on Learning Representations (ICLR)},
  year      = {2018},
}

@inproceedings{cui2019class,
  author    = {Cui, Yin and Jia, Menglin and Lin, Tsung-Yi and
               Song, Yang and Belongie, Serge},
  title     = {Class-Balanced Loss Based on Effective Number of Samples},
  booktitle = {Proceedings of the IEEE/CVF Conference on Computer Vision
               and Pattern Recognition (CVPR)},
  pages     = {9268--9277},
  year      = {2019},
  doi       = {10.1109/CVPR.2019.00949},
}

@inproceedings{loshchilov2019decoupled,
  author    = {Loshchilov, Ilya and Hutter, Frank},
  title     = {Decoupled Weight Decay Regularization},
  booktitle = {International Conference on Learning Representations (ICLR)},
  year      = {2019},
}

@inproceedings{loshchilov2017sgdr,
  author    = {Loshchilov, Ilya and Hutter, Frank},
  title     = {{SGDR}: Stochastic Gradient Descent with Warm Restarts},
  booktitle = {International Conference on Learning Representations (ICLR)},
  year      = {2017},
}

@article{polyak1992acceleration,
  author  = {Polyak, Boris T. and Juditsky, Anatoli B.},
  title   = {Acceleration of Stochastic Approximation by Averaging},
  journal = {SIAM Journal on Control and Optimization},
  volume  = {30},
  number  = {4},
  pages   = {838--855},
  year    = {1992},
  doi     = {10.1137/0330046},
}

@inproceedings{Guo2017Calibration,
  author    = {Guo, Chuan and Pleiss, Geoff and Sun, Yu and Weinberger, Kilian Q.},
  title     = {On Calibration of Modern Neural Networks},
  booktitle = {Proceedings of the 34th International Conference on Machine Learning (ICML)},
  series    = {Proceedings of Machine Learning Research},
  volume    = {70},
  pages     = {1321--1330},
  publisher = {PMLR},
  year      = {2017},
  eprint    = {1706.04599},
  archivePrefix = {arXiv},
  primaryClass  = {cs.LG}
}

@inproceedings{Hein2019Why,
  author    = {Hein, Matthias and Andriushchenko, Maksym and Bitterwolf, Julian},
  title     = {Why {ReLU} Networks Yield High-Confidence Predictions Far Away From the Training Data and How to Mitigate the Problem},
  booktitle = {Proceedings of the {IEEE/CVF} Conference on Computer Vision and Pattern Recognition (CVPR)},
  pages     = {41--50},
  year      = {2019},
  doi       = {10.1109/CVPR.2019.00013},
  eprint    = {1812.05720},
  archivePrefix = {arXiv},
  primaryClass  = {cs.LG}
}

@inproceedings{Lakshminarayanan2017DeepEnsembles,
  author    = {Lakshminarayanan, Balaji and Pritzel, Alexander and Blundell, Charles},
  title     = {Simple and Scalable Predictive Uncertainty Estimation using Deep Ensembles},
  booktitle = {Advances in Neural Information Processing Systems 30 (NIPS 2017)},
  pages     = {6402--6413},
  year      = {2017},
  eprint    = {1612.01474},
  archivePrefix = {arXiv},
  primaryClass  = {stat.ML}
}

@inproceedings{Liu2008iForest,
  author    = {Liu, Fei Tony and Ting, Kai Ming and Zhou, Zhi-Hua},
  title     = {Isolation Forest},
  booktitle = {Proceedings of the 8th {IEEE} International Conference on Data Mining ({ICDM})},
  pages     = {413--422},
  publisher = {IEEE Computer Society},
  year      = {2008},
  doi       = {10.1109/ICDM.2008.17}
}

@inproceedings{Breunig2000LOF,
  author    = {Breunig, Markus M. and Kriegel, Hans-Peter and Ng, Raymond T. and Sander, J{\"o}rg},
  title     = {{LOF}: Identifying Density-Based Local Outliers},
  booktitle = {Proceedings of the 2000 {ACM} {SIGMOD} International Conference on Management of Data},
  pages     = {93--104},
  publisher = {ACM},
  year      = {2000},
  doi       = {10.1145/342009.335388}
}

@inproceedings{Ke2017LightGBM,
  author    = {Ke, Guolin and Meng, Qi and Finley, Thomas and Wang, Taifeng and Chen, Wei and Ma, Weidong and Ye, Qiwei and Liu, Tie-Yan},
  title     = {{LightGBM}: A Highly Efficient Gradient Boosting Decision Tree},
  booktitle = {Advances in Neural Information Processing Systems 30 (NIPS 2017)},
  pages     = {3146--3154},
  year      = {2017}
}

@article{AngelopoulosBates2023,
  author = {Angelopoulos, A. N. and Bates, S.},
  title = {Conformal Prediction: A Gentle Introduction},
  journal = {Foundations and Trends in Machine Learning},
  year = {2023},
  volume = {16},
  pages = {494--591}
}

@ARTICLE{LSST2019,
  author = {{Ivezi{\'c}}, {\v{Z}}eljko and {Kahn}, Steven M. and others},
  title = {{LSST: From Science Drivers to Reference Design and Anticipated Data Products}},
  journal = {The Astrophysical Journal},
  year = 2019,
  volume = 873,
  number = 2,
  pages = 111,
  doi = {10.3847/1538-4357/ab042c},
  archivePrefix = {arXiv},
  eprint = {0805.2366},
}

@ARTICLE{Foley2013,
  author = {{Foley}, Ryan J. and {Challis}, P.~J. and {Chornock}, R. and others},
  title = {{Type Iax Supernovae: A New Class of Stellar Explosion}},
  journal = {The Astrophysical Journal},
  year = 2013,
  volume = 767,
  number = 1,
  pages = 57,
  doi = {10.1088/0004-637X/767/1/57},
  archivePrefix = {arXiv},
  eprint = {1212.2209},
}

@ARTICLE{Villar2021,
  author = {{Villar}, V. Ashley and {Cranmer}, Miles and {Berger}, Edo and others},
  title = {{A Deep-learning Approach for Live Anomaly Detection of Extragalactic Transients}},
  journal = {The Astrophysical Journal Supplement Series},
  year = 2021,
  volume = 255,
  number = 2,
  pages = 24,
  doi = {10.3847/1538-4365/ac0893},
  archivePrefix = {arXiv},
  eprint = {2103.12102},
}

@ARTICLE{Pruzhinskaya2019,
  author = {{Pruzhinskaya}, Maria V. and {Malanchev}, Konstantin L. and {Kornilov}, Matwey V. and others},
  title = {{Anomaly detection in the Zwicky Transient Facility DR3}},
  journal = {Monthly Notices of the Royal Astronomical Society},
  year = 2019,
  volume = 489,
  number = 3,
  pages = {3591--3608},
  doi = {10.1093/mnras/stz2362},
  archivePrefix = {arXiv},
  eprint = {1905.11516},
}

@misc{DiMMAD2025,
      title={In Search of the Unknown Unknowns: A Multi-Metric Distance Ensemble for Out of Distribution Anomaly Detection in Astronomical Surveys}, 
      author={Siddharth Chaini and Federica B. Bianco and Ashish Mahabal},
      year={2025},
      eprint={2510.23702},
      archivePrefix={arXiv},
      primaryClass={astro-ph.IM},
      url={https://arxiv.org/abs/2510.23702}, 
}

@ARTICLE{Gupta2025MCIF,
  author = {{Gupta}, R. and others},
  title = {{MCIF: Multi-Class Isolation Forest for transient anomaly detection}},
  journal = {arXiv e-prints},
  year = 2025,
  eprint = {2503.00000},
  note = {{\sl Update with actual MCIF reference.}}
}

@article{Chaini_2024,
   title={Light curve classification with DistClassiPy: A new distance-based classifier},
   volume={48},
   ISSN={2213-1337},
   url={http://dx.doi.org/10.1016/j.ascom.2024.100850},
   DOI={10.1016/j.ascom.2024.100850},
   journal={Astronomy and Computing},
   publisher={Elsevier BV},
   author={Chaini, S. and Mahabal, A. and Kembhavi, A. and Bianco, F.B.},
   year={2024},
   month=July, pages={100850} }

@BOOK{Vovk2005,
  author = {{Vovk}, Vladimir and {Gammerman}, Alex and {Shafer}, Glenn},
  title = {{Algorithmic Learning in a Random World}},
  year = 2005,
  publisher = {Springer},
  address = {New York},
  doi = {10.1007/b106715},
}

@INPROCEEDINGS{Romano2020APS,
  author = {{Romano}, Yaniv and {Sesia}, Matteo and {Cand{\`e}s}, Emmanuel J.},
  title = {{Classification with Valid and Adaptive Coverage}},
  booktitle = {Advances in Neural Information Processing Systems (NeurIPS)},
  year = 2020,
  archivePrefix = {arXiv},
  eprint = {2006.02544},
}

@ARTICLE{Swann4MOST,
  author = {{Swann}, Elizabeth and {Sullivan}, Mark and {Carollo}, Daniela and others},
  title = {{4MOST Consortium Survey 10: The Time-Domain Extragalactic Survey (TiDES)}},
  journal = {The Messenger},
  year = 2019,
  volume = 175,
  pages = {58--61},
  doi = {10.18727/0722-6691/5132},
  archivePrefix = {arXiv},
  eprint = {1903.02476},
}

@INPROCEEDINGS{Frohmaier4MOST,
  author = {{Frohmaier}, C. and {Sullivan}, M. and {Maguire}, K. and others},
  title = {{The 4MOST Time-Domain Extragalactic Survey (TiDES): Spectroscopic follow-up of LSST transients}},
  booktitle = {Proceedings of SPIE},
  year = 2024,
  volume = 13096,
  pages = {130963X},
  doi = {10.1117/12.3019033},
}

@INPROCEEDINGS{Schipani2020SOXS,
  author = {{Schipani}, P. and {Campana}, S. and {Claudi}, R. and others},
  title = {{Detailed design of the SOXS spectrograph for the ESO NTT}},
  booktitle = {Proceedings of SPIE},
  year = 2020,
  volume = 11447,
  pages = {1144709},
  doi = {10.1117/12.2562930},
}

@misc{IRIS_software,
author = {{Sasli}, Argyro},
title = {{IRIS: Identification and Reduction of Interesting Spectra}},
year = {2026},
howpublished = {\url{https://github.com/applecider-ml/IRIS}},
note = {Spectral preprocessing pipeline for transient classification}
}

@inproceedings{Izmailov2018SWA,
  title     = {Averaging Weights Leads to Wider Optima and Better Generalization},
  author    = {Izmailov, Pavel and Podoprikhin, Dmitrii and Garipov, Timur and
               Vetrov, Dmitry and Wilson, Andrew Gordon},
  booktitle = {Proceedings of the Thirty-Fourth Conference on Uncertainty in
               Artificial Intelligence (UAI)},
  pages     = {876--885},
  year      = {2018},
  eprint    = {1803.05407},
  archivePrefix = {arXiv},
  primaryClass  = {cs.LG}
}

@misc{yuksekgonul2023confidencereliablemodelsconsider,
      title={Beyond Confidence: Reliable Models Should Also Consider Atypicality}, 
      author={Mert Yuksekgonul and Linjun Zhang and James Zou and Carlos Guestrin},
      year={2023},
      eprint={2305.18262},
      archivePrefix={arXiv},
      primaryClass={cs.LG},
      url={https://arxiv.org/abs/2305.18262}, 
}

@inproceedings{hendrycks2019oe,
  title     = {Deep Anomaly Detection with Outlier Exposure},
  author    = {Hendrycks, Dan and Mazeika, Mantas and Dietterich, Thomas},
  booktitle = {International Conference on Learning Representations (ICLR)},
  year      = {2019},
  eprint    = {1812.04606},
  archivePrefix = {arXiv}
}

@article{Hendrycks2017,
  author = {Hendrycks, Dan and Gimpel, Kevin},
  title = {A Baseline for Detecting Misclassified and Out-of-Distribution Examples},
  journal = {ICLR},
  year = {2017}
}

@article{2019PASP..131a8001P,
  author  = {{Patterson}, M.~T. and others},
  title   = {{The Zwicky Transient Facility Alert Distribution System}},
  journal = {\pasp},
  year    = {2019},
  volume  = {131},
  number  = {995},
  pages   = {018001},
  doi     = {10.1088/1538-3873/aae904}
}

@article{Liu2020Energy,
  author = {Liu, Weitang and Wang, Xiaoyun and Owens, John and Li, Yixuan},
  title = {Energy-based Out-of-distribution Detection},
  journal = {NeurIPS},
  year = {2020}
}

@article{Lee2018Mahalanobis,
  author = {Lee, Kimin and others},
  title = {A Simple Unified Framework for Detecting Out-of-Distribution Samples and Adversarial Attacks},
  journal = {NeurIPS},
  year = {2018}
}

@misc{Liang2018,
      title={Enhancing The Reliability of Out-of-distribution Image Detection in Neural Networks}, 
      author={Shiyu Liang and Yixuan Li and R. Srikant},
      year={2020},
      eprint={1706.02690},
      archivePrefix={arXiv},
      primaryClass={cs.LG},
      url={https://arxiv.org/abs/1706.02690}, 
}

@ARTICLE{ZTF,
       author = {{Bellm}, Eric C. and {Kulkarni}, Shrinivas R. and {Graham}, Matthew J. and {Dekany}, Richard and {Smith}, Roger M. and {Riddle}, Reed and {Masci}, Frank J. and {Helou}, George and {Prince}, Thomas A. and {Adams}, Scott M. and {Barbarino}, C. and {Barlow}, Tom and {Bauer}, James and {Beck}, Ron and {Belicki}, Justin and {Biswas}, Rahul and {Blagorodnova}, Nadejda and {Bodewits}, Dennis and {Bolin}, Bryce and {Brinnel}, Valery and {Brooke}, Tim and {Bue}, Brian and {Bulla}, Mattia and {Burruss}, Rick and {Cenko}, S. Bradley and {Chang}, Chan-Kao and {Connolly}, Andrew and {Coughlin}, Michael and {Cromer}, John and {Cunningham}, Virginia and {De}, Kishalay and {Delacroix}, Alex and {Desai}, Vandana and {Duev}, Dmitry A. and {Eadie}, Gwendolyn and {Farnham}, Tony L. and {Feeney}, Michael and {Feindt}, Ulrich and {Flynn}, David and {Franckowiak}, Anna and {Frederick}, S. and {Fremling}, C. and {Gal-Yam}, Avishay and {Gezari}, Suvi and {Giomi}, Matteo and {Goldstein}, Daniel A. and {Golkhou}, V. Zach and {Goobar}, Ariel and {Groom}, Steven and {Hacopians}, Eugean and {Hale}, David and {Henning}, John and {Ho}, Anna Y.~Q. and {Hover}, David and {Howell}, Justin and {Hung}, Tiara and {Huppenkothen}, Daniela and {Imel}, David and {Ip}, Wing-Huen and {Ivezi{\'c}}, {\v{Z}}eljko and {Jackson}, Edward and {Jones}, Lynne and {Juric}, Mario and {Kasliwal}, Mansi M. and {Kaspi}, S. and {Kaye}, Stephen and {Kelley}, Michael S.~P. and {Kowalski}, Marek and {Kramer}, Emily and {Kupfer}, Thomas and {Landry}, Walter and {Laher}, Russ R. and {Lee}, Chien-De and {Lin}, Hsing Wen and {Lin}, Zhong-Yi and {Lunnan}, Ragnhild and {Giomi}, Matteo and {Mahabal}, Ashish and {Mao}, Peter and {Miller}, Adam A. and {Monkewitz}, Serge and {Murphy}, Patrick and {Ngeow}, Chow-Choong and {Nordin}, Jakob and {Nugent}, Peter and {Ofek}, Eran and {Patterson}, Maria T. and {Penprase}, Bryan and {Porter}, Michael and {Rauch}, Ludwig and {Rebbapragada}, Umaa and {Reiley}, Dan and {Rigault}, Mickael and {Rodriguez}, Hector and {van Roestel}, Jan and {Rusholme}, Ben and {van Santen}, Jakob and {Schulze}, S. and {Shupe}, David L. and {Singer}, Leo P. and {Soumagnac}, Maayane T. and {Stein}, Robert and {Surace}, Jason and {Sollerman}, Jesper and {Szkody}, Paula and {Taddia}, F. and {Terek}, Scott and {Van Sistine}, Angela and {van Velzen}, Sjoert and {Vestrand}, W. Thomas and {Walters}, Richard and {Ward}, Charlotte and {Ye}, Quan-Zhi and {Yu}, Po-Chieh and {Yan}, Lin and {Zolkower}, Jeffry},
        title = "{The Zwicky Transient Facility: System Overview, Performance, and First Results}",
      journal = {\pasp},
         year = 2019,
        month = jan,
       volume = {131},
       number = {995},
        pages = {018002},
          doi = {10.1088/1538-3873/aaecbe},
archivePrefix = {arXiv},
       eprint = {1902.01932},
 primaryClass = {astro-ph.IM},
       adsurl = {https://ui.adsabs.harvard.edu/abs/2019PASP..131a8002B}
}

@ARTICLE{ZTF2,
       author = {{Masci}, Frank J. and {Laher}, Russ R. and {Rusholme}, Ben and {Shupe}, David L. and {Groom}, Steven and {Surace}, Jason and {Jackson}, Edward and {Monkewitz}, Serge and {Beck}, Ron and {Flynn}, David and {Terek}, Scott and {Landry}, Walter and {Hacopians}, Eugean and {Desai}, Vandana and {Howell}, Justin and {Brooke}, Tim and {Imel}, David and {Wachter}, Stefanie and {Ye}, Quan-Zhi and {Lin}, Hsing-Wen and {Cenko}, S. Bradley and {Cunningham}, Virginia and {Rebbapragada}, Umaa and {Bue}, Brian and {Miller}, Adam A. and {Mahabal}, Ashish and {Bellm}, Eric C. and {Patterson}, Maria T. and {Juri{\'c}}, Mario and {Golkhou}, V. Zach and {Ofek}, Eran O. and {Walters}, Richard and {Graham}, Matthew and {Kasliwal}, Mansi M. and {Dekany}, Richard G. and {Kupfer}, Thomas and {Burdge}, Kevin and {Cannella}, Christopher B. and {Barlow}, Tom and {Van Sistine}, Angela and {Giomi}, Matteo and {Fremling}, Christoffer and {Blagorodnova}, Nadejda and {Levitan}, David and {Riddle}, Reed and {Smith}, Roger M. and {Helou}, George and {Prince}, Thomas A. and {Kulkarni}, Shrinivas R.},
        title = "{The Zwicky Transient Facility: Data Processing, Products, and Archive}",
      journal = {\pasp},
         year = 2019,
        month = jan,
       volume = {131},
       number = {995},
        pages = {018003},
          doi = {10.1088/1538-3873/aae8ac},
archivePrefix = {arXiv},
       eprint = {1902.01872},
 primaryClass = {astro-ph.IM},
       adsurl = {https://ui.adsabs.harvard.edu/abs/2019PASP..131a8003M}
}

@ARTICLE{ZTF3,
       author = {{Graham}, Matthew J. and {Kulkarni}, S.~R. and {Bellm}, Eric C. and {Adams}, Scott M. and {Barbarino}, Cristina and {Blagorodnova}, Nadejda and {Bodewits}, Dennis and {Bolin}, Bryce and {Brady}, Patrick R. and {Cenko}, S. Bradley and {Chang}, Chan-Kao and {Coughlin}, Michael W. and {De}, Kishalay and {Eadie}, Gwendolyn and {Farnham}, Tony L. and {Feindt}, Ulrich and {Franckowiak}, Anna and {Fremling}, Christoffer and {Gezari}, Suvi and {Ghosh}, Shaon and {Goldstein}, Daniel A. and {Golkhou}, V. Zach and {Goobar}, Ariel and {Ho}, Anna Y.~Q. and {Huppenkothen}, Daniela and {Ivezi{\'c}}, {\v{Z}}eljko and {Jones}, R. Lynne and {Juric}, Mario and {Kaplan}, David L. and {Kasliwal}, Mansi M. and {Kelley}, Michael S.~P. and {Kupfer}, Thomas and {Lee}, Chien-De and {Lin}, Hsing Wen and {Lunnan}, Ragnhild and {Mahabal}, Ashish A. and {Miller}, Adam A. and {Ngeow}, Chow-Choong and {Nugent}, Peter and {Ofek}, Eran O. and {Prince}, Thomas A. and {Rauch}, Ludwig and {van Roestel}, Jan and {Schulze}, Steve and {Singer}, Leo P. and {Sollerman}, Jesper and {Taddia}, Francesco and {Yan}, Lin and {Ye}, Quan-Zhi and {Yu}, Po-Chieh and {Barlow}, Tom and {Bauer}, James and {Beck}, Ron and {Belicki}, Justin and {Biswas}, Rahul and {Brinnel}, Valery and {Brooke}, Tim and {Bue}, Brian and {Bulla}, Mattia and {Burruss}, Rick and {Connolly}, Andrew and {Cromer}, John and {Cunningham}, Virginia and {Dekany}, Richard and {Delacroix}, Alex and {Desai}, Vandana and {Duev}, Dmitry A. and {Feeney}, Michael and {Flynn}, David and {Frederick}, Sara and {Gal-Yam}, Avishay and {Giomi}, Matteo and {Groom}, Steven and {Hacopians}, Eugean and {Hale}, David and {Helou}, George and {Henning}, John and {Hover}, David and {Hillenbrand}, Lynne A. and {Howell}, Justin and {Hung}, Tiara and {Imel}, David and {Ip}, Wing-Huen and {Jackson}, Edward and {Kaspi}, Shai and {Kaye}, Stephen and {Kowalski}, Marek and {Kramer}, Emily and {Kuhn}, Michael and {Landry}, Walter and {Laher}, Russ R. and {Mao}, Peter and {Masci}, Frank J. and {Monkewitz}, Serge and {Murphy}, Patrick and {Nordin}, Jakob and {Patterson}, Maria T. and {Penprase}, Bryan and {Porter}, Michael and {Rebbapragada}, Umaa and {Reiley}, Dan and {Riddle}, Reed and {Rigault}, Mickael and {Rodriguez}, Hector and {Rusholme}, Ben and {van Santen}, Jakob and {Shupe}, David L. and {Smith}, Roger M. and {Soumagnac}, Maayane T. and {Stein}, Robert and {Surace}, Jason and {Szkody}, Paula and {Terek}, Scott and {Van Sistine}, Angela and {van Velzen}, Sjoert and {Vestrand}, W. Thomas and {Walters}, Richard and {Ward}, Charlotte and {Zhang}, Chaoran and {Zolkower}, Jeffry},
        title = "{The Zwicky Transient Facility: Science Objectives}",
      journal = {\pasp},
         year = 2019,
        month = jul,
       volume = {131},
       number = {1001},
        pages = {078001},
          doi = {10.1088/1538-3873/ab006c},
archivePrefix = {arXiv},
       eprint = {1902.01945},
 primaryClass = {astro-ph.IM},
       adsurl = {https://ui.adsabs.harvard.edu/abs/2019PASP..131g8001G}
}

@ARTICLE{SEDM,
       author = {{Blagorodnova}, Nadejda and {Neill}, James D. and {Walters}, Richard and {Kulkarni}, Shrinivas R. and {Fremling}, Christoffer and {Ben-Ami}, Sagi and {Dekany}, Richard G. and {Fucik}, Jason R. and {Konidaris}, Nick and {Nash}, Reston and {Ngeow}, Chow-Choong and {Ofek}, Eran O. and {O' Sullivan}, Donal and {Quimby}, Robert and {Ritter}, Andreas and {Vyhmeister}, Karl E.},
        title = "{The SED Machine: A Robotic Spectrograph for Fast Transient Classification}",
      journal = {\pasp},
         year = 2018,
        month = mar,
       volume = {130},
       number = {985},
        pages = {035003},
          doi = {10.1088/1538-3873/aaa53f},
archivePrefix = {arXiv},
       eprint = {1710.02917},
 primaryClass = {astro-ph.IM},
       adsurl = {https://ui.adsabs.harvard.edu/abs/2018PASP..130c5003B}
}

@article{Blagorodnova_2018,
   title={The SED Machine: A Robotic Spectrograph for Fast Transient Classification},
   volume={130},
   ISSN={1538-3873},
   url={http://dx.doi.org/10.1088/1538-3873/aaa53f},
   DOI={10.1088/1538-3873/aaa53f},
   number={985},
   journal={PASP},
   publisher={IOP Publishing},
   author={Blagorodnova, Nadejda and Neill, James D. and Walters, Richard and Kulkarni, Shrinivas R. and Fremling, Christoffer and Ben-Ami, Sagi and Dekany, Richard G. and Fucik, Jason R. and Konidaris, Nick and Nash, Reston and Ngeow, Chow-Choong and Ofek, Eran O. and Sullivan, Donal O’ and Quimby, Robert and Ritter, Andreas and Vyhmeister, Karl E.},
   year={2018},
   month=feb, pages={035003} }

@ARTICLE{2019A&A...627A.115R,
       author = {{Rigault}, M. and {Neill}, J.~D. and {Blagorodnova}, N. and {Dugas}, A. and {Feeney}, M. and {Walters}, R. and {Brinnel}, V. and {Copin}, Y. and {Fremling}, C. and {Nordin}, J. and {Sollerman}, J.},
        title = "{Fully automated integral field spectrograph pipeline for the SEDMachine: pysedm}",
      journal = {\aap},
         year = 2019,
        month = jul,
       volume = {627},
          eid = {A115},
        pages = {A115},
          doi = {10.1051/0004-6361/201935344},
archivePrefix = {arXiv},
       eprint = {1902.08526},
 primaryClass = {astro-ph.IM},
       adsurl = {https://ui.adsabs.harvard.edu/abs/2019A&A...627A.115R}
}

@ARTICLE{2022PASP..134b4505K,
       author = {{Kim}, Y. -L. and {Rigault}, M. and {Neill}, J.~D. and {Briday}, M. and {Copin}, Y. and {Lezmy}, J. and {Nicolas}, N. and {Riddle}, R. and {Sharma}, Y. and {Smith}, M. and {Sollerman}, J. and {Walters}, R.},
        title = "{New Modules for the SEDMachine to Remove Contaminations from Cosmic Rays and Non-target Light: BYECR and CONTSEP}",
      journal = {\pasp},
         year = 2022,
        month = feb,
       volume = {134},
       number = {1032},
          eid = {024505},
        pages = {024505},
          doi = {10.1088/1538-3873/ac50a0},
archivePrefix = {arXiv},
       eprint = {2203.01346},
 primaryClass = {astro-ph.IM},
       adsurl = {https://ui.adsabs.harvard.edu/abs/2022PASP..134b4505K}
}

@ARTICLE{2020ApJ...895...32F,
       author = {{Fremling}, C. and {Miller}, A.~A. and {Sharma}, Y. and {Dugas}, A. and {Perley}, D.~A. and {Taggart}, K. and {Sollerman}, J. and {Goobar}, A. and {Graham}, M.~L. and {Neill}, J.~D. and {Nordin}, J. and {Rigault}, M. and {Walters}, R. and {Andreoni}, I. and {Bagdasaryan}, A. and {Belicki}, J. and {Cannella}, C. and {Bellm}, E.~C. and {Cenko}, S.~B. and {De}, K. and {Dekany}, R. and {Frederick}, S. and {Golkhou}, V.~Z. and {Graham}, M.~J. and {Helou}, G. and {Ho}, A.~Y.~Q. and {Kasliwal}, M.~M. and {Kupfer}, T. and {Laher}, R.~R. and {Mahabal}, A. and {Masci}, F.~J. and {Riddle}, R. and {Rusholme}, B. and {Schulze}, S. and {Shupe}, D.~L. and {Smith}, R.~M. and {van Velzen}, S. and {Yan}, Lin and {Yao}, Y. and {Zhuang}, Z. and {Kulkarni}, S.~R.},
        title = "{The Zwicky Transient Facility Bright Transient Survey. I. Spectroscopic Classification and the Redshift Completeness of Local Galaxy Catalogs}",
      journal = {\apj},
         year = 2020,
        month = may,
       volume = {895},
       number = {1},
          eid = {32},
        pages = {32},
          doi = {10.3847/1538-4357/ab8943},
archivePrefix = {arXiv},
       eprint = {1910.12973},
 primaryClass = {astro-ph.HE},
       adsurl = {https://ui.adsabs.harvard.edu/abs/2020ApJ...895...32F}
}

@article{Perley_2020,
   title={The Zwicky Transient Facility Bright Transient Survey. II. A Public Statistical Sample for Exploring Supernova Demographics*},
   volume={904},
   ISSN={1538-4357},
   url={http://dx.doi.org/10.3847/1538-4357/abbd98},
   DOI={10.3847/1538-4357/abbd98},
   number={1},
   journal={ApJ},
   publisher={American Astronomical Society},
   author={Perley, Daniel A. and Fremling, Christoffer and Sollerman, Jesper and Miller, Adam A. and Dahiwale, Aishwarya S. and Sharma, Yashvi and Bellm, Eric C. and Biswas, Rahul and Brink, Thomas G. and Bruch, Rachel J. and De, Kishalay and Dekany, Richard and Drake, Andrew J. and Duev, Dmitry A. and Filippenko, Alexei V. and Gal-Yam, Avishay and Goobar, Ariel and Graham, Matthew J. and Graham, Melissa L. and Ho, Anna Y. Q. and Irani, Ido and Kasliwal, Mansi M. and Kim, Young-Lo and Kulkarni, S. R. and Mahabal, Ashish and Masci, Frank J. and Modak, Shaunak and Neill, James D. and Nordin, Jakob and Riddle, Reed L. and Soumagnac, Maayane T. and Strotjohann, Nora L. and Schulze, Steve and Taggart, Kirsty and Tzanidakis, Anastasios and Walters, Richard S. and Yan, Lin},
   year={2020},
   month=nov, pages={35} }

@ARTICLE{Rehemtulla+2024,
       author = {{Rehemtulla}, Nabeel and {Miller}, Adam A. and {Jegou Du Laz}, Theophile and {Coughlin}, Michael W. and {Fremling}, Christoffer and {Perley}, Daniel A. and {Qin}, Yu-Jing and {Sollerman}, Jesper and {Mahabal}, Ashish A. and {Laher}, Russ R. and {Riddle}, Reed and {Rusholme}, Ben and {Kulkarni}, Shrinivas R.},
        title = "{The Zwicky Transient Facility Bright Transient Survey. III. BTSbot: Automated Identification and Follow-up of Bright Transients with Deep Learning}",
      journal = {\apj},
         year = 2024,
        month = sep,
       volume = {972},
       number = {1},
          eid = {7},
        pages = {7},
          doi = {10.3847/1538-4357/ad5666},
archivePrefix = {arXiv},
       eprint = {2401.15167},
 primaryClass = {astro-ph.IM},
       adsurl = {https://ui.adsabs.harvard.edu/abs/2024ApJ...972....7R}
}

@ARTICLE{2022ApJS..259...35A,
       author = {{Abdurro'uf} and {Accetta}, Katherine and {Aerts}, Conny and {Silva Aguirre}, V{\'\i}ctor and {Ahumada}, Romina and {Ajgaonkar}, Nikhil and {Filiz Ak}, N. and {Alam}, Shadab and {Allende Prieto}, Carlos and {Almeida}, Andr{\'e}s and {Anders}, Friedrich and {Anderson}, Scott F. and {Andrews}, Brett H. and {Anguiano}, Borja and {Aquino-Ort{\'\i}z}, Erik and {Arag{\'o}n-Salamanca}, Alfonso and {Argudo-Fern{\'a}ndez}, Maria and {Ata}, Metin and {Aubert}, Marie and {Avila-Reese}, Vladimir and {Badenes}, Carles and {Barb{\'a}}, Rodolfo H. and {Barger}, Kat and {Barrera-Ballesteros}, Jorge K. and {Beaton}, Rachael L. and {Beers}, Timothy C. and {Belfiore}, Francesco and {Bender}, Chad F. and {Bernardi}, Mariangela and {Bershady}, Matthew A. and {Beutler}, Florian and {Bidin}, Christian Moni and {Bird}, Jonathan C. and {Bizyaev}, Dmitry and {Blanc}, Guillermo A. and {Blanton}, Michael R. and {Boardman}, Nicholas Fraser and {Bolton}, Adam S. and {Boquien}, M{\'e}d{\'e}ric and {Borissova}, Jura and {Bovy}, Jo and {Brandt}, W.~N. and {Brown}, Jordan and {Brownstein}, Joel R. and {Brusa}, Marcella and {Buchner}, Johannes and {Bundy}, Kevin and {Burchett}, Joseph N. and {Bureau}, Martin and {Burgasser}, Adam and {Cabang}, Tuesday K. and {Campbell}, Stephanie and {Cappellari}, Michele and {Carlberg}, Joleen K. and {Wanderley}, F{\'a}bio Carneiro and {Carrera}, Ricardo and {Cash}, Jennifer and {Chen}, Yan-Ping and {Chen}, Wei-Huai and {Cherinka}, Brian and {Chiappini}, Cristina and {Choi}, Peter Doohyun and {Chojnowski}, S. Drew and {Chung}, Haeun and {Clerc}, Nicolas and {Cohen}, Roger E. and {Comerford}, Julia M. and {Comparat}, Johan and {da Costa}, Luiz and {Covey}, Kevin and {Crane}, Jeffrey D. and {Cruz-Gonzalez}, Irene and {Culhane}, Connor and {Cunha}, Katia and {Dai}, Y. Sophia and {Damke}, Guillermo and {Darling}, Jeremy and {Davidson}, Jr., James W. and {Davies}, Roger and {Dawson}, Kyle and {De Lee}, Nathan and {Diamond-Stanic}, Aleksandar M. and {Cano-D{\'\i}az}, Mariana and {S{\'a}nchez}, Helena Dom{\'\i}nguez and {Donor}, John and {Duckworth}, Chris and {Dwelly}, Tom and {Eisenstein}, Daniel J. and {Elsworth}, Yvonne P. and {Emsellem}, Eric and {Eracleous}, Mike and {Escoffier}, Stephanie and {Fan}, Xiaohui and {Farr}, Emily and {Feng}, Shuai and {Fern{\'a}ndez-Trincado}, Jos{\'e} G. and {Feuillet}, Diane and {Filipp}, Andreas and {Fillingham}, Sean P. and {Frinchaboy}, Peter M. and {Fromenteau}, Sebastien and {Galbany}, Llu{\'\i}s and {Garc{\'\i}a}, Rafael A. and {Garc{\'\i}a-Hern{\'a}ndez}, D.~A. and {Ge}, Junqiang and {Geisler}, Doug and {Gelfand}, Joseph and {G{\'e}ron}, Tobias and {Gibson}, Benjamin J. and {Goddy}, Julian and {Godoy-Rivera}, Diego and {Grabowski}, Kathleen and {Green}, Paul J. and {Greener}, Michael and {Grier}, Catherine J. and {Griffith}, Emily and {Guo}, Hong and {Guy}, Julien and {Hadjara}, Massinissa and {Harding}, Paul and {Hasselquist}, Sten and {Hayes}, Christian R. and {Hearty}, Fred and {Hern{\'a}ndez}, Jes{\'u}s and {Hill}, Lewis and {Hogg}, David W. and {Holtzman}, Jon A. and {Horta}, Danny and {Hsieh}, Bau-Ching and {Hsu}, Chin-Hao and {Hsu}, Yun-Hsin and {Huber}, Daniel and {Huertas-Company}, Marc and {Hutchinson}, Brian and {Hwang}, Ho Seong and {Ibarra-Medel}, H{\'e}ctor J. and {Chitham}, Jacob Ider and {Ilha}, Gabriele S. and {Imig}, Julie and {Jaekle}, Will and {Jayasinghe}, Tharindu and {Ji}, Xihan and {Johnson}, Jennifer A. and {Jones}, Amy and {J{\"o}nsson}, Henrik and {Katkov}, Ivan and {Khalatyan}, Dr., Arman and {Kinemuchi}, Karen and {Kisku}, Shobhit and {Knapen}, Johan H. and {Kneib}, Jean-Paul and {Kollmeier}, Juna A. and {Kong}, Miranda and {Kounkel}, Marina and {Kreckel}, Kathryn and {Krishnarao}, Dhanesh and {Lacerna}, Ivan and {Lane}, Richard R. and {Langgin}, Rachel and {Lavender}, Ramon and {Law}, David R. and {Lazarz}, Daniel and {Leung}, Henry W. and {Leung}, Ho-Hin and {Lewis}, Hannah M. and {Li}, Cheng and {Li}, Ran and {Lian}, Jianhui and {Liang}, Fu-Heng and {Lin}, Lihwai and {Lin}, Yen-Ting and {Lin}, Sicheng and {Lintott}, Chris and {Long}, Dan and {Longa-Pe{\~n}a}, Pen{\'e}lope and {L{\'o}pez-Cob{\'a}}, Carlos and {Lu}, Shengdong and {Lundgren}, Britt F. and {Luo}, Yuanze and {Mackereth}, J. Ted and {de la Macorra}, Axel and {Mahadevan}, Suvrath and {Majewski}, Steven R. and {Manchado}, Arturo and {Mandeville}, Travis and {Maraston}, Claudia and {Margalef-Bentabol}, Berta and {Masseron}, Thomas and {Masters}, Karen L. and {Mathur}, Savita and {McDermid}, Richard M. and {Mckay}, Myles and {Merloni}, Andrea and {Merrifield}, Michael and {Meszaros}, Szabolcs and {Miglio}, Andrea and {Di Mille}, Francesco and {Minniti}, Dante and {Minsley}, Rebecca and {Monachesi}, Antonela},
        title = "{The Seventeenth Data Release of the Sloan Digital Sky Surveys: Complete Release of MaNGA, MaStar, and APOGEE-2 Data}",
      journal = {\apjs},
         year = 2022,
        month = apr,
       volume = {259},
       number = {2},
          eid = {35},
        pages = {35},
          doi = {10.3847/1538-4365/ac4414},
archivePrefix = {arXiv},
       eprint = {2112.02026},
 primaryClass = {astro-ph.GA},
       adsurl = {https://ui.adsabs.harvard.edu/abs/2022ApJS..259...35A}
}

@ARTICLE{2024AJ....168...58D,
       author = {{DESI Collaboration} and {Adame}, A.~G. and {Aguilar}, J. and {Ahlen}, S. and {Alam}, S. and {Aldering}, G. and {Alexander}, D.~M. and {Alfarsy}, R. and {Allende Prieto}, C. and {Alvarez}, M. and {Alves}, O. and {Anand}, A. and {Andrade-Oliveira}, F. and {Armengaud}, E. and {Asorey}, J. and {Avila}, S. and {Aviles}, A. and {Bailey}, S. and {Balaguera-Antol{\'\i}nez}, A. and {Ballester}, O. and {Baltay}, C. and {Bault}, A. and {Bautista}, J. and {Behera}, J. and {Beltran}, S.~F. and {BenZvi}, S. and {Beraldo e Silva}, L. and {Bermejo-Climent}, J.~R. and {Berti}, A. and {Besuner}, R. and {Beutler}, F. and {Bianchi}, D. and {Blake}, C. and {Blum}, R. and {Bolton}, A.~S. and {Brieden}, S. and {Brodzeller}, A. and {Brooks}, D. and {Brown}, Z. and {Buckley-Geer}, E. and {Burtin}, E. and {Cabayol-Garcia}, L. and {Cai}, Z. and {Canning}, R. and {Cardiel-Sas}, L. and {Carnero Rosell}, A. and {Castander}, F.~J. and {Cervantes-Cota}, J.~L. and {Chabanier}, S. and {Chaussidon}, E. and {Chaves-Montero}, J. and {Chen}, S. and {Chen}, X. and {Chuang}, C. and {Claybaugh}, T. and {Cole}, S. and {Cooper}, A.~P. and {Cuceu}, A. and {Davis}, T.~M. and {Dawson}, K. and {de Belsunce}, R. and {de la Cruz}, R. and {de la Macorra}, A. and {Della Costa}, J. and {de Mattia}, A. and {Demina}, R. and {Demirbozan}, U. and {DeRose}, J. and {Dey}, A. and {Dey}, B. and {Dhungana}, G. and {Ding}, J. and {Ding}, Z. and {Doel}, P. and {Doshi}, R. and {Douglass}, K. and {Edge}, A. and {Eftekharzadeh}, S. and {Eisenstein}, D.~J. and {Elliott}, A. and {Ereza}, J. and {Escoffier}, S. and {Fagrelius}, P. and {Fan}, X. and {Fanning}, K. and {Fawcett}, V.~A. and {Ferraro}, S. and {Flaugher}, B. and {Font-Ribera}, A. and {Forero-Romero}, J.~E. and {Forero-S{\'a}nchez}, D. and {Frenk}, C.~S. and {G{\"a}nsicke}, B.~T. and {Garc{\'\i}a}, L. {\'A}. and {Garc{\'\i}a-Bellido}, J. and {Garcia-Quintero}, C. and {Garrison}, L.~H. and {Gil-Mar{\'\i}n}, H. and {Golden-Marx}, J. and {Gontcho A Gontcho}, S. and {Gonzalez-Morales}, A.~X. and {Gonzalez-Perez}, V. and {Gordon}, C. and {Graur}, O. and {Green}, D. and {Gruen}, D. and {Guy}, J. and {Hadzhiyska}, B. and {Hahn}, C. and {Han}, J.~J. and {Hanif}, M.~M.~S. and {Herrera-Alcantar}, H.~K. and {Honscheid}, K. and {Hou}, J. and {Howlett}, C. and {Huterer}, D. and {Ir{\v{s}}i{\v{c}}}, V. and {Ishak}, M. and {Jacques}, A. and {Jana}, A. and {Jiang}, L. and {Jimenez}, J. and {Jing}, Y.~P. and {Joudaki}, S. and {Joyce}, R. and {Jullo}, E. and {Juneau}, S. and {Kara{\c{c}}ayl{\i}}, N.~G. and {Karim}, T. and {Kehoe}, R. and {Kent}, S. and {Khederlarian}, A. and {Kim}, S. and {Kirkby}, D. and {Kisner}, T. and {Kitaura}, F. and {Kizhuprakkat}, N. and {Kneib}, J. and {Koposov}, S.~E. and {Kov{\'a}cs}, A. and {Kremin}, A. and {Krolewski}, A. and {L'Huillier}, B. and {Lahav}, O. and {Lambert}, A. and {Lamman}, C. and {Lan}, T. -W. and {Landriau}, M. and {Lang}, D. and {Lange}, J.~U. and {Lasker}, J. and {Leauthaud}, A. and {Le Guillou}, L. and {Levi}, M.~E. and {Li}, T.~S. and {Linder}, E. and {Lyons}, A. and {Magneville}, C. and {Manera}, M. and {Manser}, C.~J. and {Margala}, D. and {Martini}, P. and {McDonald}, P. and {Medina}, G.~E. and {Medina-Varela}, L. and {Meisner}, A. and {Mena-Fern{\'a}ndez}, J. and {Meneses-Rizo}, J. and {Mezcua}, M. and {Miquel}, R. and {Montero-Camacho}, P. and {Moon}, J. and {Moore}, S. and {Moustakas}, J. and {Mueller}, E. and {Mundet}, J. and {Mu{\~n}oz-Guti{\'e}rrez}, A. and {Myers}, A.~D. and {Nadathur}, S. and {Napolitano}, L. and {Neveux}, R. and {Newman}, J.~A. and {Nie}, J. and {Nikutta}, R. and {Niz}, G. and {Norberg}, P. and {Noriega}, H.~E. and {Paillas}, E. and {Palanque-Delabrouille}, N. and {Palmese}, A. and {Pan}, Z. and {Parkinson}, D. and {Penmetsa}, S. and {Percival}, W.~J. and {P{\'e}rez-Fern{\'a}ndez}, A. and {P{\'e}rez-R{\`a}fols}, I. and {Pieri}, M. and {Poppett}, C. and {Porredon}, A. and {Pothier}, S.},
        title = "{The Early Data Release of the Dark Energy Spectroscopic Instrument}",
      journal = {\aj},
         year = 2024,
        month = aug,
       volume = {168},
       number = {2},
          eid = {58},
        pages = {58},
          doi = {10.3847/1538-3881/ad3217},
archivePrefix = {arXiv},
       eprint = {2306.06308},
 primaryClass = {astro-ph.CO},
       adsurl = {https://ui.adsabs.harvard.edu/abs/2024AJ....168...58D}
}

@article{vanderWalt2019,
    doi = {10.21105/joss.01247},
    url = {https://doi.org/10.21105/joss.01247},
    year = {2019},
    publisher = {The Open Journal},
    volume = {4}, number = {37},
    pages = {1247},
    author = {Stéfan J. van der Walt and Arien Crellin-Quick and Joshua S. Bloom},
    title = {SkyPortal: An Astronomical Data Platform},
    journal = {Journal of Open Source Software} }

@ARTICLE{2023ApJS..267...31C,
       author = {{Coughlin}, Michael W. and {Bloom}, Joshua S. and {Nir}, Guy and {Antier}, Sarah and {du Laz}, Theophile Jegou and {van der Walt}, St{\'e}fan and {Crellin-Quick}, Arien and {Culino}, Thomas and {Duev}, Dmitry A. and {Goldstein}, Daniel A. and {Healy}, Brian F. and {Karambelkar}, Viraj and {Lilleboe}, Jada and {Shin}, Kyung Min and {Singer}, Leo P. and {Ahumada}, Tom{\'a}s and {Anand}, Shreya and {Bellm}, Eric C. and {Dekany}, Richard and {Graham}, Matthew J. and {Kasliwal}, Mansi M. and {Kostadinova}, Ivona and {Kiendrebeogo}, R. Weizmann and {Kulkarni}, Shrinivas R. and {Jenkins}, Sydney and {LeBaron}, Natalie and {Mahabal}, Ashish A. and {Neill}, James D. and {Parazin}, B. and {Peloton}, Julien and {Perley}, Daniel A. and {Riddle}, Reed and {Rusholme}, Ben and {van Santen}, Jakob and {Sollerman}, Jesper and {Stein}, Robert and {Turpin}, D. and {Wold}, Avery and {Amat}, Carla and {Bonnefon}, Adrien and {Bonnefoy}, Adrien and {Flament}, Manon and {Kerkow}, Frank and {Kishore}, Sulekha and {Jani}, Shloke and {Mahanty}, Stephen K. and {Liu}, C{\'e}line and {Llinares}, Laura and {Makarison}, Jolyane and {Olli{\'e}ric}, Alix and {Perez}, In{\`e}s and {Pont}, Lydie and {Sharma}, Vyom},
        title = "{A Data Science Platform to Enable Time-domain Astronomy}",
      journal = {\apjs},
         year = 2023,
        month = aug,
       volume = {267},
       number = {2},
          eid = {31},
        pages = {31},
          doi = {10.3847/1538-4365/acdee1},
archivePrefix = {arXiv},
       eprint = {2305.00108},
 primaryClass = {astro-ph.IM},
       adsurl = {https://ui.adsabs.harvard.edu/abs/2023ApJS..267...31C}
}

@article{Kasliwal_2019,
   title={The GROWTH Marshal: A Dynamic Science Portal for Time-domain Astronomy},
   volume={131},
   ISSN={1538-3873},
   url={http://dx.doi.org/10.1088/1538-3873/aafbc2},
   DOI={10.1088/1538-3873/aafbc2},
   number={997},
   journal={PASP},
   publisher={IOP Publishing},
   author={Kasliwal, M. M. and Cannella, C. and Bagdasaryan, A. and Hung, T. and Feindt, U. and Singer, L. P. and Coughlin, M. and Fremling, C. and Walters, R. and Duev, D. and Itoh, R. and Quimby, R. M.},
   year={2019},
   month=feb, pages={038003} }

@article{Fremling2021SNIascore,
  author       = {Fremling, Christoffer and Hall, Xander J. and Coughlin, Michael W. and Dahiwale, Aishwarya S. and Duev, Dmitry A. and Graham, Matthew J. and Kasliwal, Mansi M. and Kool, Erik C. and Mahabal, Ashish A. and Miller, Adam A. and Neill, James D. and Perley, Daniel A. and Rigault, Mickael and Rosnet, Philippe and Rusholme, Ben and Sharma, Yashvi and Shin, Kyung Min and Shupe, David L. and Sollerman, Jesper and Walters, Richard S.},
  title        = {SNIascore: Deep-learning Classification of Low-resolution Supernova Spectra},
  journal      = {ApJL},
  volume       = {917},
  number       = {1},
  pages        = {L2},
  year         = {2021},
  doi          = {10.3847/2041-8213/ac116f},
  eprint       = {2104.12980},
  archivePrefix = {arXiv},
  primaryClass = {astro-ph.IM}
}

@article{sharma2025ccsnscoremultiinputdeeplearning,
doi = {10.1088/1538-3873/adbf4b},
url = {https://dx.doi.org/10.1088/1538-3873/adbf4b},
year = {2025},
month = {mar},
publisher = {The Astronomical Society of the Pacific},
volume = {137},
number = {3},
pages = {034507},
author = {Sharma, Yashvi and Mahabal, Ashish A. and Sollerman, Jesper and Fremling, Christoffer and Kulkarni, S. R. and Rehemtulla, Nabeel and Miller, Adam A. and Aubert, Marie and Chen, Tracy X. and Coughlin, Michael W. and Graham, Matthew J. and Hale, David and Kasliwal, Mansi M. and Kim, Young-Lo and Neill, James D. and Purdum, Josiah N. and Rusholme, Ben and Singh, Avinash and Sravan, Niharika},
title = {CCSNscore: A Multi-input Deep Learning Tool for Classification of Core-collapse Supernovae Using SED-machine Spectra},
journal = {PASP}
}

@misc{applecider,
      title={Applying multimodal learning to Classify transient Detections Early (AppleCiDEr) I: Data set, methods, and infrastructure}, 
      author={Alexandra Junell and Argyro Sasli and Felipe Fontinele Nunes and Maojie Xu and Benny Border and Nabeel Rehemtulla and Mariia Rizhko and Yu-Jing Qin and Theophile Jegou Du Laz and Antoine Le Calloch and Sushant Sharma Chaudhary and Shaowei Wu and Jesper Sollerman and Niharika Sravan and Steven L. Groom and David Hale and Mansi M. Kasliwal and Josiah Purdum and Avery Wold and Matthew J. Graham and Michael W. Coughlin},
      year={2025},
      eprint={2507.16088},
      archivePrefix={arXiv},
      primaryClass={astro-ph.IM},
      url={https://arxiv.org/abs/2507.16088}, 
}

@misc{fortino2025,
      title={ABC-SN: Attention Based Classifier for Supernova Spectra}, 
      author={Willow Fox Fortino and Federica B. Bianco and Pavlos Protopapas and Daniel Muthukrishna and Austin Brockmeier},
      year={2025},
      eprint={2507.22106},
      archivePrefix={arXiv},
      primaryClass={astro-ph.IM},
      url={https://arxiv.org/abs/2507.22106}, 
}

@article{Muthukrishna2019,
doi = {10.3847/1538-4357/ab48f4},
url = {https://dx.doi.org/10.3847/1538-4357/ab48f4},
year = {2019},
month = {nov},
publisher = {The American Astronomical Society},
volume = {885},
number = {1},
pages = {85},
author = {Muthukrishna, Daniel and Parkinson, David and Tucker, Brad E.},
title = {DASH: Deep Learning for the Automated Spectral Classification of Supernovae and Their Hosts},
journal = {ApJ}
}

@ARTICLE{2024TNSAN.340....1K,
       author = {{Kasliwal}, M.~M. and {Fremling}, C. and {Yan}, L. and {Das}, K. and {Verdi}, F. and {Zmudzinas}, J. and {Martin}, C. and {Kirby}, E. and {Xue}, S. and {Ho}, L. and {Herczeg}, G. and {Wu}, X. and {Hu}, Z. and {Ji}, H. and {Matuszewski}, M. and {Bertz}, R. and {Hale}, D. and {Rodriguez}, H. and {Boden}, A. and {Dekany}, R. and {Smith}, R. and {Reiley}, D. and {Nash}, R. and {Milburn}, J. and {Neill}, D. and {Brugger}, J. and {Zarzaca}, R. and {Weber}, B. and {Shapiro}, C.},
        title = "{The First Science Spectrum with the Next Generation Palomar Spectrograph (NGPS): Classification of ZTF24abrfcqd as a Type Ia Supernova}",
      journal = {Transient Name Server AstroNote},
         year = 2024,
        month = nov,
       volume = {340},
        pages = {1},
       adsurl = {https://ui.adsabs.harvard.edu/abs/2024TNSAN.340....1K}
}

@article{Ho_2023,
doi = {10.3847/1538-4357/acc533},
url = {https://dx.doi.org/10.3847/1538-4357/acc533},
year = {2023},
month = {jun},
publisher = {The American Astronomical Society},
volume = {949},
number = {2},
pages = {120},
author = {Ho, Anna Y. Q. and Perley, Daniel A. and Gal-Yam, Avishay and Lunnan, Ragnhild and Sollerman, Jesper and Schulze, Steve and Das, Kaustav K. and Dobie, Dougal and Yao, Yuhan and Fremling, Christoffer and Adams, Scott and Anand, Shreya and Andreoni, Igor and Bellm, Eric C. and Bruch, Rachel J. and Burdge, Kevin B. and Castro-Tirado, Alberto J. and Dahiwale, Aishwarya and De, Kishalay and Dekany, Richard and Drake, Andrew J. and Duev, Dmitry A. and Graham, Matthew J. and Helou, George and Kaplan, David L. and Karambelkar, Viraj and Kasliwal, Mansi M. and Kool, Erik C. and Kulkarni, S. R. and Mahabal, Ashish A. and Medford, Michael S. and Miller, A. A. and Nordin, Jakob and Ofek, Eran and Petitpas, Glen and Riddle, Reed and Sharma, Yashvi and Smith, Roger and Stewart, Adam J. and Taggart, Kirsty and Tartaglia, Leonardo and Tzanidakis, Anastasios and Winters, Jan Martin},
title = {A Search for Extragalactic Fast Blue Optical Transients in ZTF and the Rate of AT2018cow-like Transients},
journal = {The Astrophysical Journal},
}

@misc{BOOM,
      title={BOOM and Babamul: a real-time, multi-survey, optical alert broker system operating at scale}, 
      author={Theophile Jegou du Laz and Michael W. Coughlin and Peter Bachant and Jacob E. Simones and Thomas Culino and Antoine Le Calloch and Sushant Sharma Chaudhary and Xander J. Hall and Tyler Barna and Daniel Warshofsky and Matthew Graham and Mansi M. Kasliwal and Ashish Mahabal and Joshua S. Bloom and Antonella Palmese and Frank J. Masci and Steven L. Groom and Richard Dekany and Reed L. Riddle and George Helou},
      year={2025},
      eprint={2511.00164},
      archivePrefix={arXiv},
      primaryClass={astro-ph.IM},
      url={https://arxiv.org/abs/2511.00164}, 
}

@ARTICLE{2023PASP..135f4501C,
       author = {{Coulter}, D.~A. and {Jones}, D.~O. and {McGill}, P. and {Foley}, R.~J. and {Aleo}, P.~D. and {Bustamante-Rosell}, M.~J. and {Chatterjee}, D. and {Davis}, K.~W. and {Dickinson}, C. and {Engel}, A. and {Gagliano}, A. and {Jacobson-Gal{\'a}n}, W.~V. and {Kilpatrick}, C.~D. and {Kutcka}, J. and {Le Saux}, X.~K. and {Malanchev}, K. and {Pan}, Y. -C. and {Qui{\~n}onez}, P.~J. and {Rojas-Bravo}, C. and {Siebert}, M.~R. and {Taggart}, K. and {Tinyanont}, S. and {Wang}, Q.},
        title = "{YSE-PZ: A Transient Survey Management Platform that Empowers the Human-in-the-loop}",
      journal = {\pasp},
         year = 2023,
        month = jun,
       volume = {135},
       number = {1048},
          eid = {064501},
        pages = {064501},
          doi = {10.1088/1538-3873/acd662},
archivePrefix = {arXiv},
       eprint = {2303.02154},
 primaryClass = {astro-ph.IM},
       adsurl = {https://ui.adsabs.harvard.edu/abs/2023PASP..135f4501C}
}

@ARTICLE{2020PASP..132c8001D,
       author = {{Dekany}, Richard and {Smith}, Roger M. and {Riddle}, Reed and {Feeney}, Michael and {Porter}, Michael and {Hale}, David and {Zolkower}, Jeffry and {Belicki}, Justin and {Kaye}, Stephen and {Henning}, John and {Walters}, Richard and {Cromer}, John and {Delacroix}, Alex and {Rodriguez}, Hector and {Reiley}, Daniel J. and {Mao}, Peter and {Hover}, David and {Murphy}, Patrick and {Burruss}, Rick and {Baker}, John and {Kowalski}, Marek and {Reif}, Klaus and {Mueller}, Phillip and {Bellm}, Eric and {Graham}, Matthew and {Kulkarni}, Shrinivas R.},
        title = "{The Zwicky Transient Facility: Observing System}",
      journal = {\pasp},
         year = 2020,
        month = mar,
       volume = {132},
       number = {1009},
          eid = {038001},
        pages = {038001},
          doi = {10.1088/1538-3873/ab4ca2},
archivePrefix = {arXiv},
       eprint = {2008.04923},
 primaryClass = {astro-ph.IM},
       adsurl = {https://ui.adsabs.harvard.edu/abs/2020PASP..132c8001D}
}

@article{Oke_1995,
doi = {10.1086/133562},
url = {https://doi.org/10.1086/133562},
year = {1995},
month = {apr},
publisher = {The Astronomical Society of the Pacific},
volume = {107},
number = {710},
pages = {375},
author = {Oke, J. B. and Cohen, J. G. and Carr, M. and Cromer, J. and Dingizian, A. and Harris, F. H. and Labrecque, S. and Lucinio, R. and Schaal, W. and Epps, H. and Miller, J.},
title = {THE KECK LOW-RESOLUTION IMAGING SPECTROMETER},
journal = {Publications of the Astronomical Society of the Pacific}
}

@INPROCEEDINGS{2003SPIE.4841.1657F,
       author = {{Faber}, Sandra M. and {Phillips}, Andrew C. and {Kibrick}, Robert I. and {Alcott}, Barry and {Allen}, Steven L. and {Burrous}, Jim and {Cantrall}, T. and {Clarke}, De and {Coil}, Alison L. and {Cowley}, David J. and {Davis}, Marc and {Deich}, William T.~S. and {Dietsch}, Ken and {Gilmore}, David K. and {Harper}, Carol A. and {Hilyard}, David F. and {Lewis}, Jeffrey P. and {McVeigh}, Molly and {Newman}, Jeffrey and {Osborne}, Jack and {Schiavon}, Ricardo and {Stover}, Richard J. and {Tucker}, Dean and {Wallace}, Vernon and {Wei}, Mingzhi and {Wirth}, Gregory and {Wright}, Christopher A.},
        title = "{The DEIMOS spectrograph for the Keck II Telescope: integration and testing}",
    booktitle = {Instrument Design and Performance for Optical/Infrared Ground-based Telescopes},
         year = 2003,
       editor = {{Iye}, Masanori and {Moorwood}, Alan F.~M.},
       series = {Society of Photo-Optical Instrumentation Engineers (SPIE) Conference Series},
       volume = {4841},
        month = mar,
        pages = {1657-1669},
          doi = {10.1117/12.460346},
       adsurl = {https://ui.adsabs.harvard.edu/abs/2003SPIE.4841.1657F}
}

@INPROCEEDINGS{1997SPIE.2871.1099D,
       author = {{Davies}, Roger L. and {Allington-Smith}, Jeremy R. and {Bettess}, P. and {Chadwick}, E. and {Content}, Robert and {Dodsworth}, G.~N. and {Haynes}, Roger and {Lee}, D. and {Lewis}, Ian J. and {Webster}, J. and {Atad}, E. and {Beard}, Steven M. and {Ellis}, M. and {Hastings}, Peter R. and {Williams}, Phil R. and {Bond}, Tim and {Crampton}, David and {Davidge}, Timothy J. and {Fletcher}, Murray and {Leckie}, Brian and {Morbey}, Christopher L. and {Murowinski}, Richard G. and {Roberts}, Scott and {Saddlemyer}, Leslie K. and {Sebesta}, Jerry and {Stilburn}, James R. and {Szeto}, Kei},
        title = "{GMOS: the GEMINI Multiple Object Spectrographs}",
    booktitle = {Optical Telescopes of Today and Tomorrow},
         year = 1997,
       editor = {{Ardeberg}, Arne L.},
       series = {Society of Photo-Optical Instrumentation Engineers (SPIE) Conference Series},
       volume = {2871},
        month = mar,
        pages = {1099-1106},
          doi = {10.1117/12.268996},
       adsurl = {https://ui.adsabs.harvard.edu/abs/1997SPIE.2871.1099D}
}

@inproceedings{Pogge2010MODS,
  author = {Pogge, R. W. et al.},
  title = {The Multi-Object Double Spectrographs for the Large Binocular Telescope},
  booktitle = {SPIE},
  volume = {7735},
  year = {2010}
}

@ARTICLE{Drout2014,
  author        = {{Drout}, M.~R. and {Chornock}, R. and {Soderberg}, A.~M. and 
                   {Sanders}, N.~E. and {McKinnon}, R. and {Rest}, A. and 
                   {Foley}, R.~J. and {Milisavljevic}, D. and {Margutti}, R. and 
                   {Berger}, E. and others},
  title         = {{Rapidly Evolving and Luminous Transients from Pan-STARRS1}},
  journal       = {The Astrophysical Journal},
  volume        = {794},
  number        = {1},
  pages         = {23},
  year          = {2014},
  doi           = {10.1088/0004-637X/794/1/23},
  eprint        = {1405.3668},
  archivePrefix = {arXiv},
  primaryClass  = {astro-ph.HE},
  adsurl        = {https://ui.adsabs.harvard.edu/abs/2014ApJ...794...23D}
}

@ARTICLE{Thompson2009,
  author        = {{Thompson}, Todd A. and {Prieto}, Jos{\'e} L. and 
                   {Stanek}, K.~Z. and {Kistler}, Matthew D. and 
                   {Beacom}, John F. and {Kochanek}, Christopher S.},
  title         = {{A New Class of Luminous Transients and a First 
                   Census of their Massive Stellar Progenitors}},
  journal       = {The Astrophysical Journal},
  volume        = {705},
  number        = {2},
  pages         = {1364--1384},
  year          = {2009},
  doi           = {10.1088/0004-637X/705/2/1364},
  eprint        = {0809.0510},
  archivePrefix = {arXiv},
  primaryClass  = {astro-ph},
  adsurl        = {https://ui.adsabs.harvard.edu/abs/2009ApJ...705.1364T}
}

@ARTICLE{Kochanek2012,
  author        = {{Kochanek}, C.~S. and {Szczygie{\l}}, D.~M. and 
                   {Stanek}, K.~Z.},
  title         = {{Unmasking the Supernova Impostors}},
  journal       = {The Astrophysical Journal},
  volume        = {758},
  number        = {2},
  pages         = {142},
  year          = {2012},
  doi           = {10.1088/0004-637X/758/2/142},
  eprint        = {1202.0281},
  archivePrefix = {arXiv},
  primaryClass  = {astro-ph.SR},
  adsurl        = {https://ui.adsabs.harvard.edu/abs/2012ApJ...758..142K}
}

@ARTICLE{Margutti2019,
  author        = {{Margutti}, R. and {Metzger}, B.~D. and 
                   {Chornock}, R. and {Vurm}, I. and {Roth}, N. and 
                   {Grefenstette}, B.~W. and {Savchenko}, V. and 
                   {Cartier}, R. and {Steiner}, J.~F. and 
                   {Terreran}, G. and others},
  title         = {{An Embedded X-Ray Source Shines through the 
                   Aspherical AT 2018cow: Revealing the Inner Workings 
                   of the Most Luminous and Rapidly Evolving Optical 
                   Transients}},
  journal       = {The Astrophysical Journal},
  volume        = {872},
  number        = {1},
  pages         = {18},
  year          = {2019},
  doi           = {10.3847/1538-4357/aafa01},
  eprint        = {1810.10720},
  archivePrefix = {arXiv},
  primaryClass  = {astro-ph.HE},
  adsurl        = {https://ui.adsabs.harvard.edu/abs/2019ApJ...872...18M}
}

@ARTICLE{Perley2019,
  author        = {{Perley}, D.~A. and {Mazzali}, P.~A. and {Yan}, Lin 
                   and {Cenko}, S.~B. and {Gezari}, S. and 
                   {Taggart}, K. and {Blagorodnova}, N. and 
                   {Fremling}, C. and {Mockler}, B. and 
                   {Singh}, A. and others},
  title         = {{The fast, luminous ultraviolet transient AT2018cow: 
                   extreme supernova, or disruption of a star by an 
                   intermediate-mass black hole?}},
  journal       = {Monthly Notices of the Royal Astronomical Society},
  volume        = {484},
  number        = {1},
  pages         = {1031--1049},
  year          = {2019},
  doi           = {10.1093/mnras/sty3420},
  eprint        = {1808.00969},
  archivePrefix = {arXiv},
  primaryClass  = {astro-ph.HE},
  adsurl        = {https://ui.adsabs.harvard.edu/abs/2019MNRAS.484.1031P}
}

@ARTICLE{Perets2010,
  author        = {{Perets}, H.~B. and {Gal-Yam}, A. and 
                   {Mazzali}, P.~A. and {Arnett}, D. and 
                   {Kagan}, D. and {Filippenko}, A.~V. and {Li}, W. 
                   and {Arcavi}, I. and {Cenko}, S.~B. and 
                   {Fox}, D.~B. and others},
  title         = {{A faint type of supernova from a white dwarf with 
                   a helium-rich companion}},
  journal       = {Nature},
  volume        = {465},
  number        = {7296},
  pages         = {322--325},
  year          = {2010},
  doi           = {10.1038/nature09056},
  eprint        = {0906.2003},
  archivePrefix = {arXiv},
  primaryClass  = {astro-ph.HE},
  adsurl        = {https://ui.adsabs.harvard.edu/abs/2010Natur.465..322P}
}

@misc{xuspectra,
      title={AppleCiDEr II: SpectraNet -- A Deep Learning Network for Spectroscopic Data}, 
      author={Maojie Xu and Argyro Sasli and Alexandra Junell and Felipe Fontinele Nunes and Yu-Jing Qin and Christoffer Fremling and Sam Rose and Theophile Jegou Du Laz and Benny Border and Antoine Le Calloch and Sushant Sharma Chaudhary and Hailey Markoff and Avyukt Raghuvanshi and Nabeel Rehemtulla and Jesper Sollerman and Yashvi Sharma and Niharika Sravan and Judy Adler and Tracy X. Chen and Richard Dekany and Reed Riddle and Mansi M. Kasliwal and Matthew J. Graham and Michael W. Coughlin},
      year={2025},
      eprint={2510.07215},
      archivePrefix={arXiv},
      primaryClass={astro-ph.IM},
      url={https://arxiv.org/abs/2510.07215}, 
}

@article{Barna_2025,
doi = {10.1088/1538-3873/adf578},
url = {https://doi.org/10.1088/1538-3873/adf578},
year = {2025},
month = {aug},
publisher = {The Astronomical Society of the Pacific},
volume = {137},
number = {8},
pages = {084105},
author = {Barna, Tyler and Fremling, Christoffer and Ahumada, Tomas and Andreoni, Igor and Banerjee, Smaranika and Bloom, Joshua S. and Bulla, Mattia and Chen, Tracy X. and Coughlin, Michael W. and Dietrich, Tim and Hall, Xander J. and Junell, Alexandra and Rusholme, Ben and Sollerman, Jesper and Sravan, Niharika},
title = {IIb or not IIb: A Catalog of ZTF Kilonova Imposters},
journal = {Publications of the Astronomical Society of the Pacific}
}

@ARTICLE{2022GCN.31629....1F,
       author = {{Fremling}, C. and {Perley}, D. and {Ho}, A.~Y.~Q. and {ZTF Collaboration}},
        title = "{ZTF22aabjpxh/AT2022cva: GMOS-N spectroscopy}",
      journal = {GRB Coordinates Network},
         year = 2022,
        month = feb,
       volume = {31629},
        pages = {1},
       adsurl = {https://ui.adsabs.harvard.edu/abs/2022GCN.31629....1F}
}

@ARTICLE{2025ApJ...985..124L,
       author = {{Li}, Maggie L. and {Ho}, Anna Y.~Q. and {Ryan}, Geoffrey and {Perley}, Daniel A. and {Lamb}, Gavin P. and {Nayana}, A.~J. and {Andreoni}, Igor and {Anupama}, G.~C. and {Bellm}, Eric C. and {Berger}, Edo and {Bloom}, Joshua S. and {Burns}, Eric and {Caiazzo}, Ilaria and {Chandra}, Poonam and {Coughlin}, Michael W. and {El-Badry}, Kareem and {Graham}, Matthew J. and {Kasliwal}, Mansi and {Keating}, Garrett K. and {Kulkarni}, S.~R. and {Kumar}, Harsh and {Masci}, Frank J. and {Perley}, Richard A. and {Purdum}, Josiah and {Rao}, Ramprasad and {Rodriguez}, Antonio C. and {Rusholme}, Ben and {Sarin}, Nikhil and {Sollerman}, Jesper and {Srinivasaragavan}, Gokul P. and {Swain}, Vishwajeet and {Vanderbosch}, Zachary},
        title = "{The Nature of Optical Afterglows without Gamma-Ray Bursts: Identification of AT2023lcr and Multiwavelength Modeling}",
      journal = {\apj},
         year = 2025,
        month = may,
       volume = {985},
       number = {1},
          eid = {124},
        pages = {124},
          doi = {10.3847/1538-4357/adc800},
archivePrefix = {arXiv},
       eprint = {2411.07973},
 primaryClass = {astro-ph.HE},
       adsurl = {https://ui.adsabs.harvard.edu/abs/2025ApJ...985..124L}
}

@ARTICLE{Groot2025,
       author = {{Groot}, Paul J. and {Scaringi}, Simone and {Elias-Rosa}, Nancy},
        title = "{ESO Expanding Horizons: Underluminous Thermonuclear Supernovae}",
      journal = {arXiv e-prints},
         year = 2025,
        month = dec,
          eid = {arXiv:2512.17404},
        pages = {arXiv:2512.17404},
          doi = {10.48550/arXiv.2512.17404},
archivePrefix = {arXiv},
       eprint = {2512.17404},
 primaryClass = {astro-ph.IM},
       adsurl = {https://ui.adsabs.harvard.edu/abs/2025arXiv251217404G}
}
\end{document}